\documentclass{jfm}
\usepackage{graphicx}
\usepackage{float}
\usepackage{rotating}
\usepackage[flushleft]{threeparttable}
\usepackage{epstopdf,epsfig}
\usepackage{lscape}
\usepackage{amsmath}
\usepackage{mathtools}
\usepackage{tabularx}
\usepackage{tikz}
\usepackage{centernot}
\usepackage{booktabs}
\usepackage{comment} 
\usepackage{subcaption}
\usepackage{siunitx}
\usepackage{lineno}
\usepackage{units}
\usepackage{nicefrac,xfrac}
\usepackage{soul}
\usepackage{ulem} 
\usetikzlibrary{shapes.geometric}
\usepackage{subcaption}
\usepackage{hyphenat}
\usepackage{multicol}
\usepackage{multirow}
\usepackage{xr}
\usepackage{bm}

\usepackage{enumitem}
\newlist{myenumi}{description}{10}
\setlist[myenumi]{leftmargin=23pt,itemsep=2pt,topsep=2pt,parsep=2pt}
\newlist{myenumi2}{description}{10}
\setlist[myenumi2]{leftmargin=40pt,itemsep=2pt,topsep=2pt,parsep=2pt}

\newcommand{\LL}{\textsf{L}}
\newcommand{\HH}{\textsf{H}}
\newcommand{\WW}{\textsf{W}}
\newcommand{\II}{\textsf{I}}
\newcommand{\TT}{\textsf{T}}

\renewcommand{\d}{\mbox{d}}
\providecommand\bnabla{\boldsymbol{\nabla}}
\newcommand{\uu}{\textbf{u}}
\newcommand{\xx}{\textbf{x}}

\newcommand{\flame}{%
  \begin{tikzpicture}[scale=0.015ex]
    \fill[color=black]
      (0,0) .. controls (-1.5,1.25) and (.5,2) .. (-.2,4)
            .. controls (1,2.5) and (2,.5) .. (0,0);
  \end{tikzpicture}
}

\newcommand{ \heat}{\!\flame\!}

\newcommand\AL[1]{{\color{black}#1}}

\shorttitle{Exchange flows in inclined ducts}
\shortauthor{A. Lefauve \& P. F. Linden}

\title{\Large{Buoyancy-driven exchange flows in inclined ducts}}

\author{Adrien Lefauve$^1$ and P. F. Linden$^1$}
\affiliation{$^1$Department of Applied Mathematics and Theoretical Physics, University of Cambridge \\ Centre for Mathematical Sciences, Wilberforce Road, Cambridge, CB3 0WA, UK.}

\begin{document}

\maketitle



\begin{abstract}

Buoyancy-driven exchange flows arise in the natural and built environment wherever bodies of fluids at different densities are connected by a narrow constriction.  In this paper we study these flows in the laboratory using the canonical stratified inclined duct experiment, which sustains an exchange flow  in an inclined duct of rectangular cross-section over long time periods (Meyer \& Linden, \textit{J. Fluid Mech.}, vol.~753, 2014).  We study the behaviour of these sustained stratified shear flows by focusing on three dependent variables of particular interest: the qualitative flow regime (laminar, wavy, intermittently turbulent, or fully turbulent), the mass flux (net transport of buoyancy between reservoirs), and the interfacial thickness (thickness of the layer of intermediate density between the two counter-flowing layers).  Dimensional analysis reveals five non-dimensional independent input parameters: the duct aspect ratios in the longitudinal direction $A$ and spanwise direction $B$, the tilt angle $\theta$, the Reynolds number $Re$ (based on the initial buoyancy difference driving the flow), and the Prandtl number $Pr$ (we consider both salt and temperature stratifications). After reviewing the literature and open questions on the scaling of regimes, mass flux, and interfacial thickness with $A,B,\theta,Re,Pr$, we present the first extensive, unified set of experimental data where we varied systematically all five input parameters and measured all three output variables with the same methodology. Our results in the $(\theta,Re)$ plane for five sets of $(A,B,Pr)$ reveal a variety of scaling laws, and a non-trivial dependence of all three variables on all five parameters, in addition to a sixth elusive parameter. We further develop three classes of candidate models to explain the observed scaling laws: (i) the recent volume-averaged energetics of Lefauve, Partridge \& Linden, \textit{J. Fluid Mech.}, 2019; (ii) two-layer frictional hydraulics; (iii) turbulent mixing models. Their limited quantitative success highlight the need for theoretical progress on shear-driven turbulent flows and their interfacial waves, layering, intermittency, and mixing properties.

\end{abstract}



\section{Introduction}\label{sec:intro}

Buoyancy-driven exchange flows naturally arise where relatively large bodies of fluid have different densities on either side of a relatively narrow constriction. In a gravitational field, this difference in buoyancy, usually in the horizontal direction, results in a horizontal hydrostatic pressure gradient along the constriction, of opposite sign above and below a `neutral level', a height at which the pressures on either side of the constriction are equal. This pressure gradient drives a counter-flow through the constriction, in which fluid from the negatively-buoyant reservoir flows below the neutral level towards the positively-buoyant reservoir, and conversely, with equal magnitude. Such buoyancy-driven exchange flows result in little to no net volume transport, but crucially, in a \emph{net buoyancy transport} between the reservoirs which tends to homogenise buoyancy differences in the system (i.e. towards equilibrium). In addition, \emph{irreversible mixing} often occurs across the interface between the two counter-flowing layers of fluid, creating an intermediate layer of partially mixed fluid, and partially reducing the buoyancy transport.
The net transport and mixing of the active scalar field (e.g. heat, salt, or other solutes) and of other potential passive scalars fields having different concentrations in either reservoirs (e.g. pollutants or nutrients) have a wide range of consequences of interest. 
For this reason, the study of buoyancy-driven exchange flows has a rich history. (The primary role of buoyancy being implicit throughout the paper, we will simply refer to these flows as `exchange flows'.)

Aristotle offered the first recorded explanation of the movement of salty water within the Mediterranean Sea \citep[pp.~8-9]{deacon_scientists_1971}. Ever since,  exchange flows through the straits of Gibraltar and the Bosphorus have driven much speculation and research, due to their crucial roles in the water and salt balances of the Mediterranean Sea, countering its evaporation by net volume transport and allowing its very existence (as first demonstrated experimentally by Marsigli in the 1680s \citep[Chap.~7]{deacon_scientists_1971}). More recently, it has been recognised that nutrient transport from the Atlantic partially supported primary production in Mediterranean ecosystems \citep{estrada_primary_1996}. The quantification, modelling, and discussion of the past and current impact of exchange flows in straits, estuaries, or between lakes continues to generate a vast literature (as an example, the dozens of papers with titles containing `exchange' and `Gibraltar' have attracted over 1,000 citations in total.)

Exchange flows of \emph{gases} also have a great variety of perhaps even more tangible and ancient applications to society in the `natural ventilation' of buildings  \citep{linden_fluid_1999}. It would be surprising indeed if some ice-age prehistoric \textit{Homo Sapiens} did not ponder the inflow of cold outside air and the outflow of heat or fire combustion products when choosing a cave suitable for living. More recently, engineering problems of air flow through open doorways or ventilation ducts, or the escape of gases from ruptured industrial pipes, have stimulated further research. 

More fundamentally, exchange flows are stably-stratified shear flows, a canonical class of flows widely used in the mathematical study of stratified turbulence, dating back at least to  \cite[\S~12]{reynolds_experimental_1883} and \cite{taylor_effect_1931}. Multi-layered stratified shear flows have complex hydrodynamic stability and turbulent mixing properties \citep{caulfield_multiple_1994,peltier_mixing_2003}. The  straightforward and steady forcing of exchange flows make them ideal laboratory stratified shear flows because of the ability to sustain, over long time periods, high levels of turbulent intensity and mixing representative of large-scale natural flows. 

The aim of this paper is to carry out a thorough review and exploratory study of buoyancy-driven exchange flows in inclined ducts. To do this, we will focus on the behaviour of three key variables: 
\begin{myenumi}
\item[\textnormal{(i)}\hspace{2.6ex}] the qualitative flow regime (e.g. laminar, wavy, intermittently or fully turbulent);
\item[\textnormal{(ii)}\hspace{2ex}]  the mean buoyancy transport; 
\item[\textnormal{(iii)}\hspace{1.5ex}] the mean thickness of any potential interfacial mixing layer.
\end{myenumi}

The above three variables are particularly relevant in applications  to predict exchange rates of active or passive scalars (e.g. salt, heat, pollutants, nutrients) between two different fluid bodies (e.g. rooms in a building,  seas or lakes on either sides of a strait).

However, our primary motivation in this paper is to contribute to a larger research effort into the fundamental properties of turbulence in sustained stratified shear flows of geophysical relevance. The above three variables have thus been chosen for their particular ability to be readily captured by simple laboratory techniques while encapsulating several key flow features that are currently the subject of active research, such as: interfacial `Holmboe'  waves \citep{salehipour_turbulent_2016,lefauve_structure_2018}; spatio-temporal turbulent intermittency \citep{kops_classical_2015,portwood_robust_2016,taylor_new_2016}; and layering and mixing \citep{salehipour_diapycnal_2015,zhou_diapycnal_2017,lucas_layer_2017,salehipour_self_2018}.

To achieve this aim, the remainder of the paper is organised as follows. In \S~\ref{sec:experiment} we introduce a canonical experiment ideally suited to study the rich dynamics of exchange flows, and analyse the \emph{a priori} importance of its non-dimensional input parameters. In \S~\ref{sec:review} we review the relevant experimental literature on the scaling laws of our variables of interest and highlight the limitations of previous research in order to motivate our study. In \S~4 we present our experimental results and scaling laws. In \S~5 we explain some of these results with a variety of models, and we conclude in \S~6.

\section{The experiment} \label{sec:experiment}

\subsection{Setup and notation}\label{sec:setup}

\begin{figure}
    \centering
        \includegraphics[width=0.7\textwidth]{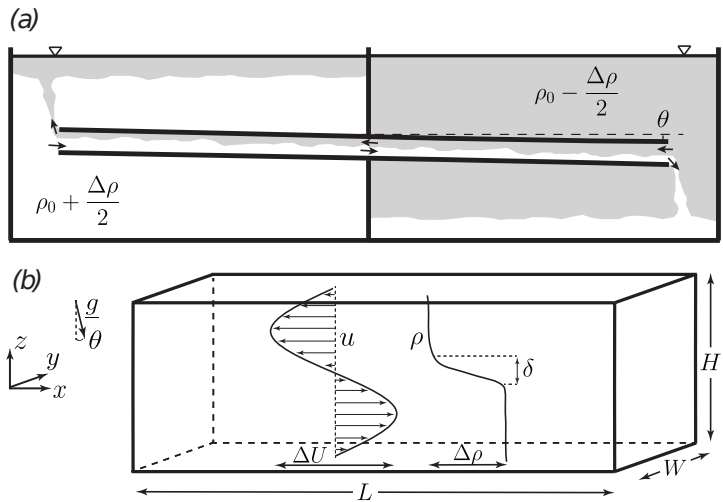}
    \caption{\emph{(a)} The Stratified Inclined Duct (SID), in which an exchange flow takes place through a rectangular duct connecting two reservoirs at densities $\rho_0\pm \Delta\rho/2$ and inclined at an angle $\theta$ from the horizontal. \emph{(b)} Notation (in dimensional units). The $x$ and $z$ axes are respectively aligned with the horizontal and vertical of the duct (hence $-z$ makes an angle $\theta$ with gravity, here $\theta>0$). The duct has dimensions $L \times W \times H$. The streamwise velocity $u$ has typical peak-to-peak magnitude $\Delta U$. The density stratification $\rho$ has magnitude $\Delta \rho$, with an interfacial layer of typical thickness $\delta$. }
  \label{fig:setup}
\end{figure}

The stratified inclined duct experiment (hereafter abbreviated `SID') is sketched in figure \ref{fig:setup}\emph{(a)}. This conceptually simple experiment consists of two reservoirs initially filled with aqueous solutions of different densities $\rho_0 \pm \Delta \rho/2$, connected by a long rectangular duct that can be tilted at an angle $\theta$ from the horizontal. At the start of the experiment, the duct is opened, initiating a brief transient gravity current. Shortly after, at $t=0$, an exchange flow starts and is sustained through the duct for long periods of time, until the accumulation of fluid of a different density from the other reservoir reaches the ends of the duct and the experiment is stopped at $t=T$ (typically after several minutes and many duct transit times). This exchange flow has at least four qualitatively different \emph{flow regimes}, based on the experimental parameters, as we discuss in more detail in \S~\ref{sec:review-reg}. 

Our notation is shown in figure \ref{fig:setup}\emph{(b)} and largely follows that of \cite{lefauve_structure_2018,lefauve_regime_2019}.  The duct has length $L$, height $H$, and width $W$. The streamwise $x$ axis is aligned along the duct, the spanwise $y$ axis is aligned across the duct, making the $z$ axis tilted at an angle $\theta$ from the vertical (resulting in a non-zero streamwise projection of gravity $g\sin\theta$). The angle $\theta$ is defined to be positive when the bottom end of the duct sits in the reservoir of lower density, as shown here. The velocity vector field is $\uu(x,y,z,t)=(u,v,w)$ along $x,y,z$, and the density field is $\rho(x, y, z, t)$. All spatial coordinates are centred in the middle of the duct, such that $(x,y,z,t) \in [-L/2,L/2]\times [-W/2, W/2]\times [-H/2,H/2]\times [0,T]$.

Next, we define two integral scalar quantities of particular interest in exchange flows:
\begin{myenumi}
\item[\textnormal{(i)} \hspace{0.6ex}  $Q$\hspace{0.5ex}]  the \emph{volume flux} as the volumetric flow rate averaged over the duration of an experiment
\begin{equation}  
Q \equiv \langle |u| \rangle_{x,y,z,t} \, , \label{definition_volume_flux}
\end{equation} 
where $\langle |u| \rangle_{x,y,z,t} \equiv 1/(LWHT)\int_{0}^{T}\int_{-H/2}^{H/2}\int_{-W/2}^{W/2}\int_{-L/2}^{L/2} |u| \, \d x \, \d y \, \d z \, \d t$. The volume flux $Q>0$ measures the magnitude of the exchange flow between the two reservoirs. It is different from the \emph{net} (or `barotropic') volume flux $\langle u \rangle_{x,y,z,t} \approx 0$, since, to a good approximation, the volume of fluid in each reservoirs is conserved during an experiment (assuming the levels of the free surface in each reservoir are carefully set before the start of the experiment).

\item[\textnormal{(ii)}  \hspace{0.2ex} $Q_m$ \hspace{0ex}] the \emph{mass flux} as the net flow rate of mass averaged of the duration of an experiment
\begin{equation}  
Q_m \equiv \frac{2}{\Delta \rho}\langle (\rho-\rho_0) u \rangle_{x,y,z,t} \, ,\label{definition_mass_flux}
\end{equation}
which is equivalent to a \emph{buoyancy flux} up to a multiplicative constant $g$. By definition $0 < Q_m \le Q$. The first inequality holds since, in our notation, negatively-buoyant fluid ($\rho_0< \rho \le \rho_0 + \Delta \rho/2$) flows on average to the right ($u>0$) and conversely. The second inequality would be an equality in the absence of molecular diffusion inside the duct (i.e. if all fluid moving right had density $\rho_0+\Delta \rho/2$ and conversely). In any real flow, laminar (and potentially turbulent) diffusion at the interface are responsible for an interfacial layer of intermediate density $|\rho-\rho_0| <\Delta \rho/2$ of finite thickness $\delta >0$ (figure~\ref{fig:setup}\emph{(b)}). 
\end{myenumi}

\subsection{Non-dimensionalisation}

A total of seven parameters are believed to play important roles in the SID: four geometrical parameters: $L$, $H$, $W$, $\theta$, and three dynamical parameters: the reduced gravity $g' \equiv g \Delta \rho/\rho_0$ (under the Boussinesq approximation $0<\Delta \rho / \rho_0 \ll 1 $),  the kinematic viscosity of water ($\nu = 1.05\times 10^{-6}$~m${}^2$~s$^{-1}$) and the molecular diffusivity of the stratifying agent (active scalar) $\kappa$. In this paper, we will primarily consider salt stratification ($\kappa_S = 1.50\times 10^{-9}$~m${}^2$~s$^{-1}$), but will also discuss temperature stratification ($\kappa_T = 1.50\times 10^{-7}$~m${}^2$~s$^{-1}$). From these seven parameters having two dimensions (of length and time), we construct five independent non-dimensional parameters below.

The first three non-dimensional parameters are geometrical: $\theta$, and the aspect ratios of the duct in the longitudinal and spanwise direction, respectively:
\begin{equation}\label{definition-A-B}
A \equiv \frac{L}{H} \quad \textrm{and} \quad B \equiv \frac{W}{H},
\end{equation}
We choose to non-dimensionalise lengths by the length scale $H/2$, defining the non-dimensional position vector as $\tilde{\xx} \equiv \xx/(H/2)$ such that $(\tilde{x}, \tilde{y},\tilde{z})\in [-A,A]\times[-B,B]\times[-1,1]$. As an exception, we choose to non-dimensionalise the typical thickness of the interfacial density layer by $H$, for consistency with other definitions in the literature: $\tilde{\delta}\equiv \delta/H$,  such that $\tilde{\delta} \in [0,1]$. 

The last two non-dimensional parameters are dynamical. We define an `input' Reynolds number based on the velocity scale $\sqrt{g'H}$ and length scale $H/2$:
\begin{equation} \label{definition_Re}
Re \equiv \frac{\sqrt{g'H}H}{2\nu} = \frac{\sqrt{gH^3}}{2\nu}\sqrt{\frac{\Delta \rho}{\rho_0}}.
\end{equation}
Consequently, we non-dimensionalise the velocity vector as $\tilde{\uu} \equiv \uu /\sqrt{g'H}$, and time by the advective time unit $\tilde{t}\equiv 2\sqrt{g'/H}t$ (hereafter abbreviated ATU). We define our last parameter, the Prandtl number, as the ratio of the momentum to active scalar diffusivity:
\begin{equation} \label{definition_Pr}
Pr \equiv \frac{\nu}{\kappa}.
\end{equation}
where $\kappa$ takes the value $\kappa_S$ or $\kappa_T$ depending on the type of stratification (salt or temperature, giving respectively $Pr=700$ and $Pr=7$). Finally, we define the non-dimensional Boussinesq density field as $\tilde{\rho} \equiv (\rho - \rho_0)/(\Delta \rho /2)$, such that  $\tilde{\rho} \in [-1,1]$. 

We now reformulate the aim of this paper (introduced in \S~\ref{sec:intro}) more specifically as: exploring the behaviour of flow regimes, mass flux $\tilde{Q}_m$, and interfacial layer thickness $\tilde{\delta}$ in the five-dimensional space of non-dimensional input parameters $(A,B,\theta,Re,Pr)$.

In the next section we address the dimensional scaling of the velocity in the experiment. By discussing the \emph{a priori} influence of the input parameters identified above on the velocity scale in this problem, we will provide a basis for subsequent scaling arguments in the paper.

\subsection{Scaling of the velocity}\label{sec:scaling-of-vel}

Having constructed our Reynolds number \eqref{definition_Re} using the velocity scale $\sqrt{g'H}$, here we show that it is  the relevant velocity scale to use in such exchange flows. As sketched in figure~\ref{fig:setup}\emph{(b)}, we define the typical peak-to-peak velocity as $\Delta U$.  This velocity scale is not set by the experimenter as an input parameter, rather it is chosen by the flow as an output parameter. From dimensional analysis, we  write 
\begin{equation}
    \frac{\Delta U}{2} = \sqrt{g'H}f_{\Delta U}(A,B,\theta,Re,Pr).
\end{equation} 
In order to show that our Reynolds number \eqref{definition_Re} and our non-dimensionalisation the velocity by $\sqrt{g'H}$ are relevant (and such that $\tilde{u} \in [-1,1]$), we will show below that we indeed expect $\Delta U/2 \sim \sqrt{g'H}$ and $f_{\Delta U}(A,B,\theta,Re,Pr) \sim 1$. Although some aspects of this discussion can be found in \cite{lefauve_structure_2018,lefauve_regime_2019}, the importance of this dimensional analysis for this paper justifies the more detailed discussion that we offer below.

The velocity scale $\Delta U$ in quasi-steady state results from a dynamical balance in the steady, horizontal momentum equation under the Boussinesq approximation (in dimensional units)
\begin{equation}\label{momentum_x_total}
\underbrace{ \vphantom{\frac{\rho-\rho_0}{\rho_0} g \sin \theta} \uu \cdot \bnabla u}_{\mathrm{inertial \, (I)}} = \underbrace{\vphantom{\frac{\rho-\rho_0}{\rho_0} g \sin \theta} -(1/\rho_0) \p_x p}_{\mathrm{hydrostatic \, (H)}} +  \underbrace{ \vphantom{\frac{\rho-\rho_0}{\rho_0}} g \sin \theta (\rho-\rho_0)/\rho_0}_{\mathrm{gravitational  \, (G)}} + \underbrace{\vphantom{\frac{\rho-\rho_0}{\rho_0} g \sin \theta} \nu \bnabla^2 u}_{\mathrm{viscous \, (V)}},
\end{equation}
In addition to the standard inertial (I) and viscous (V) terms, this equation highlights the two distinct `forcing' mechanisms in SID flows:

\begin{myenumi}
\item[\textnormal{(H)}\hspace{2.0ex}] a \emph{hydrostatic} longitudinal pressure gradient, the minimal ingredient for exchange flow, resulting from each end of the duct sitting in reservoirs at different densities. This hydrostatic pressure in the duct increases linearly with depth $\p_x p = g\cos\theta\Delta \rho/(4L) z$, driving a flow in opposite directions on either side of the neutral level $z=0$: $-(1/\rho_0)\p_x p = g'\cos\theta /(4L)z$;

\item[\textnormal{(G)}\hspace{1.9ex}] a \emph{gravitational}  body force reinforcing the flow by the the acceleration of the positively-buoyant layer upward (to the left in figure~\ref{fig:setup}) and of the negatively-buoyant layer downward (to the right) when the tilt angle is positive $g\sin\theta>0$ (the focus of this paper), and conversely when the tilt angle is negative.
\end{myenumi}
Rewriting \eqref{momentum_x_total} in non-dimensional form and ignoring multiplicative constants, we obtain
\begin{equation}\label{momentum_x_non-dim}
\underbrace{\vphantom{\frac{(\Delta U)^2 }{L}} (\Delta U)^2 \, \tilde{\uu}\cdot \tilde{\bnabla}\tilde{u}}_{\mathrm{I}} 
\ \sim \ \underbrace{\vphantom{\frac{(\Delta U)^2}{L}} ( g' H \cos \theta) \tilde{z}}_{\mathrm{H}} \ + \ \underbrace{\vphantom{\frac{(\Delta U)^2}{L}} (g'L \sin \theta) \tilde{\rho}}_{\mathrm{G}} \ + \ \underbrace{\vphantom{\frac{(\Delta U)^2}{L}} (\nu \Delta U \ell^{-2} L) \tilde{\bnabla^2}\tilde{u}}_{\mathrm{V}},
\end{equation}
where $\ell$ is the smallest length scale of density gradients ($\ell =\delta$ in laminar flows, and $\ell \ll \delta$ in turbulent flows).

To simplify this complex `four-way' balance, it is instructive to consider the four possible `two-way' dominant balances to deduce four possible  scalings for $\Delta U$ (ignoring constants and assuming $\cos \theta \approx 1$ since the focus of this paper is on small angles).
\begin{myenumi}
\item[\textnormal{(IH)}\hspace{1.0ex}] \emph{The inertial-hydrostatic  balance}. First, we can neglect the gravitational (G) term  with respect to the hydrostatic (H) term if $g'H\cos\theta \gg g'L\sin \theta$, i.e. when the tilt angle of the duct $\theta$ is much smaller than its `geometrical' angle:
    \begin{equation}
        0<\theta \ll \alpha 
    \end{equation}
    where we define the geometrical angle as
        \begin{equation} \label{def-alpha}
     \alpha \equiv \tan^{-1}(A^{-1})
    \end{equation}
    Second, we can neglect the viscous (V) term if $g'H \gg \nu \Delta U \ell^{-2}L$, i.e. if the Reynolds number is larger than  $Re \gg HL/\ell^2$. This corresponds to 
        \begin{equation}
        Re \gg A
    \end{equation}
    in laminar flow (ignoring the case $B \ll 1$ for simplicity), and to a larger lower bound in turbulent flows.
    Under these conditions, balancing I and H gives the scaling $\Delta U \sim \sqrt{g'H}$, i.e. $f_{\Delta U} \sim 1$, which corresponds to our choice in \S~\ref{sec:setup}. 

\item[\textnormal{(IG)}\hspace{1.3ex}] \emph{The  inertial-gravitational balance}. Using analogous arguments, if $\theta \gg \alpha$ and $Re \gg HL/\ell^2$, we expect the scaling $\Delta U \sim \sqrt{g'L\sin\theta}$, i.e. $f_{\Delta U}(A,\theta) \sim \sqrt{A\sin\theta} \gg 1$. 

\item[\textnormal{(HV)}\hspace{1ex}] \emph{The  hydrostatic-viscous  balance}. If $\theta \ll \alpha$ and $Re \ll A$, we expect  $f_{\Delta U}(A,B,Re) \sim A^{-1}Re \ll 1$ (some dependence on $B$ being unavoidable in such a viscous flow).

\item[\textnormal{(GV)}\hspace{1.0ex}] \emph{The  gravitational-viscous balance}. If $\theta \gg \alpha$ and $Re \ll A$, we expect  $f_{\Delta U} (B,\theta,Re) \sim \sin\theta Re \ll A$.
    
\end{myenumi}

\begin{figure}
    \centering
        \includegraphics[width=0.70\textwidth]{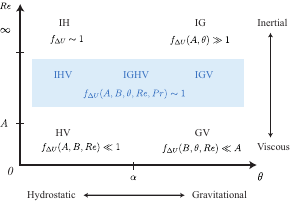}
    \caption{Summary of the scaling analysis of $\Delta U$  based on the four two-way dominant balances of the streamwise momentum equation \eqref{momentum_x_non-dim}. In each corner of the $(\theta,Re)$ plane, the IH, IG, HV, and GV scalings predict $f_{\Delta U}\equiv\Delta U/(2\sqrt{g'H})$ on either extreme side of $\theta = \alpha \equiv \tan^{-1}(A^{-1})$ and $Re=  A$. The region of practical interest studied in this paper is shown in blue. Although no \emph{a priori} `two-way' balance allows us to determine accurately the scaling of $f_{\Delta U}(A,B,\theta,Re,Pr)$ in this region, the theory of hydraulic control and empirical data show that $f_{\Delta U}\sim 1$, as in the IH scaling (see text). }
  \label{fig:balance}
\end{figure}

Figure~\ref{fig:balance} summarises the above analysis and the following conclusions.

\begin{myenumi}
\item[\textnormal{(i)}\hspace{2.6ex}] The parameters $A$, $\theta$ and $Re$ play particularly important roles in SID flows, since the variation of $\theta$ and $Re$ above or below thresholds set by $A$ can alter the scaling of $\Delta U$ (i.e. $f_{\Delta U}$). The parameter $B$ appears less important in this respect (except in narrow ducts where $B \ll 1$ and the $Re$ threshold becomes $AB^{-2}$). 
\item[\textnormal{(ii)}\hspace{2ex}]   At low tilt angles $0<\theta \ll \alpha$, $f_{\Delta U}$ increases from $\ll 1$ when $Re \ll A$ to $\sim 1$ when  $Re \gg A$. At high enough $Re$, $f_{\Delta U}$ likely retains a dependence on $A,B,Re$ due to turbulence (the constant `IH' scaling being a singular limit for $Re\rightarrow \infty$).
\item[\textnormal{(iii)}\hspace{1.5ex}] At high tilt angles $\theta \gg \alpha$ and Reynolds number $Re\gg A$, $f_{\Delta U}$ \emph{should} increase well above $1$, and likely retains a dependence on $A,B,\theta, Re$ (the `IG' scaling being a singular limit for $Re\rightarrow \infty$).
\item[\textnormal{(iv)}\hspace{1.5ex}] The blue rectangle in figure~\ref{fig:balance} represents the region of interest in most exchange flows of practical interest and in this paper. In this region, three or four physical mechanisms must be considered simultaneously (IHV, IGHV or IGV). Since few flows satisfy $\theta \ll \alpha$ or $\gg \alpha$, we consider that in general $f_{\Delta U} =f_{\Delta U}(A,B,\theta,Re,Pr)$ (the $Pr$ dependence reflects the fact that the active scalar can no longer be neglected at high $Re$ due to its effect on turbulence and mixing).
\item[\textnormal{(v)}\hspace{2.0ex}] Although the above `two-way' balances do not allow us to confidently guess the scaling of $f_{\Delta U}$ in the blue region, theoretical arguments and empirical evidence support $f_{\Delta U}(A,B,\theta,Re,Pr) \sim 1$ for IHV, IGHV and IGV flows. This in agreement  with the IH scaling and with (ii), but in disagreement with (iii). 
\end{myenumi}

The last conclusion (v) is supported by the theoretical concept of \emph{hydraulic control} of two-layer exchange flows, which dates back to \cite{stommel_control_1953,wood_selective_1968,wood_lock_1970} and was formalised mathematically by \cite{armi_hydraulics_1986,lawrence_hydraulics_1990,dalziel_two-layer_1991}. In steady, inviscid, irrotational, hydrostatic (i.e. `IH') exchange flows, the `composite Froude number' $G$ is unity, which using our notation and assuming streamwise invariance of the flow ($\p_x =0$), reads: 
\begin{equation}
   G^2 =  4\frac{\langle u^2 \rangle_{x,y,z,t}}{\sqrt{g'H}} = 1 \quad \Longrightarrow \quad \langle |\tilde{u}| \rangle_{x,y,z,t} = \tilde{Q} = \frac{1}{2}.
\end{equation}
Such exchange flows are called \emph{maximal}: the phase speed of long interfacial gravity waves $\sqrt{g'H}$ `controls' the flow at sharp changes in geometry (on either ends of the duct), and sets the maximal non-dimensional volume flux to $\tilde{Q}=\nicefrac{1}{2}$. 

In `plug-like' hydraulic flows  ($Re\rightarrow \infty$), the velocity in each layer $\Delta U/2$ is equal to its layer-average $Q$, giving an upper bound $f_{\Delta U}=\tilde{Q} =\nicefrac{1}{2}$. By contrast, in real-life finite-$Re$ flows, the peak $\Delta U/2$ is larger than the average $Q$ (typically by a factor $\approx 2$), such that the upper bound is $f_{\Delta U} \approx 2\tilde{Q} \approx 1$. This upper bound remains approximately valid throughout the blue region of figure~\ref{fig:balance}. We thus answer the question motivating this section: our choice of non-dimensionalising $\uu$ by $\sqrt{g'H}\approx \Delta U/2$ in order to have $|\tilde{\uu}| \lesssim 1$ is indeed relevant to SID flows.

Henceforth, we drop the tildes and, unless explicitly stated otherwise, use non-dimensional variables throughout.

\section{Literature review}\label{sec:review}

In this section we review the experimental literature on the questions of flow regimes (\S~\ref{sec:review-reg}), mass flux (\S~\ref{sec:review-Qm}), and interfacial layer thickness (\S~\ref{sec:review-delta}). For conciseness, we limit the discussion to the most relevant results of each study, and give further details about the geometry and parameters $A,B,\theta,Re,Pr$ used by each in appendix~\ref{sec:appendix_table} (table~\ref{tab:review} provides a synthesis of this literature review). We then highlight the main limitations of previous research to further motivate the paper in \S~\ref{sec:limitations}.

\subsection{Flow regimes} \label{sec:review-reg}

\cite{macagno_interfacial_1961} (MR61) constitutes, to our knowledge, the first experimental study in a setup similar to the SID. MR61 used dye visualisations to describe four qualitatively different regimes:
\begin{myenumi}
\item[\textnormal{L}\hspace{2.5ex}]  `uniform laminar motion with straight streamlines';
\item[\textnormal{W}\hspace{1.5ex}]  `laminar motion with regular waves';
\item[\textnormal{I}\hspace{3ex}] `incipient turbulence, with waves which break and start to show irregularity and randomness';
\item[\textnormal{T}\hspace{2.5ex}]  `pronounced turbulence and active mixing across the interface'
\end{myenumi}
MR61 mapped the above regimes together with measurements of the interfacial stress and mixing coefficients in the plane $(F_*,R_*)$, where    $F_* \approx 2\sqrt{2}Q$  and $R_* \approx 4QRe$ (using our notation) are `effective' Froude and Reynolds numbers. Arguing that flows above a certain  $R_*$ and $F_*$ would be unstable, they proposed and experimentally verified that the `transition curves' separating the flow regimes and the iso-curves of interfacial shear and mixing coefficients scaled with $R_* \, F_* \approx Re \, Q^2$ (i.e. these curves are $Re \,Q^2 =$~const.). As we have seen in \S~\ref{sec:scaling-of-vel}, $Q$ is in reality a dependent variable, not an input parameter. This confusion in MR61 comes from the fact that their setup (which they attribute to Helmholtz) differs from the SID in that they were (to some extent) able to prescribe the volume flux $Q$ by controlling the inflow of salt water by a piston communicating with one of the reservoirs (their system was closed, i.e. it had no free surface). They varied $Q$ together with $\theta$ in non-trivial ways in order to reach target values $R_*,F_*$. They did not appear to realise that the flow was hydraulically controlled and that $\theta$ and $Re$ were the relevant independent input parameters.

\cite{wilkinson_buoyancy_1986} (W86) used shadowgraph and observed regime transitions similar to MR61 in a horizontal, circular pipe: `shear-induced instabilities [...] initially in the form of cusp-like waves, but as the shear was further increased, Kelvin-Helmholtz billows were seen to grow and collapse creating a turbulent shear layer'. He suggested a scaling in $Re$ alone, independent of $A$: laminar flow under $Re<2450$, `interfacial waves radiating in both directions' for $Re\in[2600,2700]$, and turbulence for $Re>2700$, but his experiments were limited in number ($\approx 18$).

\cite{kiel_buoyancy_1991} (K91) used shadowgraph and laser sheet visualisations \AL{at larger $Re$}, and classified the regimes differently: laminar;  turbulent with $\delta<1$; and turbulent with $\delta=1$. Using a semi-empirical model that we will describe in \S~\ref{sec:review-Qm}, he proposed regime transitions scaling with $A\tan\theta$, independently of $Re$, i.e. the opposite of W86.

\cite{meyer_stratified_2014} (ML14) used shadowgraph visualisations, and (unaware of MR61) described the following four regimes (see their figure~3): 
\begin{myenumi}
\item[\textnormal{$\LL$}\hspace{3.8ex}]  laminar flow with a thin, flat density interface; 
\item[\textnormal{$\HH$}\hspace{3.5ex}]   mostly laminar flow with quasi-periodic waves on the density interface, identified as Holmboe waves;
\item[\textnormal{$\II$}\hspace{4.5ex}] spatio-temporally intermittent turbulence with small-scale structures and noticeable mixing between the two layers; 
\item[\textnormal{$\TT$}\hspace{3.5ex}] statistically-steady turbulent flow with a thick interfacial density layer.
\end{myenumi}
This is the nomenclature adopted by LPL19 and that we adopt in this paper.
The only difference between the MR61 and ML14 nomenclatures lies in the letter characterising the wavy regime (W in MR61 and $\HH$ in ML14), simply because MR61 observed Holmboe waves before they were explained by \cite{holmboe_behavior_1962}. 
ML14 mapped these regimes in the $(\theta, Re)$ plane for two different $A=15,\, 30$.  They argued that, because the flow was hydraulically controlled, the `excess kinetic energy' gained by the flow at $\theta>0$ should be dissipated turbulently (we identify this energy as the square of the `IG' velocity scaling $g'L\sin\theta$ of \S~\ref{sec:scaling-of-vel}). By non-dimensionalising this excess `IG' energy by $(\nu/H)^2$, ML14 proposed and verified experimentally that regime transitions scale with the Grashof number
\begin{equation} \label{definition-Gr}
Gr \equiv \frac{g'L\sin\theta}{(\nu/H)^2} \approx 4A\theta Re^2, 
\end{equation}
assuming $\sin \theta \approx \theta$. This scaling has two limitations: the `IG' energy  does not explain the transitions at $\theta=0$, and its non-dimensionalisation by $(\nu/H)^2$ lacks a physical basis.

\cite{lefauve_regime_2019} (LPL19) repeated the shadowgraph observations of ML14 in a smaller duct ($H=45$~mm vs $H=100$~mm) with otherwise equal parameters $(A,B,Pr)=(30,1,700)$ and mapped the regimes in the $(\theta,Re)$ plane. Surprisingly, their observations do not support the $Gr=4\times 10^7$ curve proposed by ML14 for the $\II$ to $\TT$ transition. This suggests that $H$ plays a major role other than through the non-dimensional parameters $A,B,Re$, and therefore the existence of at least an additional non-dimensional parameter that we have hitherto overlooked. LPL19 observed two distinct scalings: a $\theta Re^2 $ scaling for $\theta \lesssim \alpha$, in agreement with ML14), and a $\theta Re $ scaling for $\theta \gtrsim \alpha$, which they explained using a physical model. They developed from first principles detailed energy budgets which they applied to 16 experiments in which the full density field and three-component velocity field were simultaneously measured in a three-dimensional volume of the duct at high spatio-temporal resolution (for visualisations of flow fields in all four regimes, see their figures~2-3). They showed, theoretically and experimentally, that for $\theta \gtrsim \alpha$ (for so called `forced flows'), the  time- and volume-average rate of dissipation of kinetic energy could be predicted {\it a priori} as
\begin{equation} \label{definition-s2}
\langle \mathsf{s}_{ij}\mathsf{s}_{ij}\rangle_{x,y,z,t}  \approx \frac{1}{16}\theta Re,
\end{equation}
where $\mathsf{s}$ is the non-dimensional strain rate tensor and $\sin\theta \approx \theta$ was assumed since their focus was on small angles.  They showed that because the magnitude of streamwise velocities and wall shear stresses were bounded by hydraulic control, the requirement of high strain rates at high $\theta Re$ caused transitions to increasingly three-dimensional (turbulent) flow regimes. The $\theta Re$ scaling of energy dissipation matched the observed  regime transitions in `forced flow' ($\theta \gtrsim \alpha$), but the $\theta Re^2$ transition scaling in `lazy flows' ($\theta \lesssim \alpha$ and in particular $\theta=0$) remains unexplained.  

In summary, there is good consensus in the literature on the nomenclature of flow regimes observed in the SID and, in particular, on the fact that the flow becomes increasingly disorganised and turbulent with $A$, $\theta$ and $Re$. At a fixed $\theta \ge 0^\circ$, all flow regimes ($\LL,\HH,\II,\TT$) can be visited by increasing $Re$, and conversely at fixed $Re$ and increasing $\theta$ (W86, K91, ML14, LPL19). Both K91 and ML14 observed regime transitions scaling with $A\tan\theta=\tan\theta/\tan\alpha$ (or $A\theta$ for small angles), i.e. $A$ controls the $\theta$ scaling. However, the scaling in $Re$ is subject to debate, and may change on either side of $\theta \approx \alpha$ (LPL19). These conclusions are illustrated schematically in figure~\ref{fig:lit_sum}\emph{(a)} (the interrogation marks denote open questions)

\subsection{Mass flux} \label{sec:review-Qm}

\cite{leach_experimental_1975} (LT75) measured $Q_m =0.23$ in horizontal circular pipes for high Reynolds number $Re=O(10^4-10^5)$, and $Pr=1$ and $700$ (respectively CO$_2$/air and salt/fresh water). Surprisingly, they observed no dependence on $A,Re,Pr$.

\cite{mercer_experimental_1975} (MT75) \AL{reported dramatic} non-monotonicity of $Q_m(A,\theta)$ (we reproduce some of their data in appendix~\ref{sec:kiel-data}, figure~\ref{fig:kiel}\emph{(b)}). They measured $Q_m \approx 0.2-0.3$ at $\theta = 0^\circ$ (in agreement with LT75), increasing to $Q_m \approx 0.4$ at $\theta \approx \alpha/2$, and decreasing to $Q_m \approx 0.01-0.1$ at $\theta=90^\circ$. \AL{In a   small additional set of experiments at $\theta=30^\circ$ in a larger pipe ($Re=2\times 10^4$ \emph{vs} $2\times10^3$, and $A=6$), they observed dependence on $Re$ even in the `very high' range $Re \in [300A,3000A]$ (though it might be due to subtle differences in apparatus).}

W86 developed a Bernoulli model in a horizontal circular pipe which predicted an upper bound of $Q=\pi/8\approx 0.39$ (non-dimensionalised using the pipe diameter), making the analogy with the hydraulic control arguments in \cite{wood_lock_1970} who predicted $Q=0.5$ in rectangular ducts. Including viscous boundary layers at the circular walls, he predicted and verified experimentally a monotonic increase of $Q$ with $A^{-1}Re$ (as the thickness of boundary layers decreases): $Q_m=0.13$ at $Re \approx 20A$ to $Q_m = 0.35$ at $Re \approx 500A$ (larger than LT75), in agreement with the dimensional analysis of \S~\ref{sec:scaling-of-vel} (conclusion (ii)). 

K91 developed an inviscid Bernoulli model in an inclined duct for two counter-flowing layers of equal thickness and predicted $Q \approx  \sqrt{(4/9)\cos\theta + A\sin\theta}$. In agreement with our dimensional analysis in \S~\ref{sec:scaling-of-vel}, this expression predicts a transition from an `inertial-hydrostatic' (IH) balance at $0<\theta \ll \alpha$ with $Q\approx 2/3$ to an `inertial-gravitational' (IG) balance at $\theta \gg \alpha$ with  $Q \approx \sqrt{A\sin\theta}$. \AL{K91 showed, however,  that this `IG' scaling could only be observed experimentally when communication and mixing between the two counter-flowing layers was artificially suppressed by a rigid `splitter plate' along the duct.} He argued that the non-realisation of the IG scaling  was due to a turbulent transition occurring when the IG scaling for $Q$ `that potentially exists'  exceeds a threshold dependent on the `stabilising effect of $g'\cos\theta$' (i.e. the maximal $Q=\nicefrac{1}{2}\sqrt{\cos\theta}$).  He summarised this argument in a `geometric Richardson number' $Ri_G$, whose inverse we interpret as being the square ratio of the `potential' to the maximal $Q$
\begin{equation} \label{definition-RiG}
Ri_G^{-1} \equiv \bigg( \frac{\sqrt{(4/9)\cos\theta + A\sin\theta}}{(1/2)\sqrt{\cos\theta}}\bigg)^2 =  \frac{16}{9} +4A\tan\theta = \frac{16}{9} + 4\frac{\tan\theta}{\tan\alpha}.
\end{equation}
K91 obtained a good empirical collapse of MT75's $Q_m$ data (figure~\ref{fig:kiel}\emph{(c)}), as well as of his own $Q_m$ data (figure~\ref{fig:kiel}\emph{(e-f)}) with $Ri_G$. The mass flux peaks at $Ri_G \approx 0.2$, i.e. $\theta \approx \alpha/2$, and decays at larger tilt angles as the flow regime is increasingly turbulent. \AL{In agreement with W86's arguments, K91 reported independence of his results with $Re$ above $Re>400A$ (figure~\ref{fig:kiel}\emph{(a})), and intentionally focused on these high $Re$ throughout.}

ML14 observed monotonic increase of $Q_m(\theta)$ with $Q_m\approx 0.2-0.3$ at $\theta=0^\circ$ and $Q_m\approx 0.5$ at $\theta=\alpha/2$. They did not comment on the hint of non-monotonic behaviour ($Q_m\lesssim 0.5$) suggested by their data at $\theta\approx2\alpha$.

LPL19 observed (in passing) clear non-monotonic behaviour of $Q_m(\theta,Re)$. Their data are well fitted by a hyperbolic paraboloid in the $\log \theta-\log Re$ plane, where $Q_m=$ const. curves are hyperbolas, with $Q_m\approx 0.5$ along the major axis $\theta Re^{3/2}=$~100 ($\theta$ in radians), and $Q_m$ decays on either side of it.

\begin{figure}
    \centering
        \includegraphics[width=0.9\textwidth]{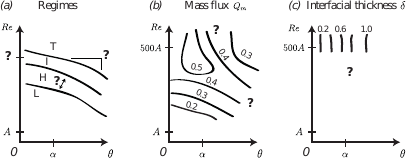}
    \caption{Synthesis of the literature review on the  \AL{idealised} behaviour of the \emph{(a)} flow regimes (\S~\ref{sec:review-reg}), \emph{(b)} mass flux  (\S~\ref{sec:review-Qm}), and  \emph{(c)} interfacial layer thickness (\S~\ref{sec:review-delta})  with respect to $A,\theta,Re$. The axes have logarithmic scale. Interrogation marks refer to open questions (see text for more details).}
  \label{fig:lit_sum}
\end{figure}

In summary, we can infer that the mass flux has a complicated non-monotonic behaviour in $A,\theta,Re$ as sketched in figure~\ref{fig:lit_sum}\emph{(b)}. While the dependence on $Re$ is clear at $Re<500A$ (MR61, W86, ML14 and LPL19) due to the influence of viscous boundary layers, it is still debated at $Re>500A$ (MT75 and ML14 argue in favour, K91 argues against). The mass flux $Q_m$ reaches a maximum  $Q_m \approx 0.4-0.5$ at $\theta\approx \alpha/2$ and `high enough' $Re$ (MT75, K91, ML14, LPL19), and decays on either side for smaller/larger $\theta$ and $Re$ (W86, LPL19) in a poorly-studied fashion.

\subsection{Interfacial layer thickness} \label{sec:review-delta}

K91 is, to our knowledge, the only thorough experimental study of $\delta(A,\theta)$. K91 performed conductivity probe measurements and observed monotonic increase with both $A$ and $\theta$ (figure~\ref{fig:kiel}\emph{(f)}) \AL{and independence on $Re$}. He found good collapse with $Ri_G$  (figure~\ref{fig:kiel}\emph{(g)}). The interfacial mixing layer is turbulent and thick ($\delta \approx 0.4-0.7$) when the maximal $Q_m$ is reached at $Ri_G \approx 0.2$ ($\theta \approx \alpha/2$), and it fills the whole duct height ($\delta=1$) at $Ri_G<0.1$ ($\theta \gtrsim 2\alpha$), see figure~\ref{fig:lit_sum}\emph{(c)}. At even larger tilt angles, the mean vertical density gradient $|\rho(z=1)-\rho(z=-1)|/2$ drops below 1 (this `extreme' turbulent scenario falls outside the scope of this paper). The behaviour of $\delta$ at low $Re<500A$ has not yet been studied.

\subsection{Limitations of previous research} \label{sec:limitations}

The above experimental literature (summarised in figure~\ref{fig:lit_sum}) provides much quantitative insight to complement the qualitative predictions of the scaling analysis in \S~\ref{sec:scaling-of-vel} (summarised in figure~\ref{fig:balance}). 

However, many aspects of the scaling of regimes, $Q_m$ and $\delta$ with $A,B,\theta,Re,Pr$ remain open questions (symbolised by the interrogation marks in figure~\ref{fig:lit_sum}). As extreme examples, the effects of $Re$ on $\delta$, and the effects of $B$ and $Pr$ on all three variables have not been studied at all.

Moreover, despite our above efforts to unify their findings, these past studies of the SID experiment inherently provide a somewhat fragmented view of the problem due to the following limitations (made clear by table~\ref{tab:review}):

\begin{myenumi}
\item[\textnormal{(i)}\hspace{2.6ex}]   they used slightly different setups and geometries (e.g. presence \emph{vs} absence of free surfaces in the reservoirs, rectangular ducts \emph{vs} circular pipes), and slightly different measuring methodologies (e.g. for $Q_m$);

\item[\textnormal{(ii)}\hspace{2ex}]  only one study (K91) addressed the interdependence of the three variables of interest (regime, $Q_m$, $\delta$), while the remaining studies measured either only regimes (MR61), only $Q_m$ (LT75, MT75), or both (ML14, LPL19);

\item[\textnormal{(iii)}\hspace{1.5ex}]  they focused on the variation of a single parameter (MR61), two parameters (W86, K91, LPL19), or at most three parameters (MT75, ML14) in which case the third parameter took only two different values;

\item[\textnormal{(iv)}\hspace{1.5ex}] they studied limited regions of the parameter space, and it is difficult to confidently interpolate results obtained by different setups in different regions (such as $Re<500A$ and $>500A$).
\end{myenumi}

The experimental results in the next section  attempt to overcome the above limitations by providing a more unified view of the problem.

\section{Experimental results} \label{sec:results}

In order to make progress on the scaling of flow regimes, $Q_m$ and $\delta$ with $A,B,\theta,Re,Pr$, we obtained a comprehensive set of experimental data using an identical setup, measuring all three dependent variables with the same methodology (described in appendix~\ref{sec:method}), and varying all five independent parameters in a systematic fashion. We introduce the different duct geometries and data sets used in \S~\ref{sec:ducts}, and present our results on flow regimes in \S~\ref{sec:results-reg}, on mass flux in \S~\ref{sec:results-Qm}, and on interfacial layer thickness in \S~\ref{sec:results-delta}.

\subsection{Data sets} \label{sec:ducts}

All experimental data presented in the following were obtained in the stratified inclined duct (SID) setup sketched in figure~\ref{fig:setup}. We used four different duct geometries and two types of stratification (salt and temperature) to obtain the following five distinct data sets, listed in table~\ref{tab:sid_geo_param}:

\begin{myenumi2}
\item[\textnormal{LSID}\hspace{3.5ex}]  (L for Large) with height $H=100$~mm, and $A=30$, $B=1$;

\item[\textnormal{HSID}\hspace{3.5ex}] (H for Half) which only differs from the LSID (the `control' geometry) in that it is half the length: $A=15$ (highlighted in bold in table~\ref{tab:sid_geo_param}); 

\item[\textnormal{mSID}\hspace{3.5ex}]  (m for mini) which only differs from the LSID in its height $H=\SI{45}{\milli\meter}$ (in bold), but \emph{should} yield identical data to LSID for identical $\theta,Re,Pr$ since $H$ \emph{should} only play a role through the non-dimensional parameters $A$, $B$ (identical in LSID and mSID) and $Re$. Our emphasis on `should' denotes the fact that we have already discussed in the literature review \S~\ref{sec:review-reg} that this was already observed \emph{not} to be the case: the $\II$ to $\TT$ regime transition curve of ML14 in LSID did not overlap with that of LPL19 in mSID;

\item[\textnormal{tSID}\hspace{4ex}]  (t for tall)  which differs from the HSID primarily in its tall spanwise aspect ratio  $B=\nicefrac{1}{4}$ (and, secondarily, in a marginally smaller height $H=90$~mm);

\item[\textnormal{mSID  \heat}\hspace{1ex}]  (m for mini and  \heat for temperature)  which differs from the mSID in that the stratification was achieved by different reservoir water temperatures (hence $Pr=7$), as opposed to different salinities in the above data sets (where $Pr=700$). This limited the density difference $\Delta \rho$ achieved, reflected in the lower $Re$.
\end{myenumi2}

Table~\ref{tab:sid_geo_param} also lists, for each data set, the range of variation of $\theta$ and $Re$, and the number of data points, i.e. distinct $(\theta,Re)$ couples for which we have data on regime, $Q_m$, and  $\delta$.

\begin{table}
\setlength{\tabcolsep}{4pt}
\centering
\caption{The five data sets used in this paper, using four duct geometries (abbreviated LSID, HSID, mSID, tSID) with different dimensional heights $H$, lengths $L=AH$ and widths $W=BH$, and two types of stratification (salt and temperature). We emphasise  in bold the resulting differences in the `fixed' non-dimensional parameters $A,B,Pr$ with respect to the `control' geometry (top row). We also emphasise the difference in $H$ between LSID and mSID, to test whether or not $H$ plays a role other than through the non-dimensional parameters $A,B,Re$. We also list the range  of $\theta,Re$ explored, and the number of regime, $Q_m$ and $\delta$ data points obtained in the $(\theta,Re)$ plane. Some of these data have been published or discussed in some form in ML14 (denoted by $^*$) and LPL19 (denoted by $^\dagger$) and are reused here with their permission for further analysis. Measurements of $Q_m$ and $\delta$ were not practical with heat stratification (hence the - symbol, see text for more details). Total: 886 individual experiments and 1545 data points. \vspace{0.3cm}}
\label{tab:sid_geo_param}
\renewcommand*{\arraystretch}{1.5}
\begin{tabular}{l  c c c c  c c c  c c c c}
\xdef\tempwidth{\the\linewidth}
 &  \multicolumn{2}{c}{Duct scale} & \multicolumn{3}{c}{Fixed params.} & \multicolumn{2}{c}{Varied params.} & \multicolumn{3}{c}{Number of data points} \\
\ \ \ Name &  $H$ &  Cross-section & $A$ & $B$ & $Pr$ & $\theta \, (^\circ)$ & $Re \, (\times 10^3)$  & regime & $Q_m$ & $\delta$  \\ 
\midrule
 LSID  & 100 & \begin{minipage}{0.07\textwidth}\begin{center}\begin{tikzpicture}
\draw[draw=black,line width=1pt] (11.1,5.5) rectangle ++(1,1);
\end{tikzpicture}\end{center}\end{minipage} &  30 & 1 & 700 & $[-1,4]$ & $[2,20]$   & 173$^*$ & 20$^*$ & 115 \\ \vspace{0.3cm}

 HSID  & 100 &  \begin{minipage}{0.07\textwidth}\begin{center}\begin{tikzpicture}
\draw[draw=black,line width=1pt] (11.1,5.5) rectangle ++(1,1);
\end{tikzpicture}\end{center}\end{minipage}  &  \textbf{15} & 1 & 700 & $[0,4]$ & $[1,20]$   & 74$^*$ & 34$^*$ & 58 \\ \vspace{0.3cm}

 mSID &  \textbf{45}  & \begin{minipage}{0.07\textwidth}\begin{center}
    \begin{tikzpicture}
\draw[draw=black,line width=1pt] (11.1,5.5) rectangle ++(0.45,0.45);
\end{tikzpicture}\end{center}\end{minipage}  & 30 & 1 & 700 & $[-1,6]$ & $[0.3,6]$  & 360$^\dagger$ & 162$^\dagger$ & 91 \\ \vspace{0.3cm}

 tSID &  90 &  \begin{minipage}{0.07\textwidth}\begin{center}\begin{tikzpicture}
\draw[draw=black,line width=1pt] (11.1,5.5) rectangle ++(0.225,0.9);
\end{tikzpicture}\end{center}\end{minipage}  &  \textbf{15} & \textbf{\large{\sfrac{1}{4}}} & 700 & $[-1,3]$ & $[3,15]$ & 131 & 92 &  87 \\ \vspace{0.3cm}

 mSID \heat & \textbf{45} &\begin{minipage}{0.07\textwidth}\begin{center} \begin{tikzpicture}
\draw[draw=black,line width=1pt] (11.1,5.5) rectangle ++(0.45,0.45);
\end{tikzpicture}\end{center}\end{minipage}   &  30 & 1 & \textbf{7} &  $[0,10]$ & $[0.3,1.5]$  & 148 & - & - \\

\end{tabular}
\end{table}

Note that the regime and $Q_m$ data of the top three data sets have already been published in some form by \cite{meyer_stratified_2014} (ML14, denoted by $^*$) and \cite{lefauve_regime_2019} (LPL19, denoted by $^\dagger$), as discussed in the literature review \S~\ref{sec:review-reg}-\ref{sec:review-Qm}. However, ML14 plotted their LSID and HSID data together (regimes in the $(\theta,Gr)$ plane, see their figure~7 and $Q_m$ as a function of $\theta$ alone, see their figure~8) and did not investigate their potential differences, while LPL19 only commented in passing on a fit of the $Q_m$ data in the $(\theta,Re)$ plane (see their figure~9). The individual reproduction and thorough discussion of these data alongside more recent data using a  unified non-dimensional approach will be key to this paper. 
All five data sets have been used in the PhD thesis of \cite{lefauve_waves_2018} (especially in Chapters 3 and 5), and the detailed parameters of all experiments are tabulated in his appendix~A for completeness.

Our focus on \emph{long ducts}, evidenced by our choice of $A=15$ and $30$, reflects our primary motivation, i.e. flows relevant to geophysical and environmental applications, which are typically largely horizontal ($\theta \approx 0^\circ$) and stably stratified in the vertical (as opposed to the different case of vertical exchange flow with $\theta =90^\circ$).
As we have seen in \S~\ref{sec:review}, the SID experiment conveniently exhibits all possible flow regimes, including high levels of turbulence and mixing, between $\theta=0^\circ$ and a few $\alpha$ at most. In long ducts (large $A$), $\alpha\equiv \tan^{-1}(A^{-1})$ is therefore small enough to allow us to study all the key dynamics of sustained stratified flows while keeping $\theta$ small enough for these flows to remain largely horizontal and relevant to our motivation.

As a result of this focus on long ducts, in the remainder of the paper we  make the approximation that
\begin{equation} \label{small-angle-approx}
    \cos \theta \approx 1 \quad \text{and} \quad \sin \theta \approx \theta, 
\end{equation}
as  in our previous discussion of the results of  ML14 and LPL19 in \S~\ref{sec:review}. This approximation is accurate to better than $2~\%$ for the angles considered in our data sets ($\theta \le 10^\circ$). Unless explicitly specified by the $^\circ$ symbol, $\theta$ will now be expressed in radians (typically in scaling laws).

\subsection{Flow regimes} \label{sec:results-reg}

\begin{figure}
    \centering
    \begin{subfigure}[b]{0.495\textwidth}§
      \mbox{\emph{(a)} \quad LSID  } \\ 
        \includegraphics[width=\textwidth]{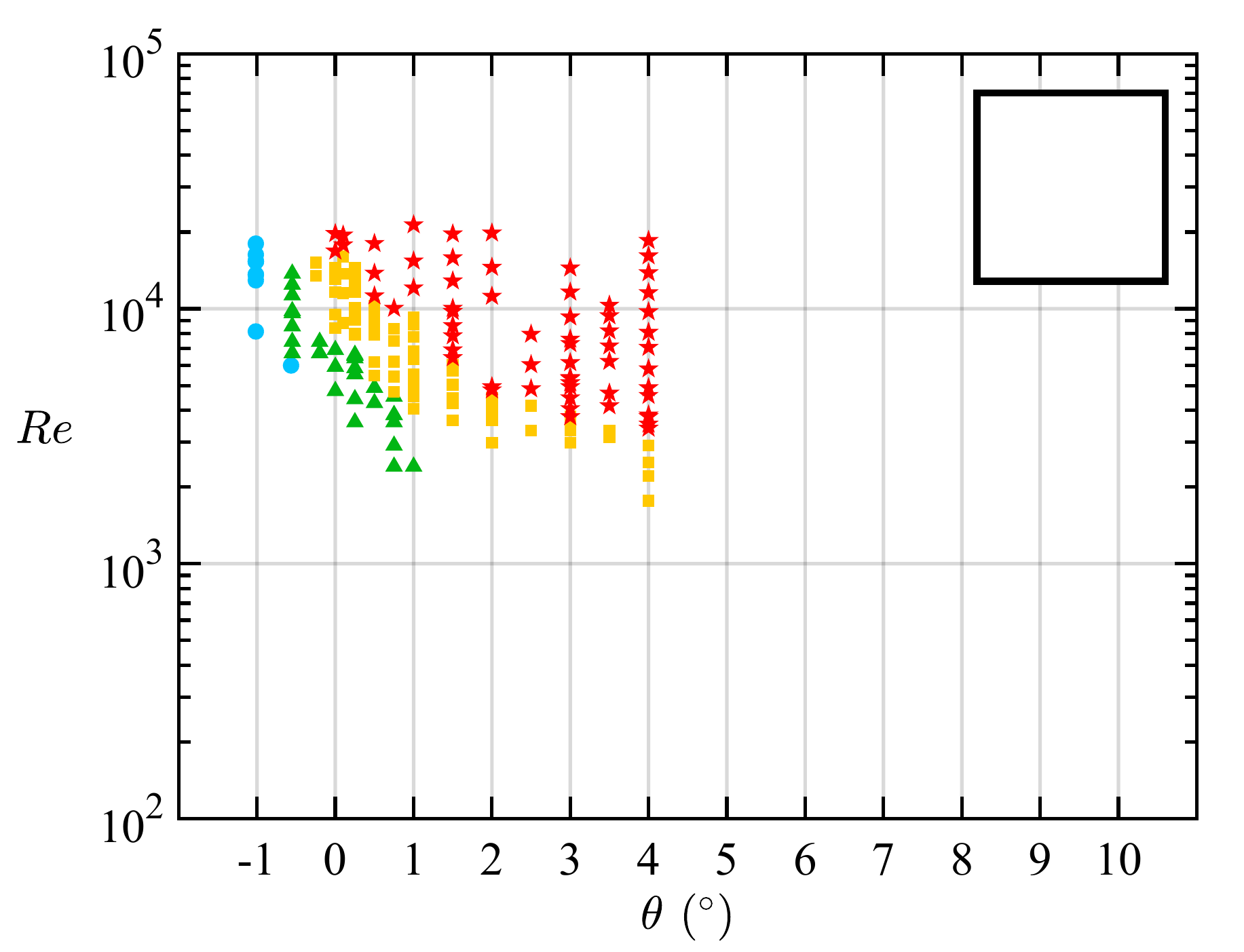}
    \end{subfigure}
        \begin{subfigure}[b]{0.495\textwidth}
            \mbox{\emph{(b)} \quad HSID   } \\ 
        \includegraphics[width=\textwidth]{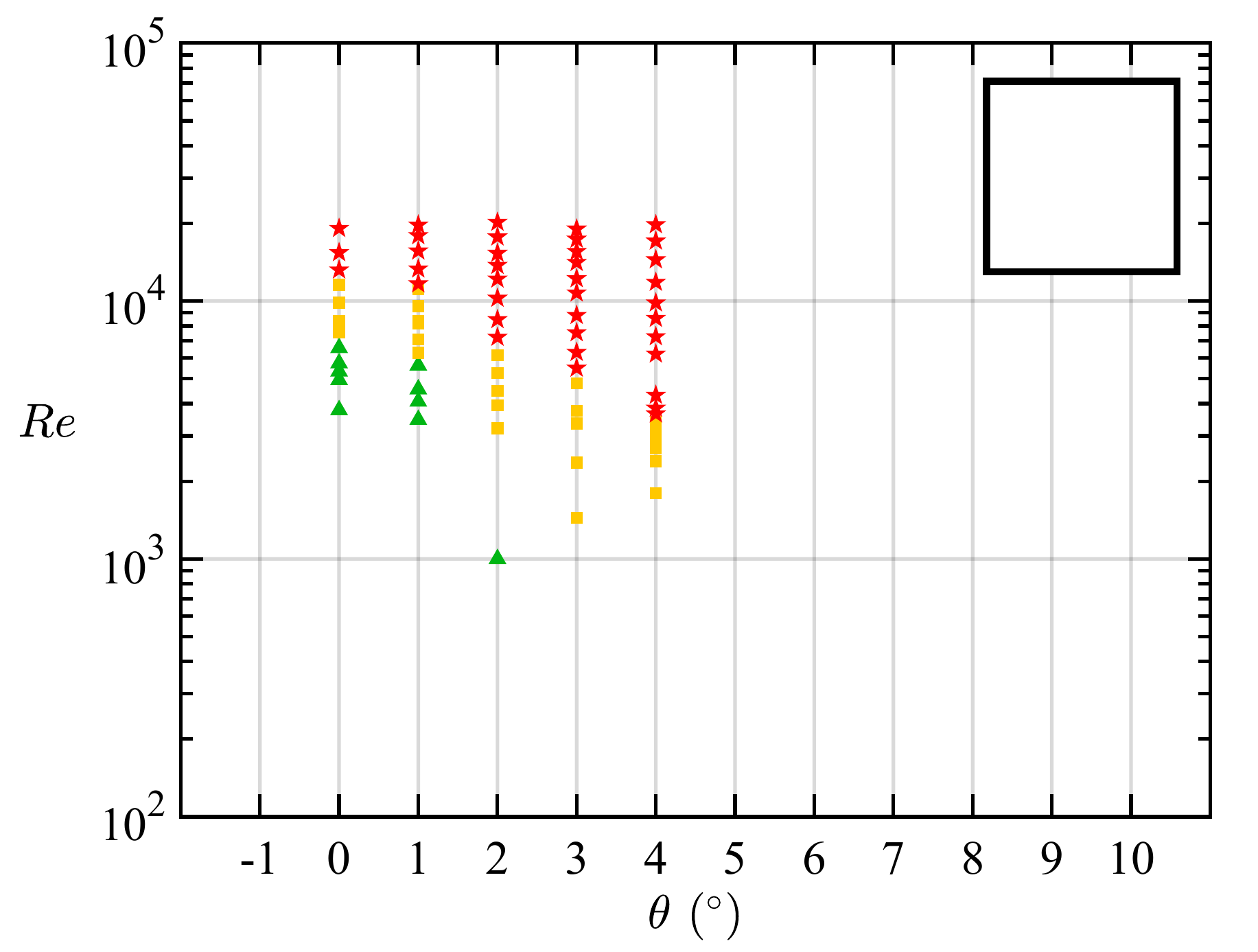}
    \end{subfigure}
\\
\vspace{0.5cm}
    \begin{subfigure}[b]{0.495\textwidth}
        \mbox{\emph{(c)} \quad mSID} \\ 
        \includegraphics[width=\textwidth]{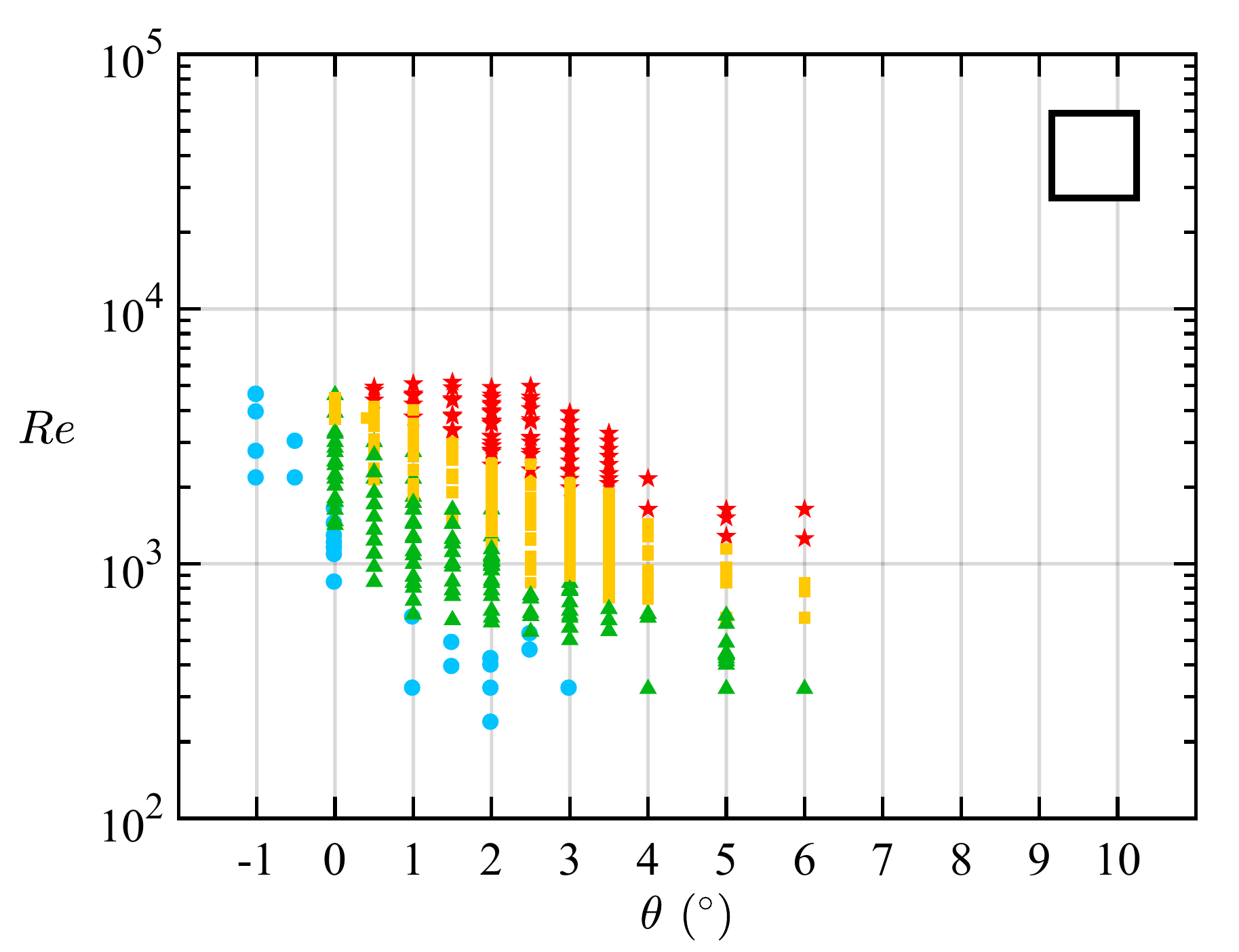}
         \end{subfigure}
    \begin{subfigure}[b]{0.495\textwidth}
            \mbox{\emph{(d)} \quad tSID } \\ 
        \includegraphics[width=\textwidth]{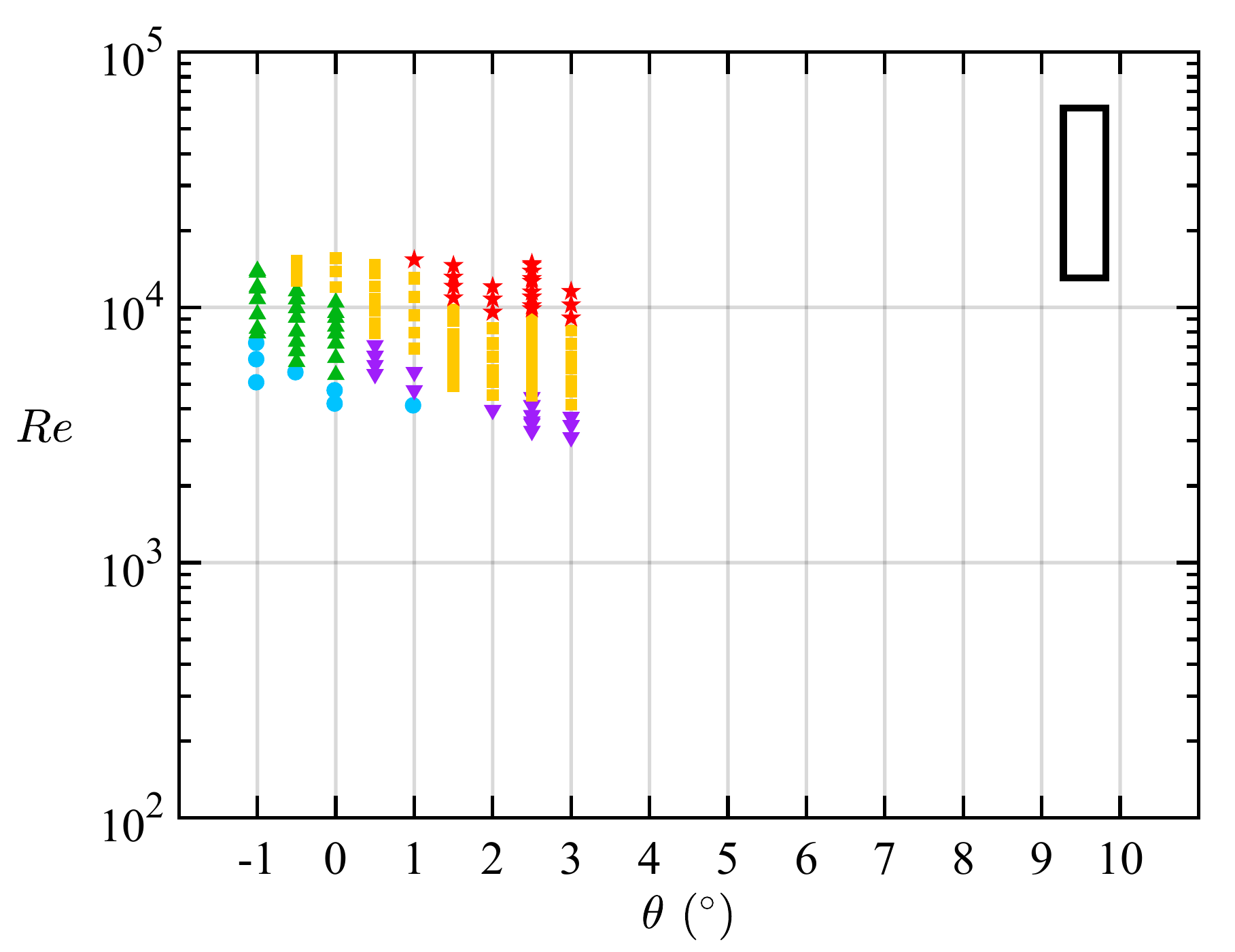}
    \end{subfigure}
\\
\vspace{0.5cm}
    \begin{subfigure}[b]{0.495\textwidth}
          \mbox{\emph{(e)} \quad mSID  \heat} \\ 
        \includegraphics[width=\textwidth]{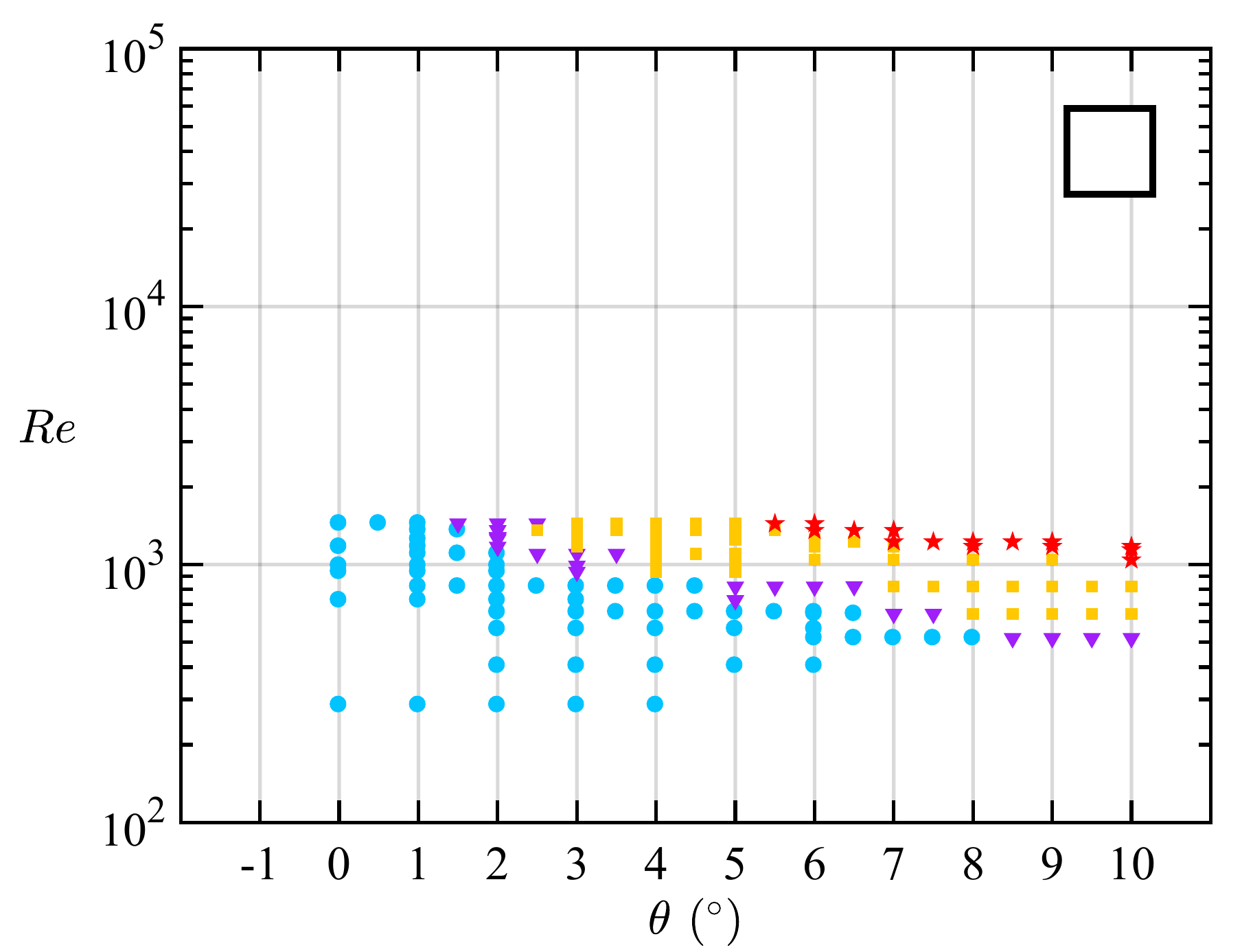}
    \end{subfigure}
            \begin{subfigure}[b]{0.495\textwidth}
        \hspace{1.2cm}
        \includegraphics[width=0.17\textwidth]{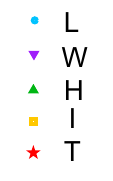} \vspace{2cm}   
    \end{subfigure}
    \caption{Regime diagrams in the $(\theta,Re)$ plane (lin-log scale) using the five data sets of table~\ref{tab:sid_geo_param} (the scaled cross-section of each duct is sketched for comparison in the top right corner of each panel). The error in $\theta$ is of order $\pm 0.2^\circ$ and is slightly larger than the symbol size, whereas the error in $Re$ is much smaller than the symbol size, except in \emph{(e)} at small $Re$. 
    }
  \label{fig:regime-diagrams}
\end{figure}

The $\LL,\HH,\II,\TT$ flow regimes were determined following the ML14 nomenclature as in appendix~\ref{sec:method_regimes} (except for a new regime which we discuss in the next paragraph). Figure~\ref{fig:regime-diagrams} shows the resulting regime maps in the $(\theta,Re)$ plane corresponding to the five data sets.

\begin{figure}
\centering
    \begin{subfigure}[b]{0.495\textwidth}
     \mbox{\emph{(a)} \quad  LSID } \\ 
        \includegraphics[width=\textwidth]{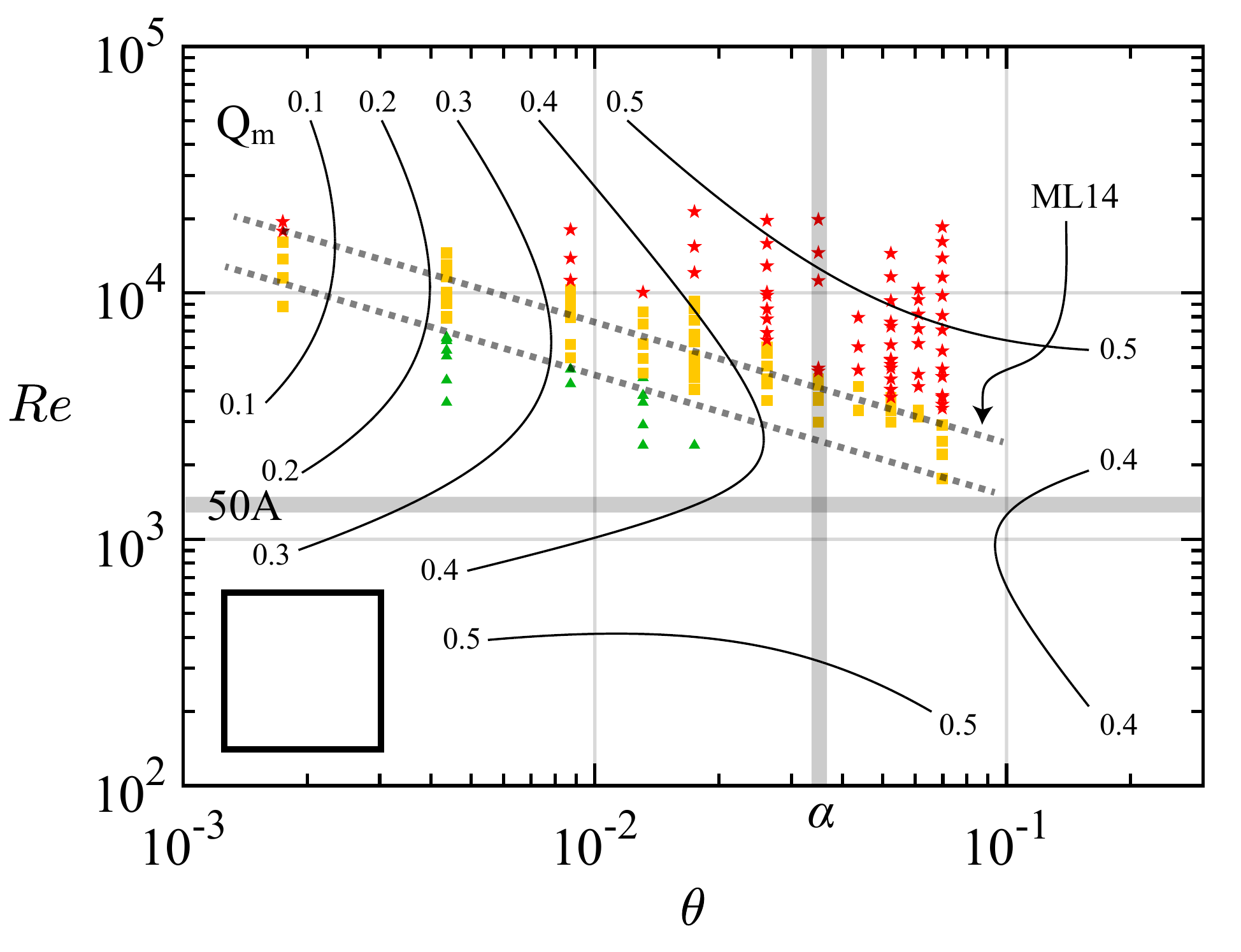}
    \end{subfigure}
    \begin{subfigure}[b]{0.495\textwidth}
          \mbox{\emph{(b)} \quad  HSID } \\ 
        \includegraphics[width=\textwidth]{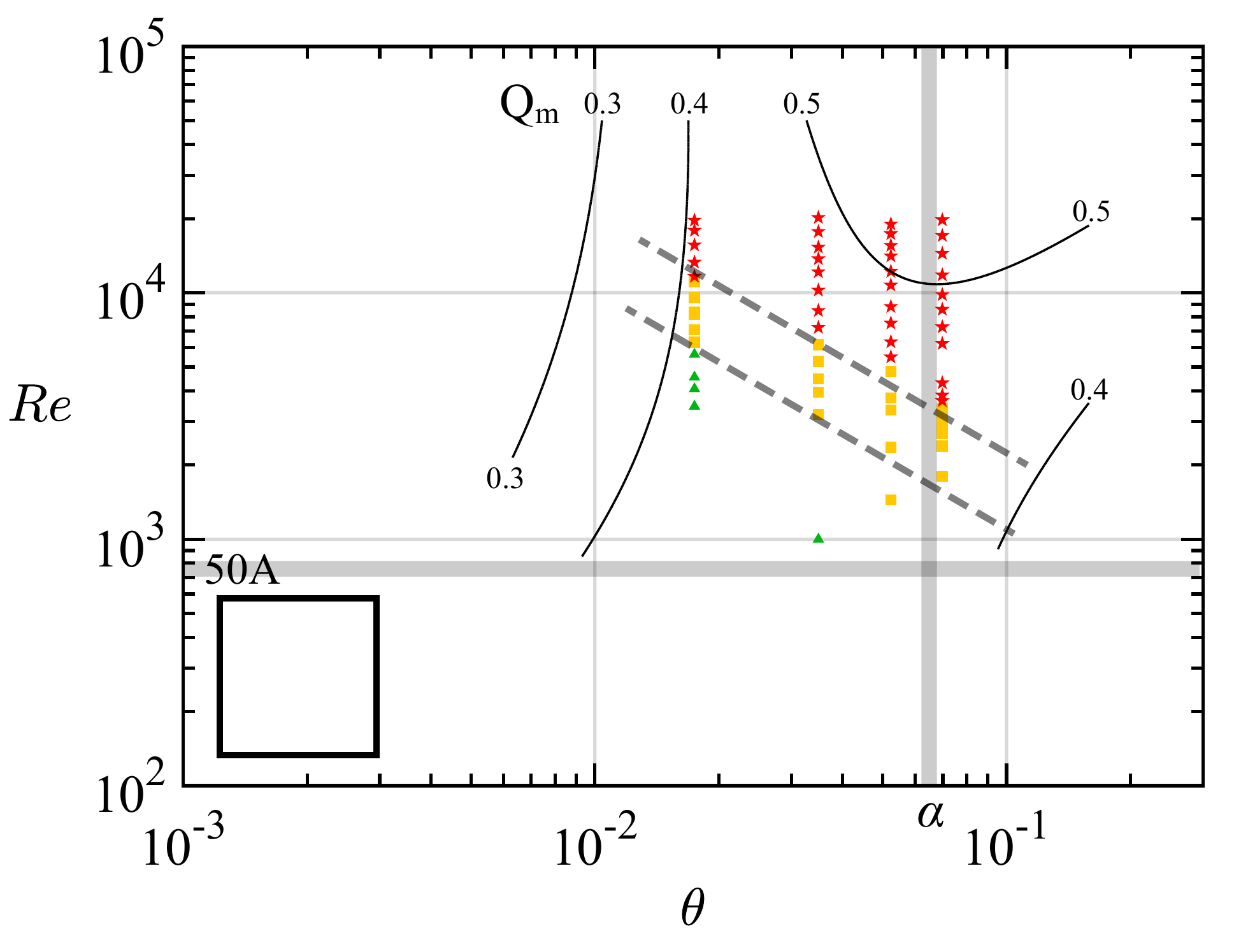}
    \end{subfigure}
\\
\vspace{0.5cm}
    \begin{subfigure}[b]{0.495\textwidth}
    \mbox{\emph{(c)} \quad  mSID } \\ 
        \includegraphics[width=\textwidth]{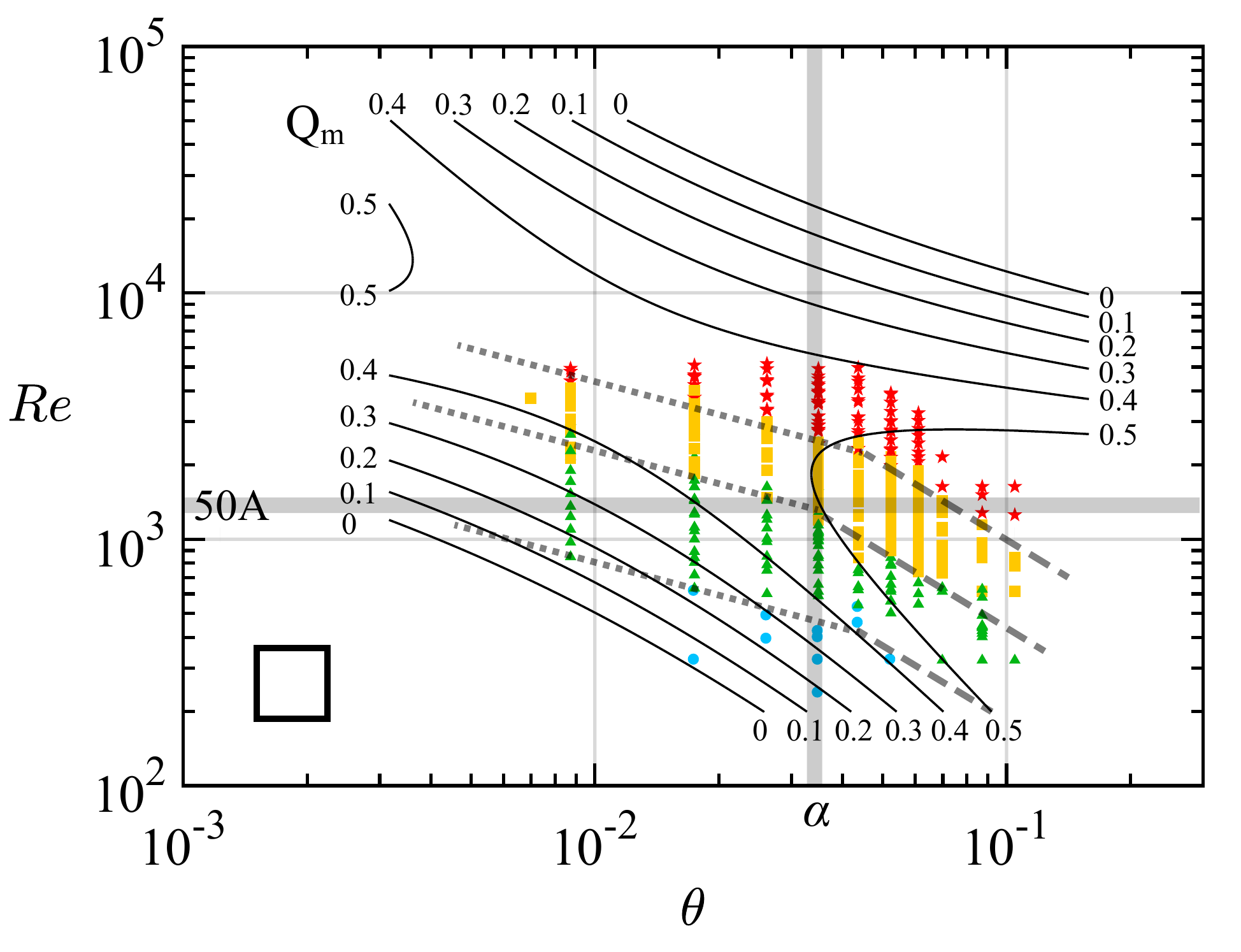}
    \end{subfigure}
    \begin{subfigure}[b]{0.495\textwidth}
         \ \     \mbox{\emph{(d)} \quad  tSID  } \\ 
        \includegraphics[width=\textwidth]{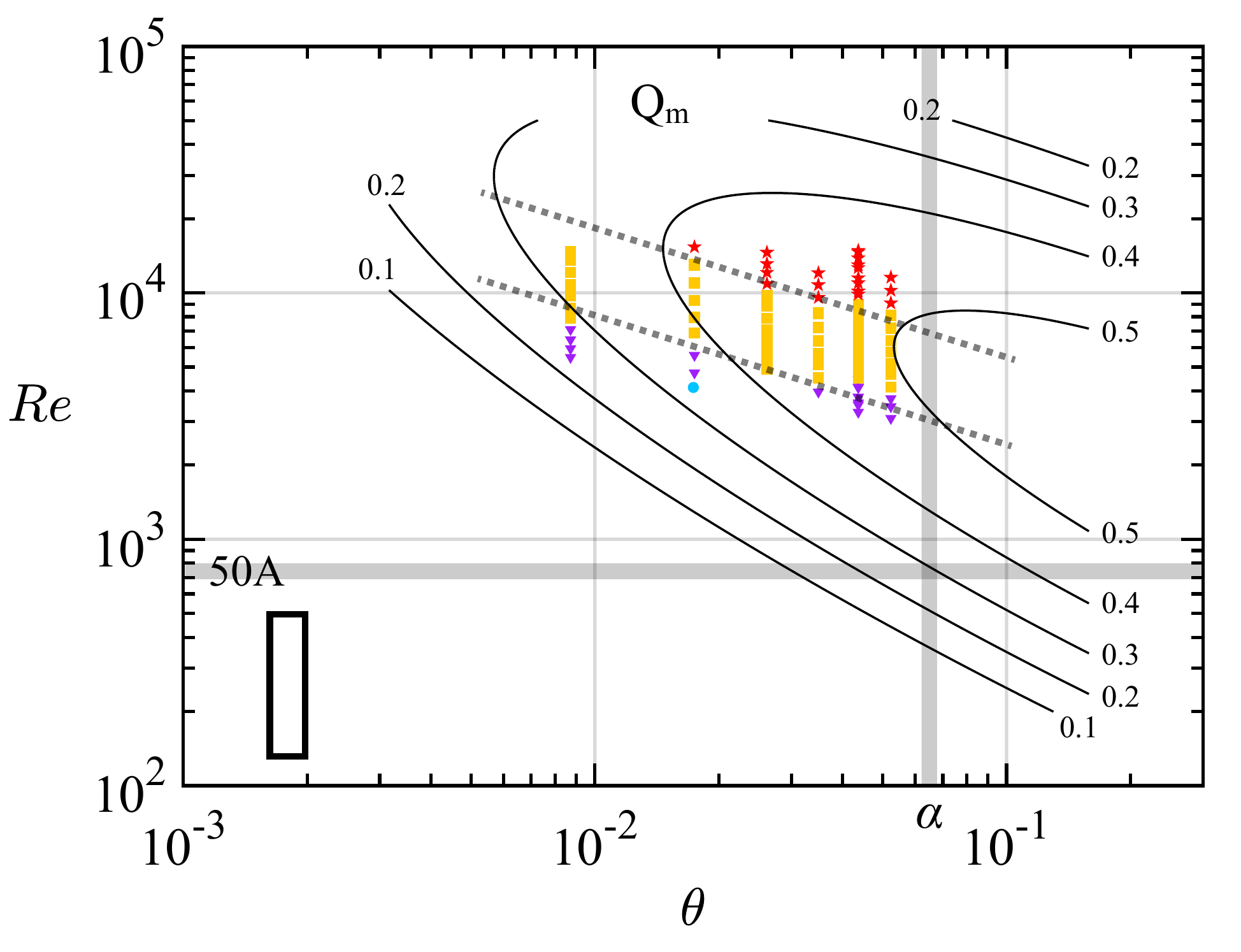}
    \end{subfigure}
\\
\vspace{0.5cm}
    \begin{subfigure}[b]{0.495\textwidth}
            \mbox{\emph{(e)} \quad mSID  \heat } \\ 
        \includegraphics[width=\textwidth]{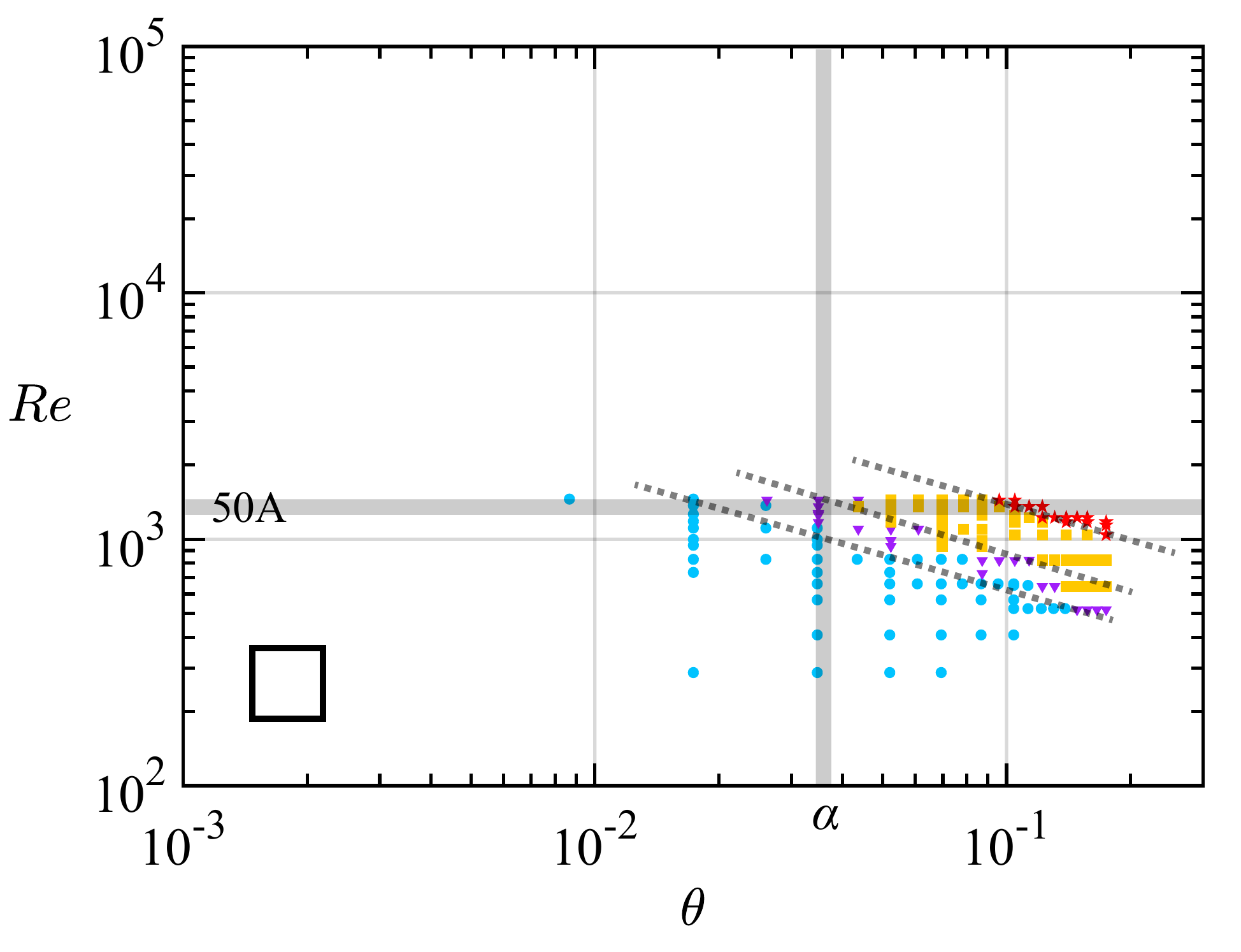}
    \end{subfigure}
    \begin{subfigure}[b]{0.495\textwidth}
      \hspace{0.5cm} \includegraphics[width=0.8\textwidth]{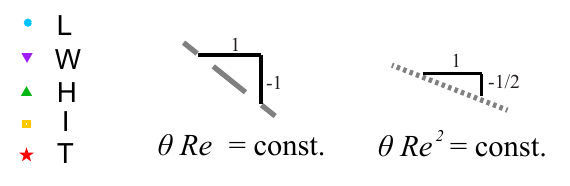} \vspace{2cm}
    \end{subfigure}

    \caption[dummy]{Regime and $Q_m$ in the $(\theta,Re)$ plane (log-log scale, thus only containing the regime and $Q_m$ data of figure~\ref{fig:regime-diagrams} for which $\theta>0^\circ$). The dashed and dotted lines represent the power law scalings $\theta Re=$ const. and $\theta Re^2=$ const., respectively. The gray shadings represent the special threshold values of interest $\theta=\alpha$ and $Re=50A$. The ML14 arrow in panel \emph{(a)} denotes the $\II\rightarrow\TT$ transition curve identified by ML14 (see \eqref{definition-Gr}). Black contours in panels \emph{(a-d)} represent the fit to the $Q_m$ data (see \S~\ref{sec:results-Qm}), representing \emph{(a)} 20 data points (coefficient of determination $R^2=0.56$), \emph{(b)} 34 points ($R^2=0.81$), \emph{(c)} 162 points ($R^2=0.80$), and \emph{(d)} 92 points ($R^2=0.86$) }
  \label{fig:regime-Re-sintheta}
\end{figure}

First, we note the introduction of a `new' $\WW$ regime in the tSID and mSID$ \heat$ data (panels~\emph{(d,e)}). This $\WW$ (wave) regime is similar to the $\HH$ (Holmboe) regime, but describes interfacial waves which could not be recognised as Holmboe waves in shadowgraphs. These waves were of two types. First, in the tSID geometry at positive angles $\theta>0$, the waves did not exhibit the distinctive `cusped' shape of Holmboe waves and the waves appeared to be generated at the ends of the duct and to decay as they travel inside the duct. Second, in the mSID \heat larger-amplitude, tilde-shaped internal waves were observed across most of the height of the duct, contrary to Holmboe waves which are typically confined to a much  thinner interfacial region. Further discussion of these waves falls outside the scope of this paper, but can be found in \cite[\S\S~3.2.3-3.2.4]{lefauve_waves_2018} (hereafter abbreviated L18). This new observation highlights the richness of the flow dynamics in the SID experiment. However, for the purpose of this paper, the $\HH$ and $\WW$ regimes are sufficiently similar in their characteristics (mostly laminar flow with interfacial waves) that we group them under the same regime for the purpose of discussing regime transitions.

The main observation of figure~\ref{fig:regime-diagrams} is that the transitions between regimes can be described as simple curves in the $(\theta, Re)$ plane that do not overlap (or `collapse') between the five data sets. The slope and location of the transitions varies greatly between panels: the difference between the LSID and HSID data (panels \emph{(a,b)}) is due to $A$, the difference between the HSID and tSID data  (panels \emph{(b,d)}) is due to $B$, and the difference between the mSID and mSID~\heat data  (panels \emph{(c,e)}) is due to $Pr$. 

However one of the most surprising differences is that between LSID and mSID data (panel \emph{(a,c)}), due to the dimensional height of the duct $H$, already observed in LPL19. It is reasonable to expect that this $H$-effect is responsible for the main differences between the LSID/HSID/tSID data and the mSID/mSID \heat data. In other words $H$ is the main reason why the LSID/HSID/tSID transitions curves  lie \emph{well above}  those for mSID/mSID~\heat, i.e. the same transitions occur at higher $Re$ for larger $H$. The factor of $\approx 2$ quantifying this observation suggests that a Reynolds number built using a length scale identical in all data sets (rather than $H/2$) would better collapse the data; however our dimensional analysis does not provide such a length scale.

We conclude that the transitions between flow regimes can be described by  hyper-surfaces depending on all five parameters $A, B,\theta,Re, Pr$ because their projections onto the $(\theta,Re)$ plane for different $A,B,Pr$ do not overlap. This dependence of flow regimes on all five parameters was not immediately obvious from our dimensional analysis which concerned the scaling of the velocity $f_{\Delta U}$ alone (\S~\ref{sec:scaling-of-vel} and figure~\ref{fig:balance}). \AL{Furthermore, the existence of at least another important non-dimensional parameter related to $H$ is a major result that could not be predicted \emph{a priori}.}

Let us now investigate in more detail the scaling of regime transitions with respect to $\theta$ and $Re$, for which we have much higher density of data than for $A,B,Pr$. We replot the $\theta>0$ data of figure~\ref{fig:regime-diagrams} using a log-log scale in figure~\ref{fig:regime-Re-sintheta}  (each panel corresponding to the respective panel of figure~\ref{fig:regime-diagrams}). To guide the eye to the two main types of regime transition scalings observed in these data, we also plot two families of lines: dashed lines with a $\theta Re=$ const. scaling, and dotted lines with a $\theta Re^2=$ const. scaling. We also show using grey shading special values of interest: $Re=50A$, and $\theta=\alpha$.  The latter was highlighted as particularly relevant in the literature review \S~\ref{sec:review} (notably as the boundary between lazy and forced flows in LPL19). Although W86 and K91 quoted $Re=500A$ as a threshold beyond which the effects of viscosity should be negligible on the turbulence in the SID, we believe that $Re=50A$ is a physically justifiable threshold beyond which the influence of the top and bottom walls of the duct becomes negligible. In the absence of turbulent diffusion, laminar flow in the duct is significantly affected by the top and bottom walls if the interfacial and wall 99~\% boundary layers overlap in the centre of the duct ($x=0$), which occurs for $Re < 50$ (L18, \S~5.2.3). If, on the other hand, $Re\gg 50 A$ ($Re=500A$ being a potential threshold), the top and bottom wall laminar boundary layers (as well as the side wall laminar boundary layers, assuming that $B \centernot\ll 1$) do not penetrate deep into the `core' of the flow (however at these $Re$, we expect interfacial turbulence to dominate the core of the flow). Note that black contours representing a fit of the $Q_m$ data are superimposed in panels~\emph{(a-d)}; these will be discussed in the next section.

We observed regime transitions scaling with $\theta Re^2=$ const. (dotted lines) in  LSID, tSID and mSID \heat (panels~\emph{(a,d,e)}), and with  $\theta Re=$  const. (dashed lines) in HSID (panel~\emph{(b)}). In mSID (panels~\emph{(c)}), we observe these two different scalings:  $\theta Re^2$ for $\theta \lesssim \alpha$ (lazy flows) and $\theta Re$ for $\theta \gtrsim \alpha$ (forced flows), as previously observed by LPL19. LPL19 physically substantiated the $\theta Re$ scaling in forced flows, but not the $\theta Re^2$ scaling in lazy flows. Furthermore, these five data sets show that this dichotomy in scalings between lazy and forced flows does not extend to all other geometries; indeed lazy flows in the HSID exhibit a $\theta Re$ scaling and forced flows in the mSID \heat exhibit a $\theta Re^2$ scaling. These observations further highlight the complexity of the scaling of regime transitions with $A,B,\theta,Re,Pr$.

\subsection{Mass flux} \label{sec:results-Qm}

Mass fluxes were determined using the same salt balance methodology as ML14 described in appendix~\ref{sec:method_Qm}. 

In figure~\ref{fig:MF-diagrams} we plot the $Q_m$ data for mSID (full symbols) and tSID (open symbols) as a function of $Re$ for all the available $\theta$ (from $\theta=-1^\circ$ in panel~\emph{(a)} to $\theta=3.5^\circ$ in panel~\emph{(j)}). The colour of each symbol denotes the regime as in figures~\ref{fig:regime-diagrams}-\ref{fig:regime-Re-sintheta} and the error bars denote the uncertainty about the precise duration $T$ of the `steady' flow of interest in an experiment (used to average the volume flux and obtain $Q_m$, see appendix~\ref{sec:method_Qm} for more details). We do not plot the LSID and HSID data in this figure because they are sparser and do not have error bars (these data were collected by ML14).

\begin{figure}
    \centering
    \begin{subfigure}[b]{0.47\textwidth}
      \ \ \mbox{\emph{(a)}  \qquad  $\theta=-1^\circ$ } \\ 
        \includegraphics[width=\textwidth]{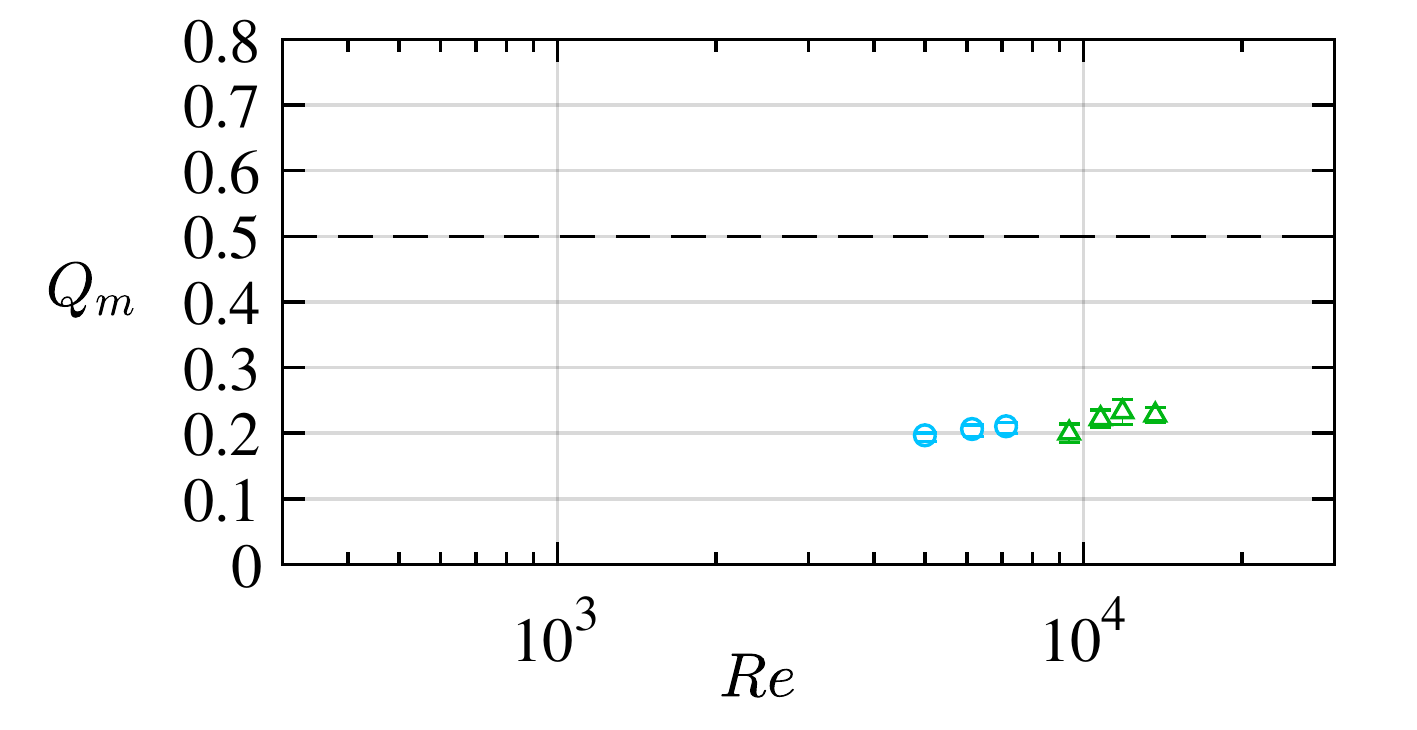}
    \end{subfigure}
    \begin{subfigure}[b]{0.47\textwidth}
         \ \     \mbox{\emph{(b)}  \qquad $\theta=-0.5^\circ$} \\ 
        \includegraphics[width=\textwidth]{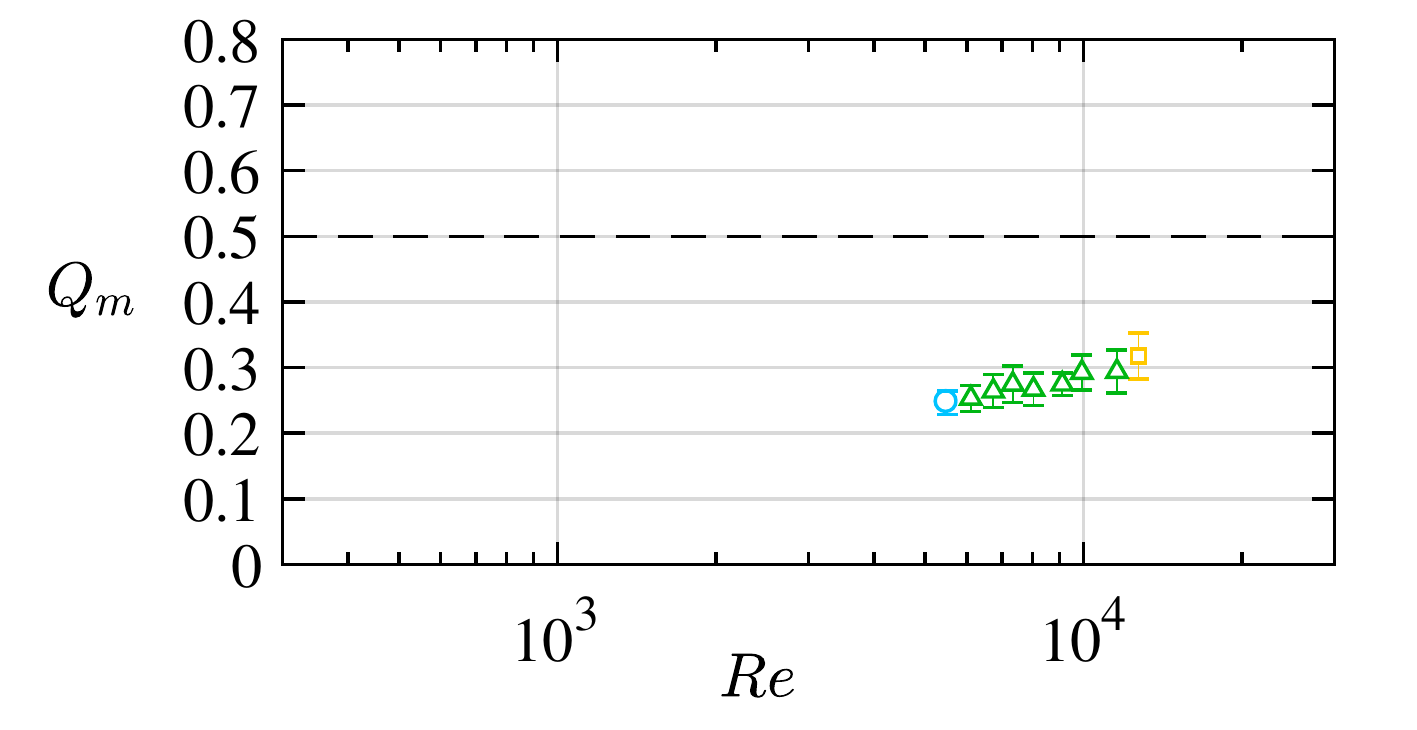}
    \end{subfigure}
\\
    \centering
    \begin{subfigure}[b]{0.47\textwidth}
      \ \ \mbox{\emph{(c)}  \qquad $\theta=0^\circ$ } \\ 
        \includegraphics[width=\textwidth]{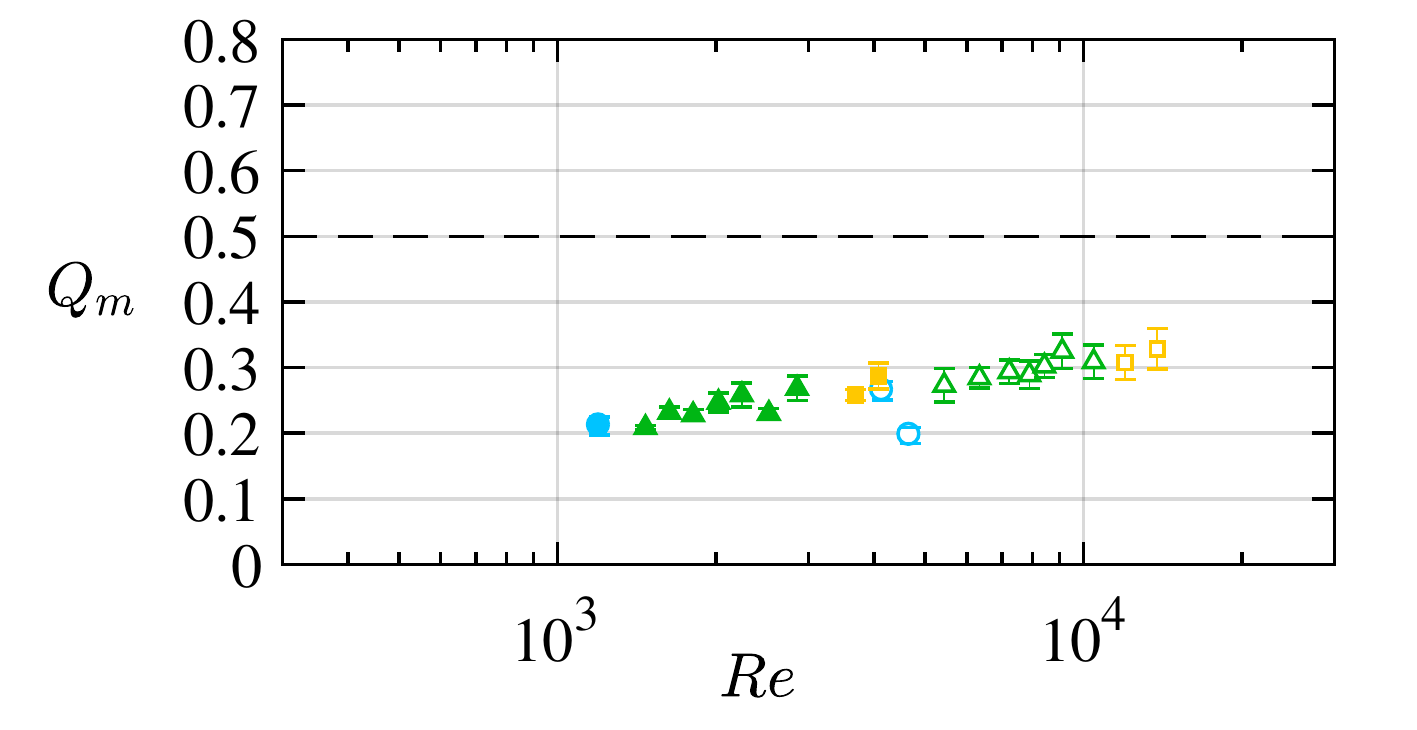}
    \end{subfigure}
    \begin{subfigure}[b]{0.47\textwidth}
         \ \     \mbox{\emph{(d)} \qquad $\theta=-0.5^\circ$} \\ 
        \includegraphics[width=\textwidth]{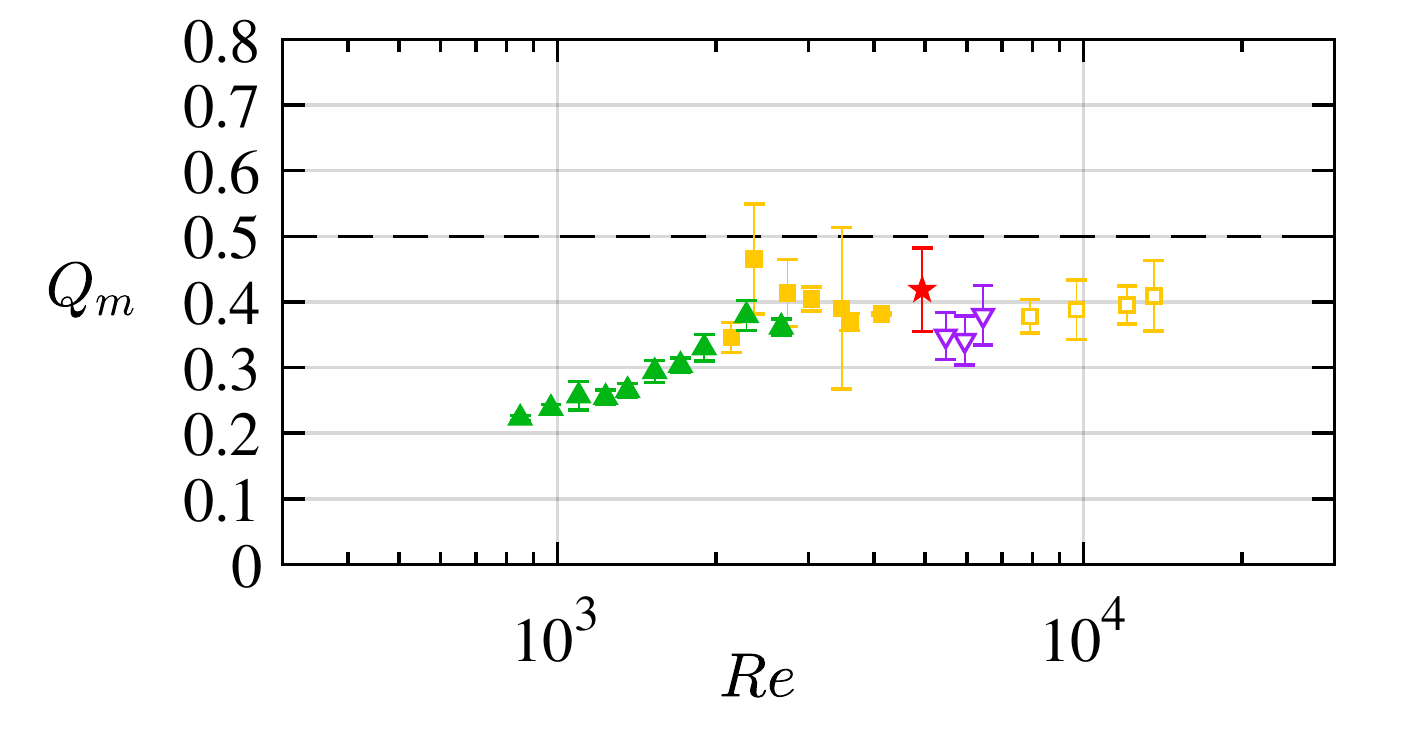}
    \end{subfigure}
\\
    \centering
    \begin{subfigure}[b]{0.47\textwidth}
      \ \ \mbox{\emph{(e)} \qquad $\theta=1^\circ$} \\ 
        \includegraphics[width=\textwidth]{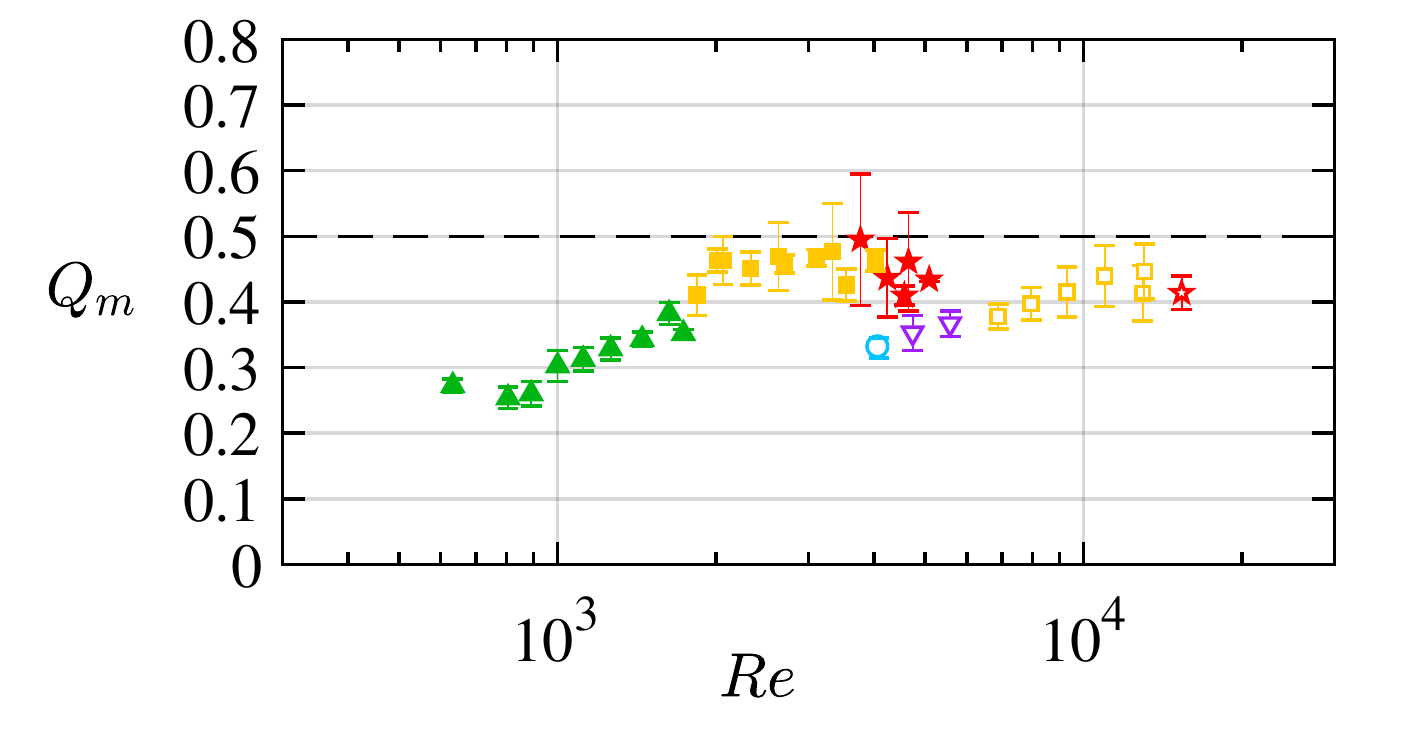}
    \end{subfigure}
    \begin{subfigure}[b]{0.47\textwidth}
         \ \     \mbox{\emph{(f)}  \qquad $\theta=1.5^\circ$} \\ 
        \includegraphics[width=\textwidth]{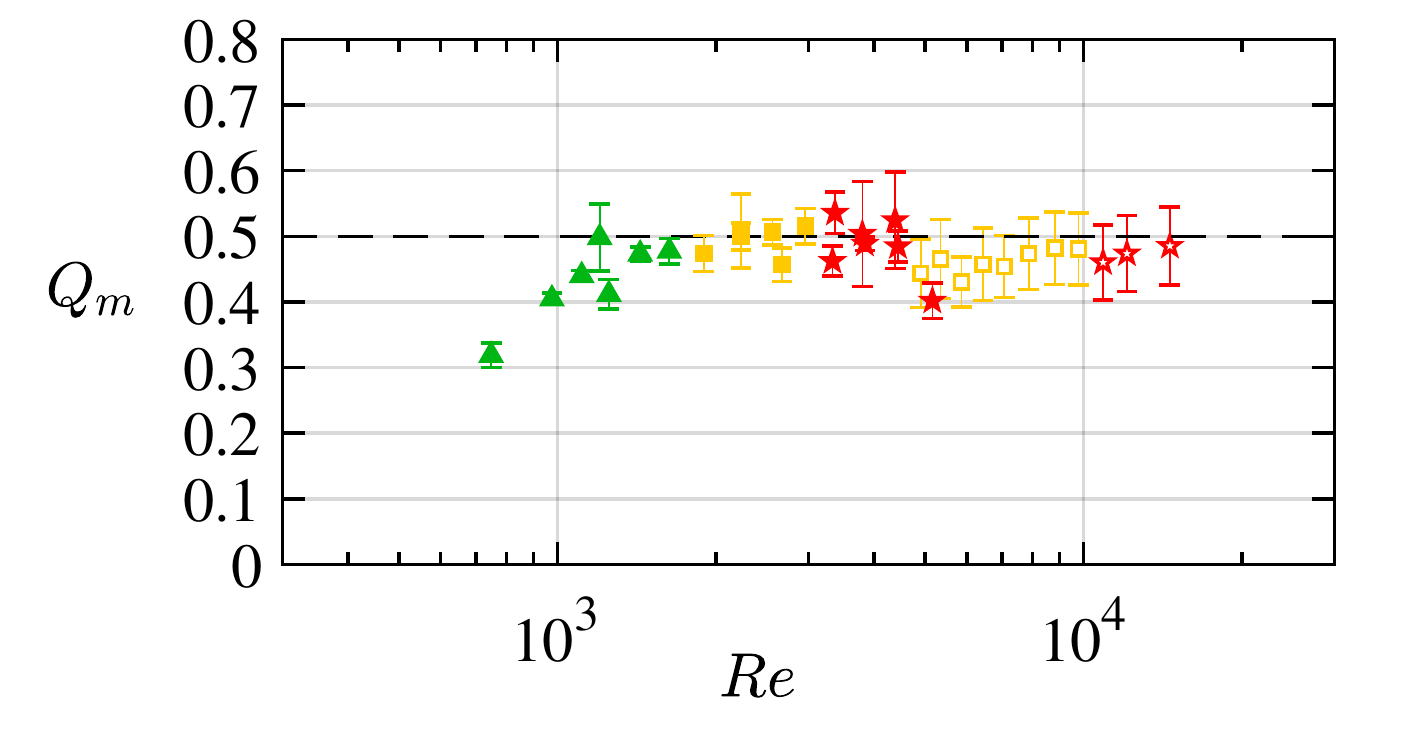}
    \end{subfigure}
    \\
    \centering
    \begin{subfigure}[b]{0.47\textwidth}
      \ \ \mbox{\emph{(g)} \qquad $\theta=2^\circ$ } \\ 
        \includegraphics[width=\textwidth]{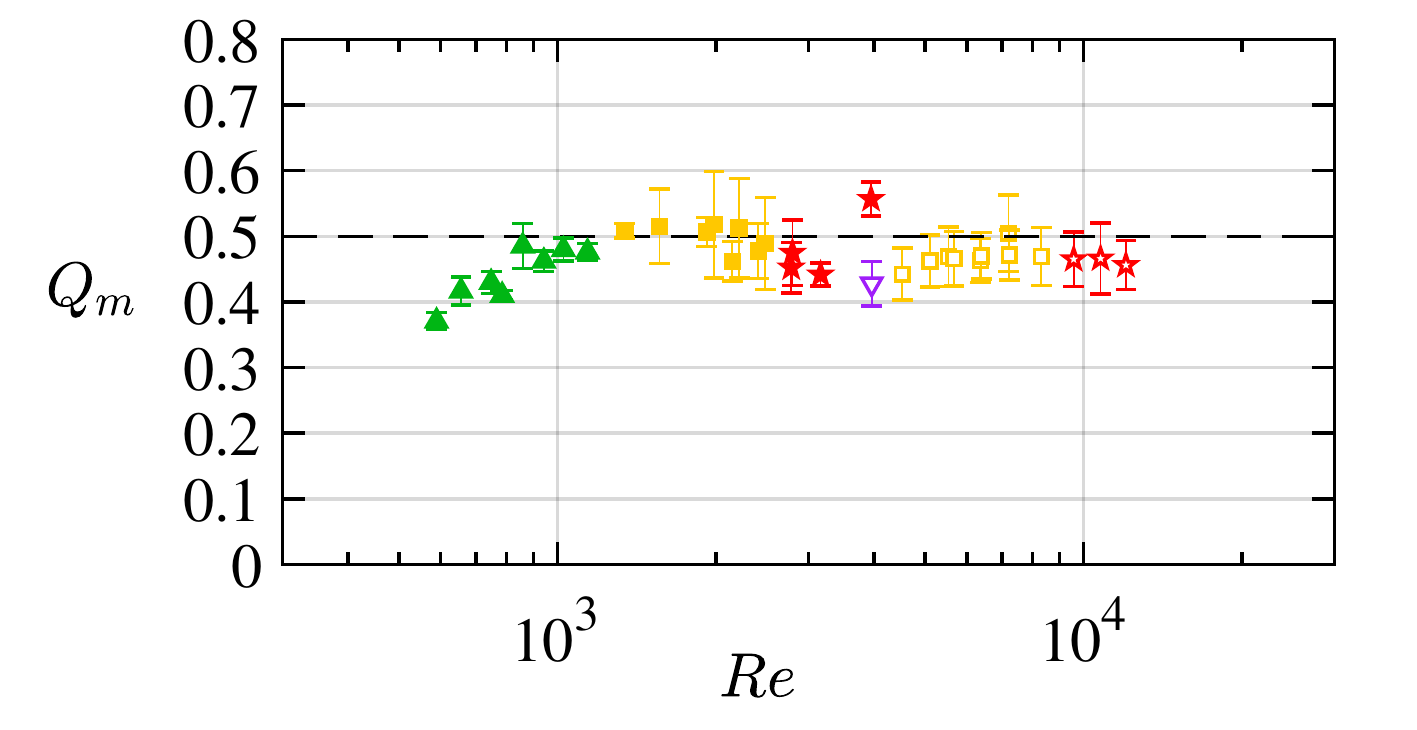}
    \end{subfigure}
    \begin{subfigure}[b]{0.47\textwidth}
         \ \     \mbox{\emph{(h)}  \qquad $\theta=2.5^\circ$} \\ 
        \includegraphics[width=\textwidth]{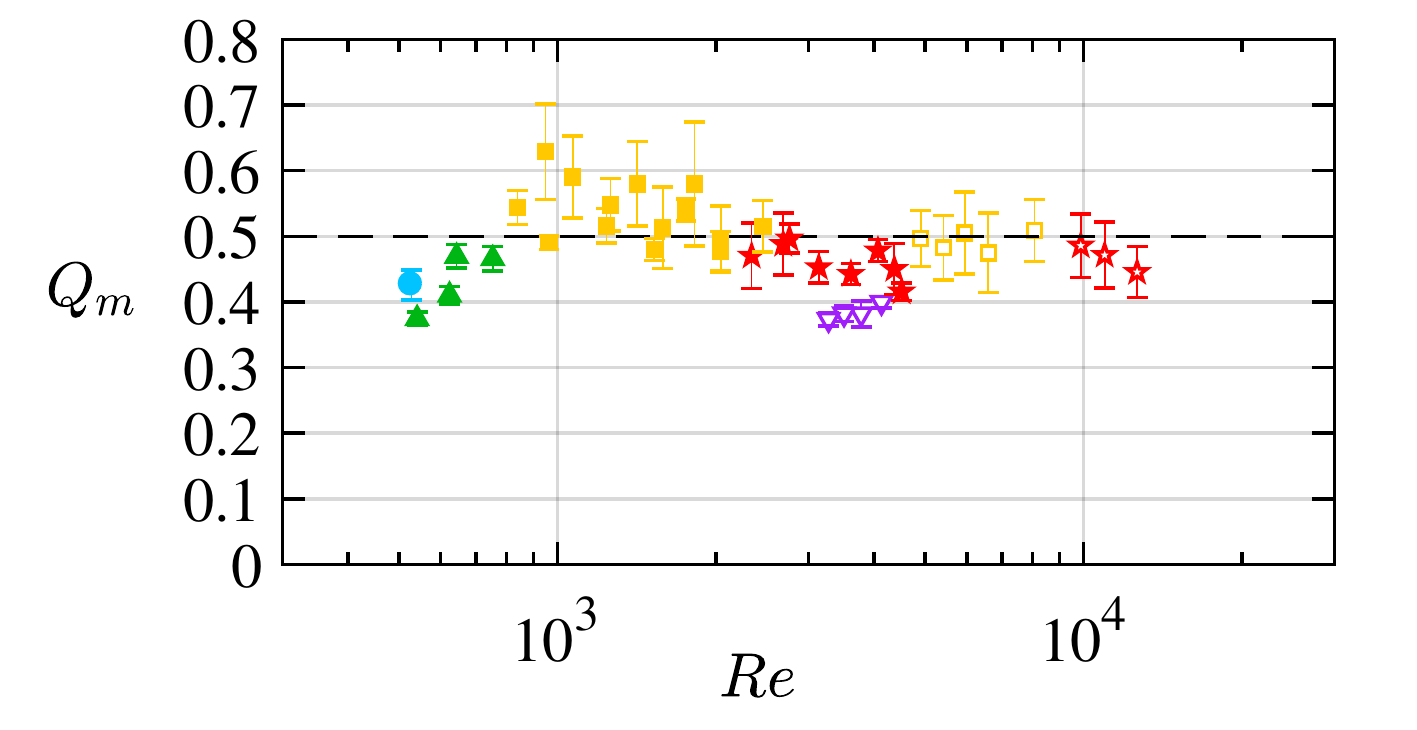}
    \end{subfigure}
    \\
    \begin{subfigure}[b]{0.47\textwidth}
      \ \ \mbox{\emph{(i)} \qquad $\theta=3^\circ$ } \\ 
        \includegraphics[width=\textwidth]{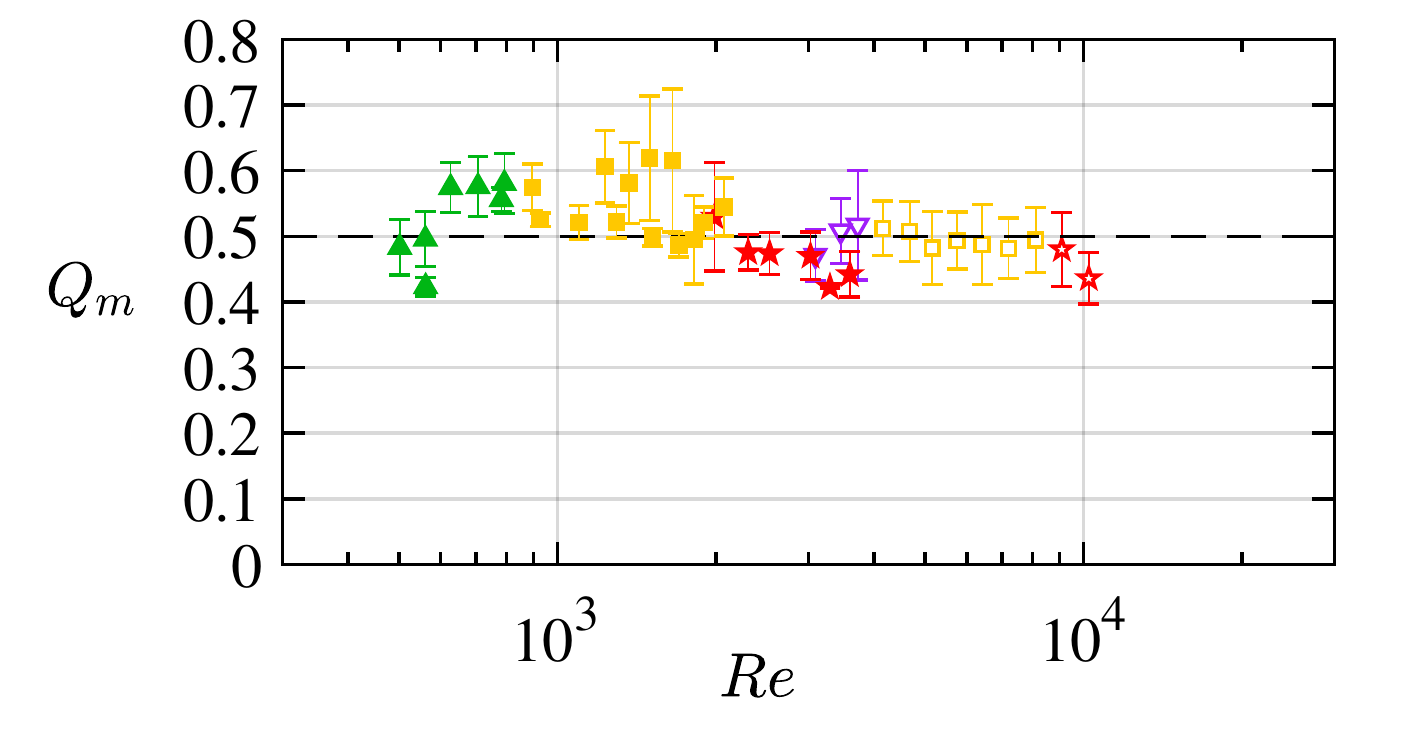}
    \end{subfigure}
        \begin{subfigure}[b]{0.47\textwidth}
              \ \ \mbox{\emph{(j)} \qquad $\theta=3.5^\circ$ } \\ 
        \includegraphics[width=\textwidth]{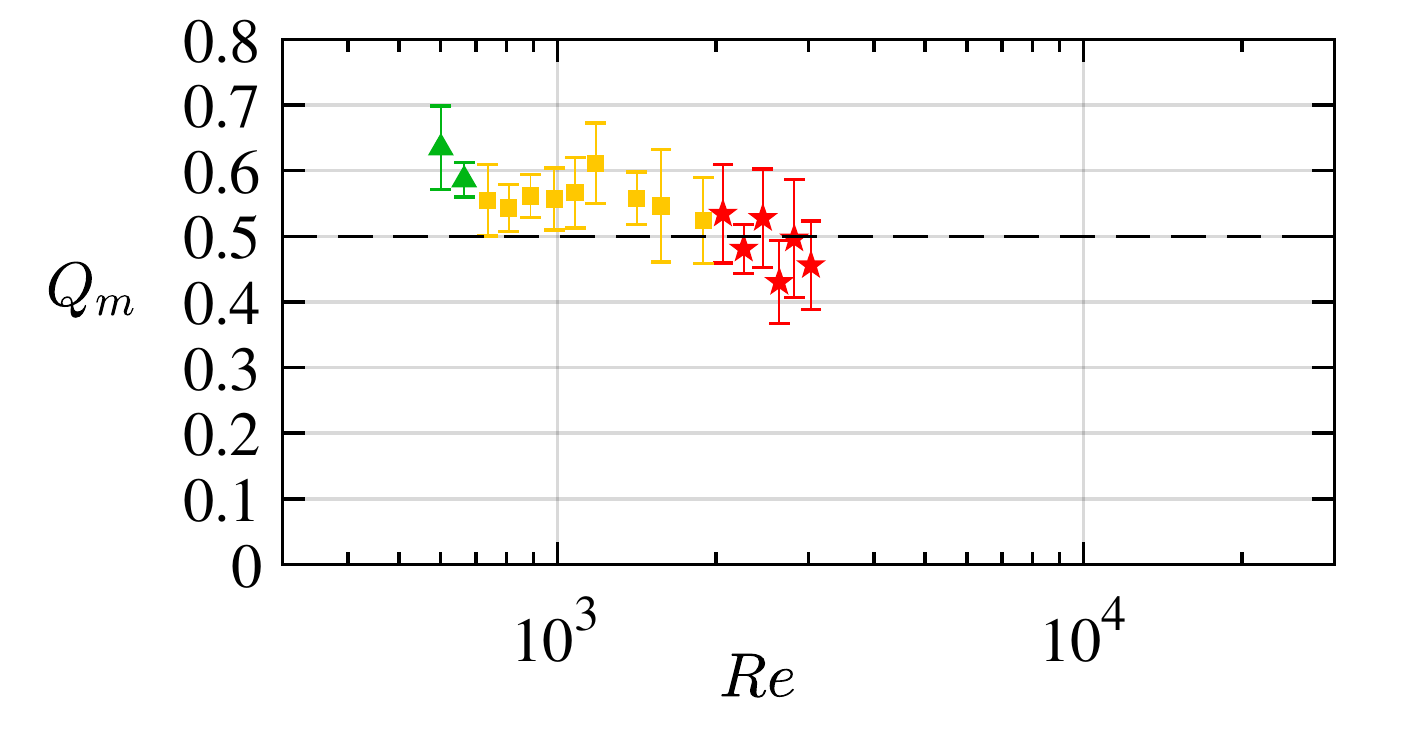}
    \end{subfigure}
    \caption[MF data for the tSID]{Mass flux for the mSID data set (full symbols) and tSID data set (open symbols) for as a function of $Re$ for various $\theta\in[-1^\circ,3.5^\circ]$ by $0.5^\circ$ increments (\emph{a-j}). The symbol colour denotes the regime as in figures~\ref{fig:regime-diagrams} and \ref{fig:regime-Re-sintheta}. The mass flux $Q_m$ is computed using the average estimation of the run time, and the error bars denote the uncertainty in this estimation (see appendix~\ref{sec:method_Qm}).}
  \label{fig:MF-diagrams}
\end{figure}

At low angles $\theta \lesssim 1^\circ < \alpha$ (where $\alpha \approx 2^\circ$ in mSID and $4^\circ$ in tSID) we observe low values $Q_m \approx 0.2-0.3$ in the $\LL$ and $\HH$ regimes. At higher angles $\theta \approx \alpha-2\alpha$ we observe convergence to the hydraulic limit $Q_m \rightarrow 0.5$  (denoted by the dashed line), as discussed in \S~\ref{sec:scaling-of-vel}, which coincides with the $\II$ and $\TT$ regimes. We also note that this hydraulic limit is not a strict upper bound in the sense that we observe values up to $Q_m=0.6$ in some experiments (some error bars even going to $0.7$).

Moreover, the non-monotonic behaviour of $Q_m$ with respect to both $\theta$ and $Re$ is  particularly clear in the mSID data: at `large' angles $\theta \gtrsim \alpha \approx 2^\circ$, $Q_m$ drops with both  $\theta$ and $Re$. All of these observations are in qualitative agreement with the literature (\S~\ref{sec:review-Qm}). However, similarly to the regime data of \S~\ref{sec:results-reg}, we also note that the mSID and tSID data do not collapse with $Re$: all tSID data (open symbols) are shifted to larger $Re$ compared to mSID data (full symbols). This again suggests that $H$ plays an important role not presently captured by our five non-dimensional parameters, and that a Reynolds number based on a length scale independent of $H$ would better collapse the data.

To gain more insight into the scaling of $Q_m$ and its relation to the flow regimes, we superimpose on the regime data of figure~\ref{fig:regime-Re-sintheta}\emph{(a-d)} black contours representing the least-squares fit of our four $Q_m$ data sets using the following quadratic form: 
\begin{eqnarray} \label{MF_contour_fit}
Q_m (\theta, Re) &=& 
\Gamma_{00} + \Gamma_{10}\log \theta + \Gamma_{20} ( \log \theta)^2 + \Gamma_{01} \log Re + \Gamma_{02} (\log Re)^2 + \Gamma_{11} \log \theta \log Re  \nonumber \\
&=& 
\begin{bmatrix} 
\log \theta & \log Re & 1 
\end{bmatrix} 
\underbrace{\begin{bmatrix} 
\Gamma_{20} & \Gamma_{11}/2 & \Gamma_{10}/2 \\
\Gamma_{11}/2 & \Gamma_{02} & \Gamma_{01}/2 \\ 
\Gamma_{10}/2 & \Gamma_{01}/2 & \Gamma_{00}\\
\end{bmatrix}}_{\bm{\Gamma}}
\begin{bmatrix} 
\log \theta\\ 
\log Re \\ 
1 
\end{bmatrix}.
\end{eqnarray}
This is the general equation of a conic section, well suited to the non-monotonic behaviour observed above, and $\bm{\Gamma}$ is commonly referred to as the matrix of the quadratic equation. 

These contours describe hyperbolas ($\det \bm{\Gamma}<0$) for LSID, HSID and mSID (panels~\emph{(a,b,c)}), and concentric ellipses ($\det \bm{\Gamma}>0$) for tSID (panel~\emph{(d)}). The hydraulic limit $Q_m \approx 0.5$ is reached either at the saddle point of the hyperbolas (panels~\emph{(a,b,c)}), or at the centre of the ellipses (panels~\emph{(d)}), and, encouragingly, no $Q_m=0.6$ contour exists here. 

We again note that these four data sets do not collapse in the $(\theta,Re)$ plane. For example, the angle at which this maximum $Q_m$ is achieved is a modest $\theta = 0.3\alpha$ in mSID (panel~\emph{(c)}) but appears much larger in tSID. The eigenvectors of $\bm{\Gamma}$ for each data set reveal that the major axis of these conic sections has equation $\theta Re^{\gamma}$ where $\gamma=2.6,0.3,1.5,1.2$ respectively for panels~\emph{(a,b,c,d)} (a larger exponent $\gamma$ represents a larger dependence on $Re$, hence a more horizontal axis). 

The exponent $\gamma$ characterising the slope of the major axis is roughly of the same order as the exponent characterising the lines of regime transition (which is 1 for the $\theta Re$ scaling, and 2 for the $\theta Re^2$ scaling), suggesting that both phenomena (regime transition and non-monotonic behaviour of $Q_m$) are linked. However, this agreement is not quantitative except in mSID (panel~\emph{(c)}) where $\gamma=1.5$ is precisely the average of the two different regime transition exponents. This general lack of correlation suggest that the relationship between regimes and $Q_m$ in the SID is not straightforward and dependent on the geometry.

\subsection{Interfacial layer thickness} \label{sec:results-delta}

Interfacial layer thickness was determined using the non-intrusive shadowgraph imaging technique (in salt experiments only). Shadowgraph is particularly suited to detect peaks in the vertical curvature of the density field $\p_{zz} \rho$ which we define as the edges of the interfacial density layer, as explained in appendix~\ref{sec:method_delta}.

\begin{figure}
 \centering
    \begin{subfigure}[b]{0.32\textwidth}
        \qquad \qquad \quad \mbox{$\downarrow$ \  $\theta=1^\circ$ } \\ 
    \end{subfigure}
    \begin{subfigure}[b]{0.32\textwidth}
      \qquad \qquad \quad \mbox{$\downarrow$ \  $\theta=2^\circ$ } \\ 
    \end{subfigure}
        \begin{subfigure}[b]{0.32\textwidth}
            \qquad \qquad \quad \mbox{$\downarrow$ \  $\theta=3^\circ$ } \\ 
    \end{subfigure}
    \\
    \centering
    \begin{subfigure}[b]{0.32\textwidth}
       LSID $\rightarrow$ \hspace{2.4cm}  \mbox{\emph{(a)} } \\ 
        \includegraphics[width=\textwidth]{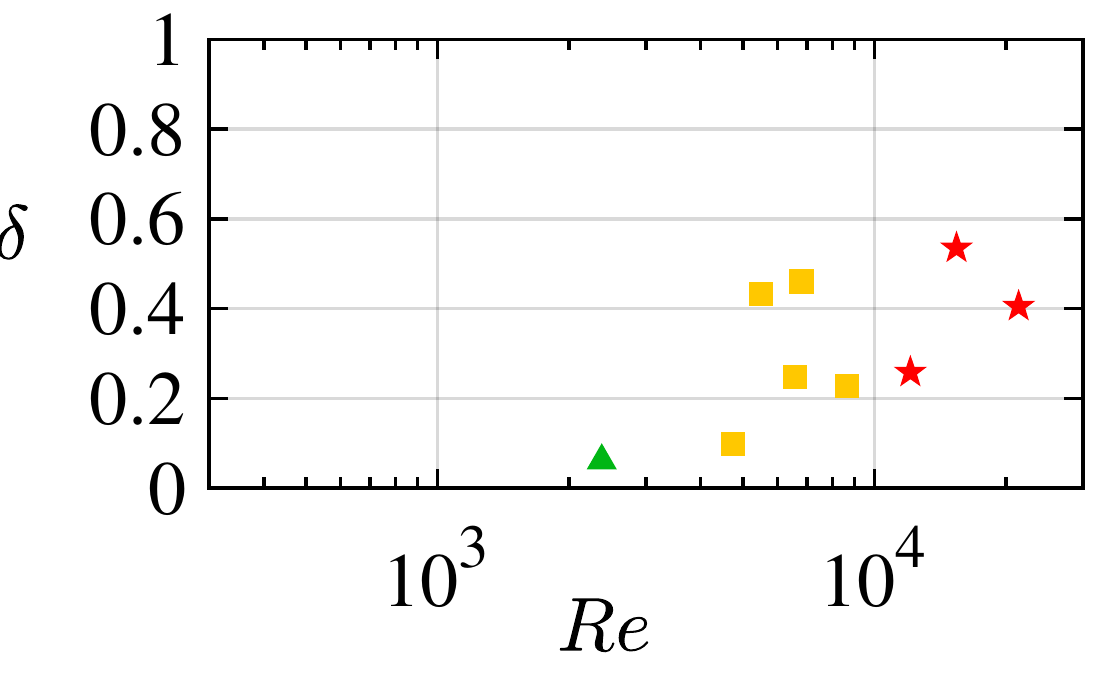}
    \end{subfigure}
    \begin{subfigure}[b]{0.32\textwidth}
           \hspace{3.7cm}   \mbox{\emph{(b)} } \\ 
        \includegraphics[width=\textwidth]{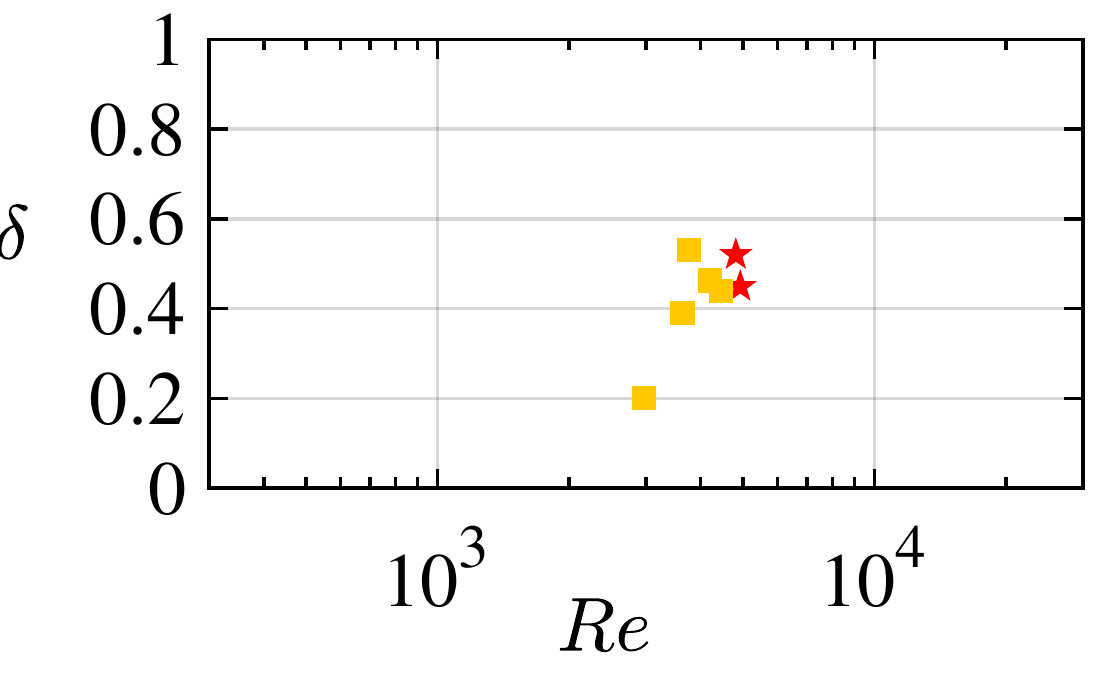}
    \end{subfigure}
        \begin{subfigure}[b]{0.32\textwidth}
             \hspace{3.7cm}   \mbox{\emph{(c)} } \\ 
        \includegraphics[width=\textwidth]{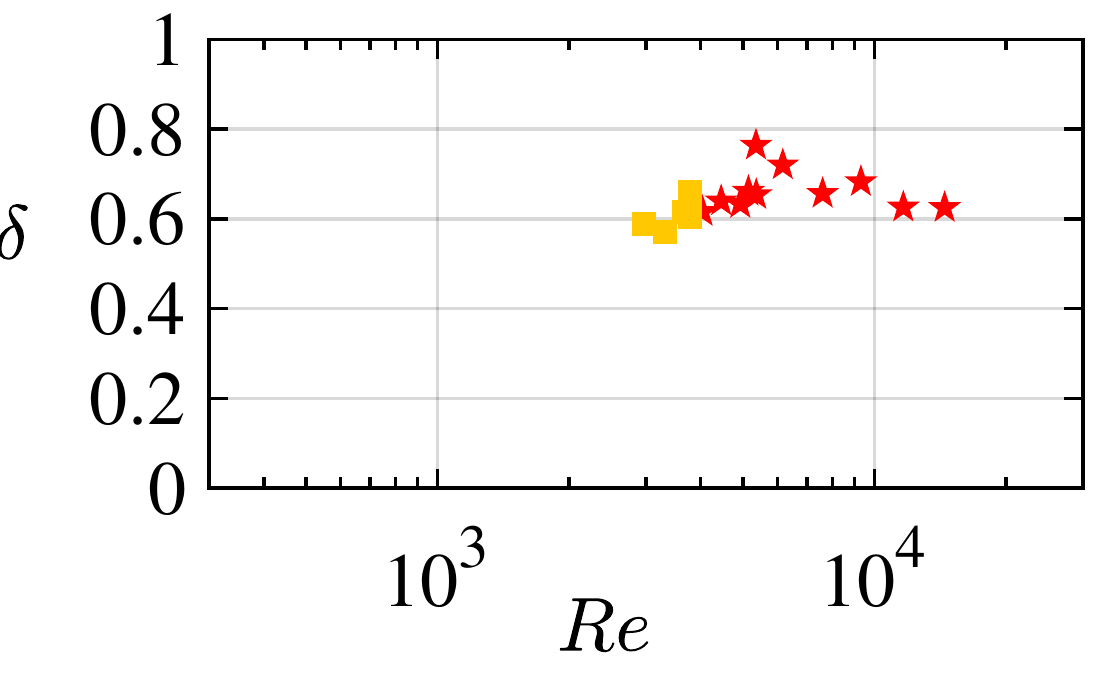}
    \end{subfigure}
\\
    \centering
     \begin{subfigure}[b]{0.32\textwidth}
      HSID $\rightarrow$ \hspace{2.4cm} \mbox{\emph{(d)}   }  \\
        \includegraphics[width=\textwidth]{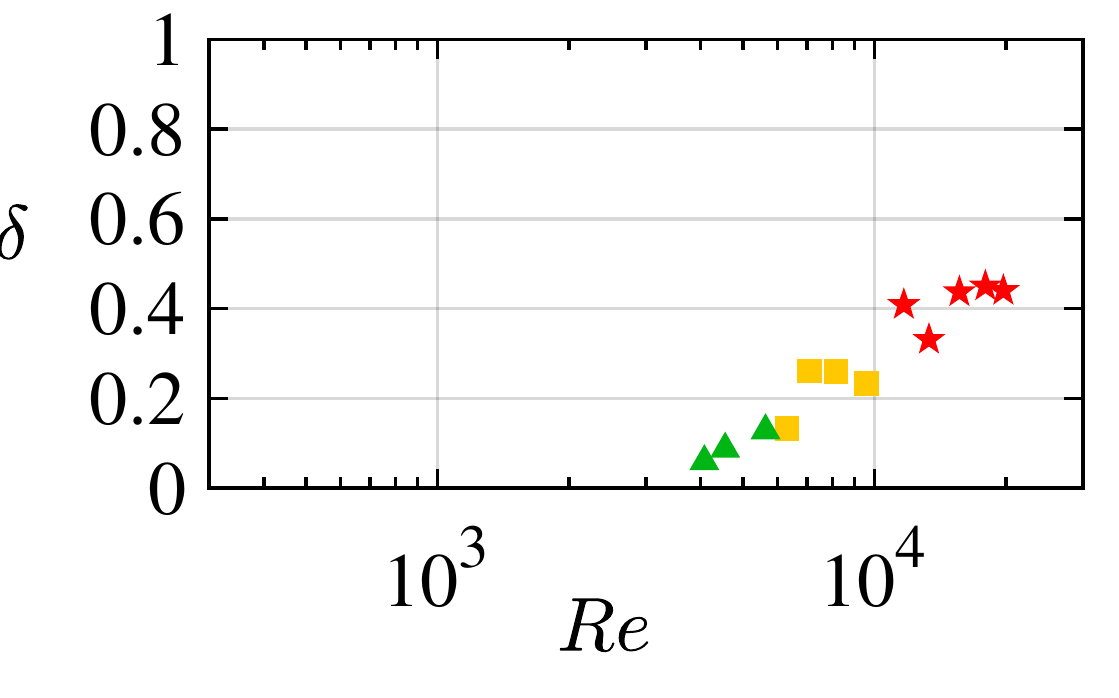}
    \end{subfigure}
    \begin{subfigure}[b]{0.32\textwidth}
           \hspace{3.7cm}   \mbox{\emph{(e)} } \\ 
        \includegraphics[width=\textwidth]{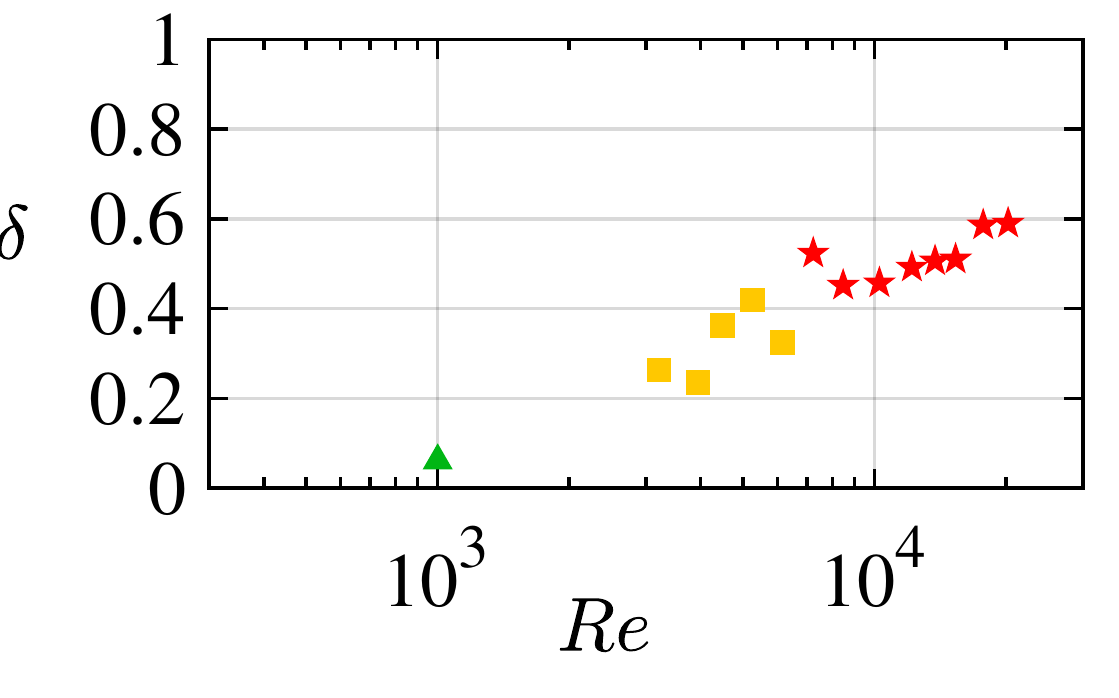}
    \end{subfigure}
        \begin{subfigure}[b]{0.32\textwidth}
             \hspace{3.7cm}   \mbox{\emph{(f)}} \\ 
        \includegraphics[width=\textwidth]{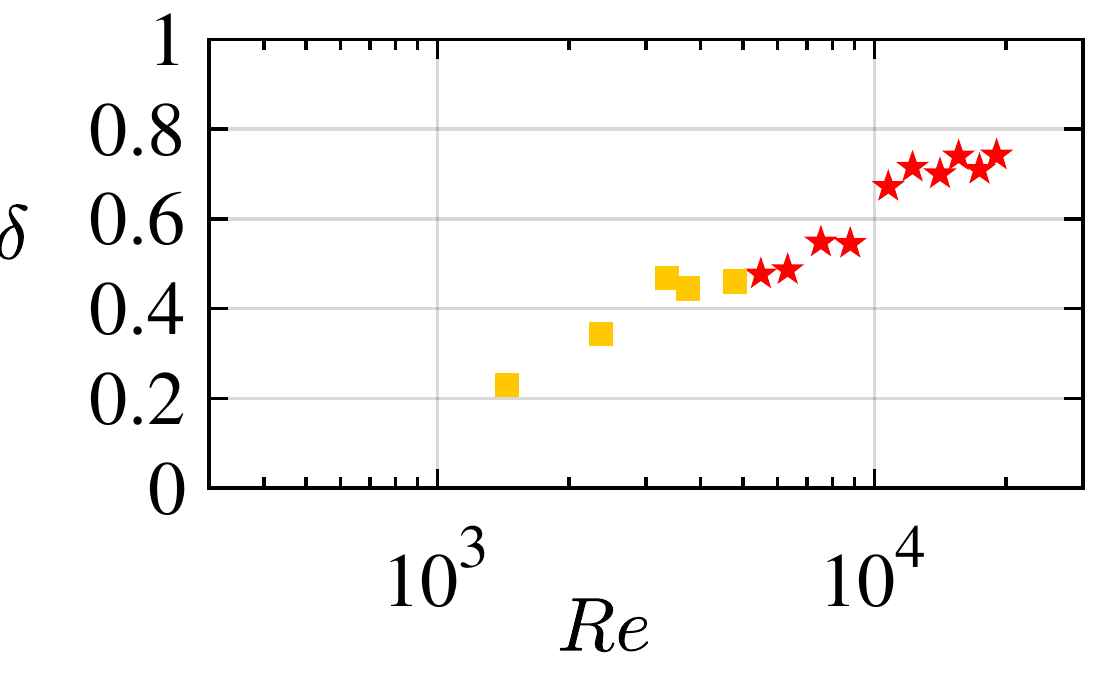}
    \end{subfigure}
\\
     \begin{subfigure}[b]{0.32\textwidth}
        mSID $\rightarrow$ \hspace{2.4cm}    \mbox{\emph{(g)}   } \\ 
        \includegraphics[width=\textwidth]{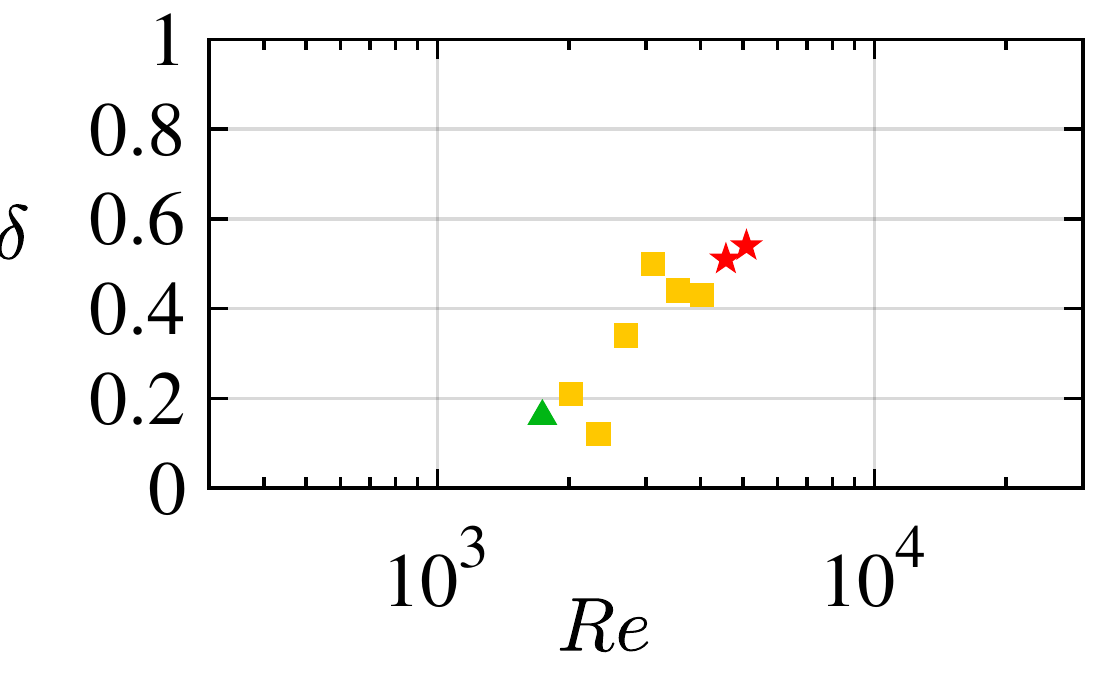}
    \end{subfigure}
    \begin{subfigure}[b]{0.32\textwidth}
          \hspace{3.7cm}    \mbox{\emph{(h)} } \\ 
        \includegraphics[width=\textwidth]{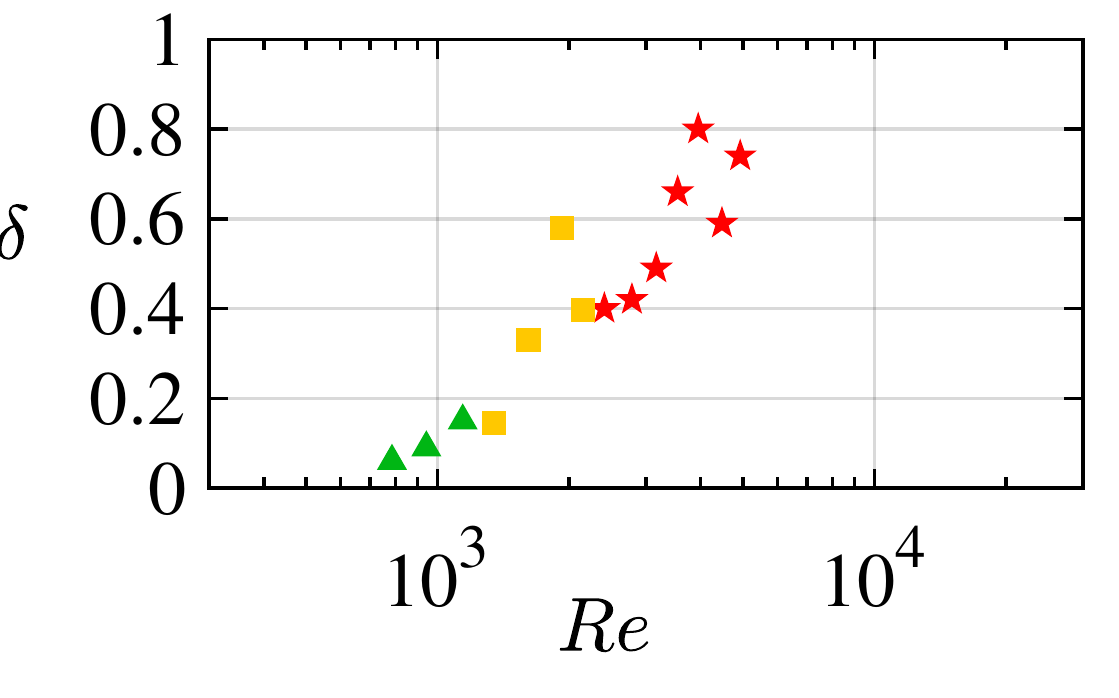}
    \end{subfigure}
        \begin{subfigure}[b]{0.32\textwidth}
        \hspace{3.7cm}      \mbox{\emph{(i)}  } \\ 
        \includegraphics[width=\textwidth]{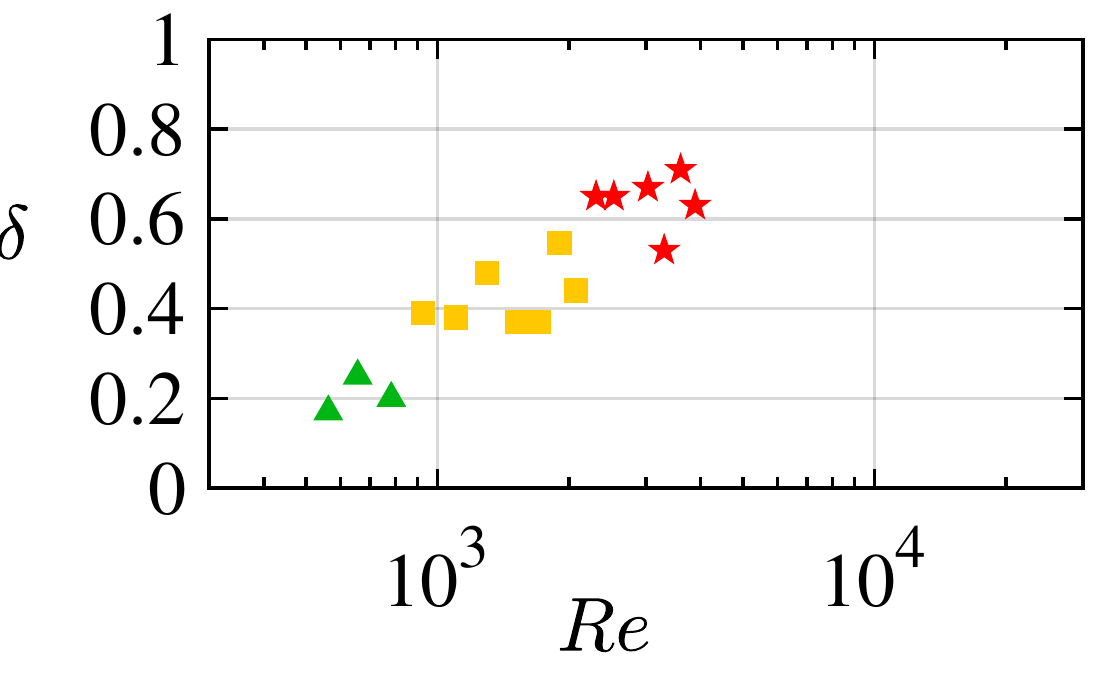}
    \end{subfigure}
    \\
    \centering
      \begin{subfigure}[b]{0.32\textwidth}
     tSID $\rightarrow$ \hspace{2.4cm}   \mbox{\emph{(j)}  } \\ 
        \includegraphics[width=\textwidth]{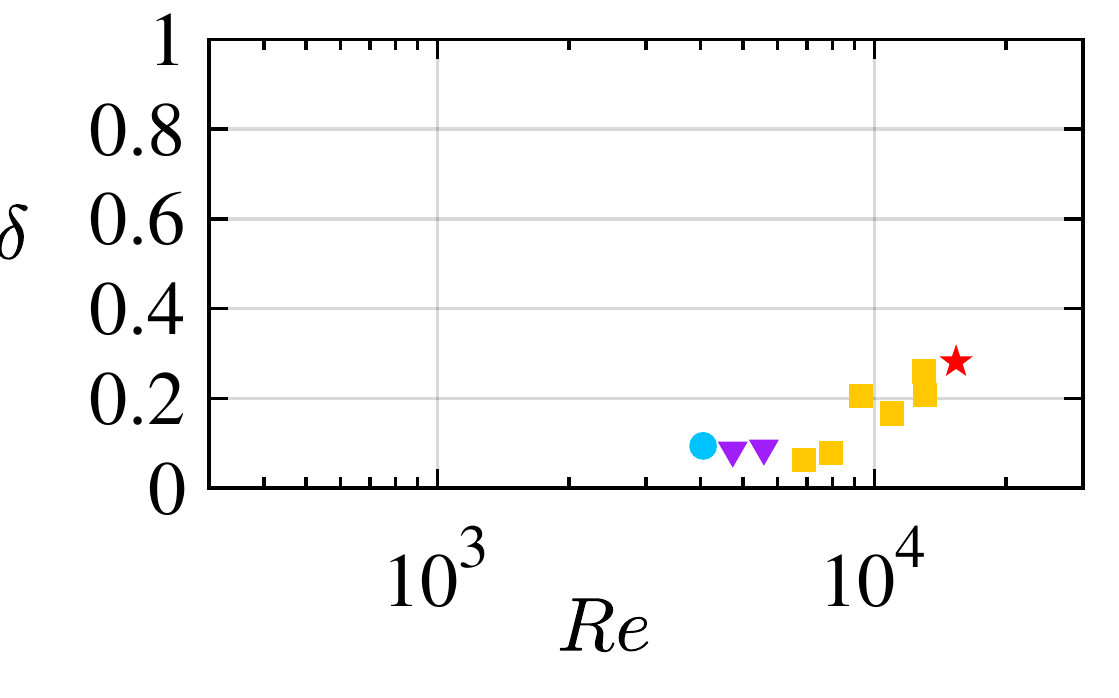}
    \end{subfigure}
    \begin{subfigure}[b]{0.32\textwidth}
          \hspace{3.7cm}   \mbox{\emph{(k)}  } \\ 
        \includegraphics[width=\textwidth]{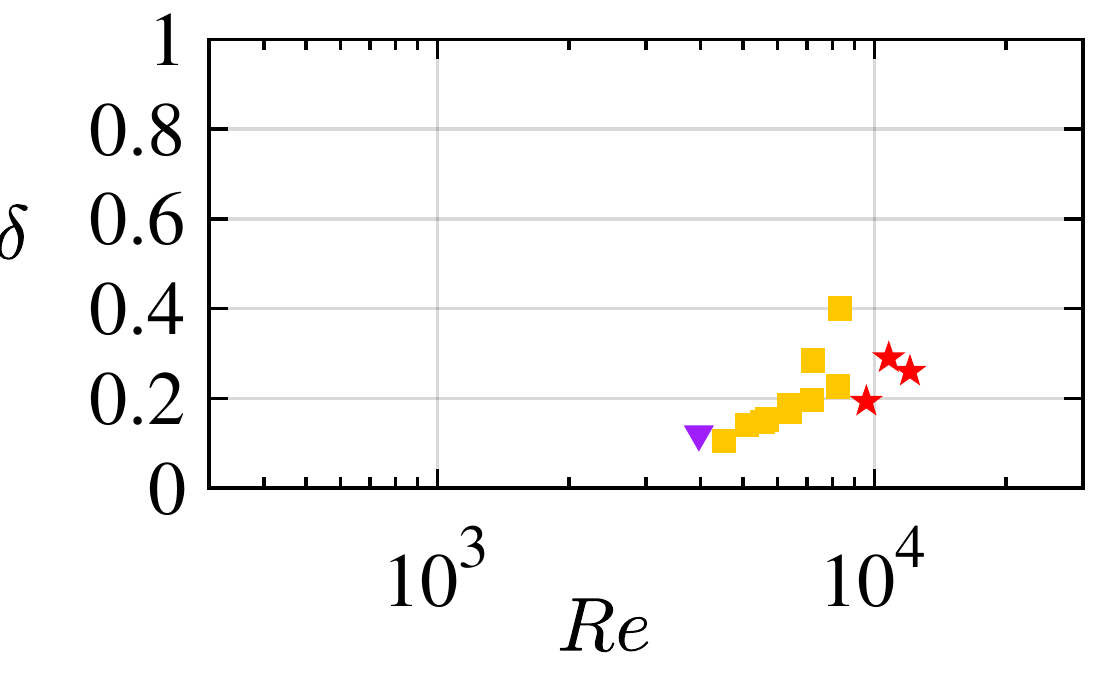}
    \end{subfigure}
        \begin{subfigure}[b]{0.32\textwidth}
          \hspace{3.7cm}      \mbox{\emph{(l)}  } \\ 
        \includegraphics[width=\textwidth]{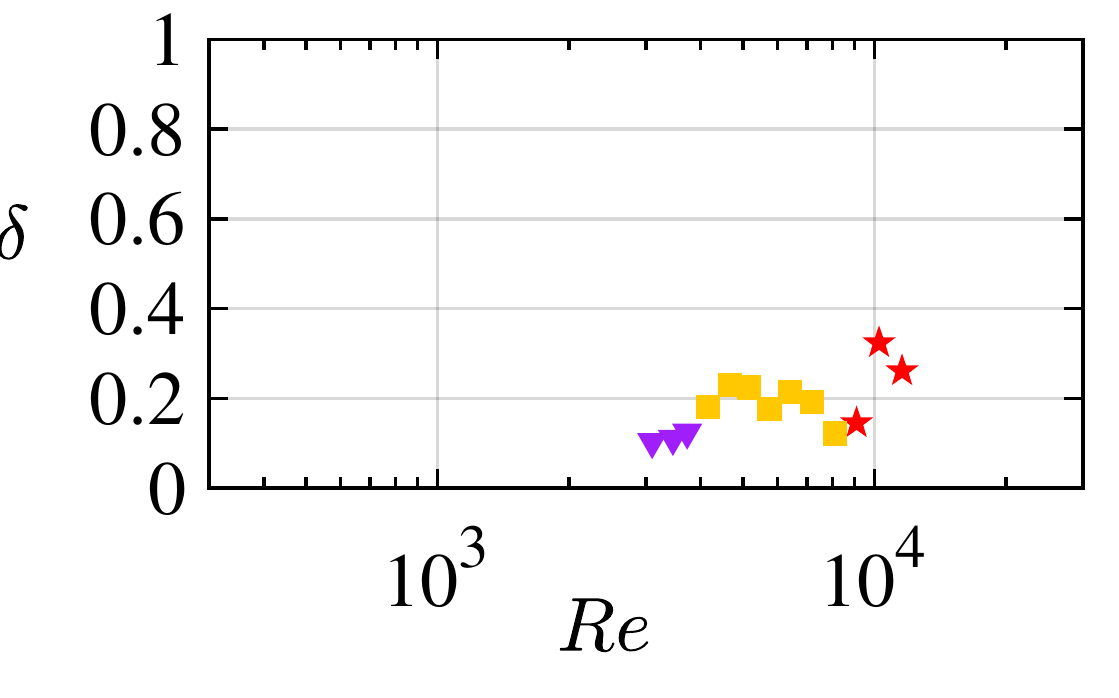}
    \end{subfigure}
    \caption{Interfacial density layer thickness $\delta(Re)$ in salt experiments for three selected angles $\theta=1^\circ, 2^\circ, 3^\circ$ (only a fraction of the available data) and for the four duct geometries: \emph{(a-c)} LSID, \emph{(d-f)} HSID, \emph{(g-i)} mSID, \emph{(j-l)} tSID. Symbol colour denotes flow regime as in previous figures. }
  \label{fig:delta-diagrams}
\end{figure}

%
%
%
\begin{figure}
    \centering
    \begin{subfigure}[b]{0.495\textwidth}
    \mbox{\emph{(a)} \quad LSID  } \\ 
        \includegraphics[width=0.95\textwidth]{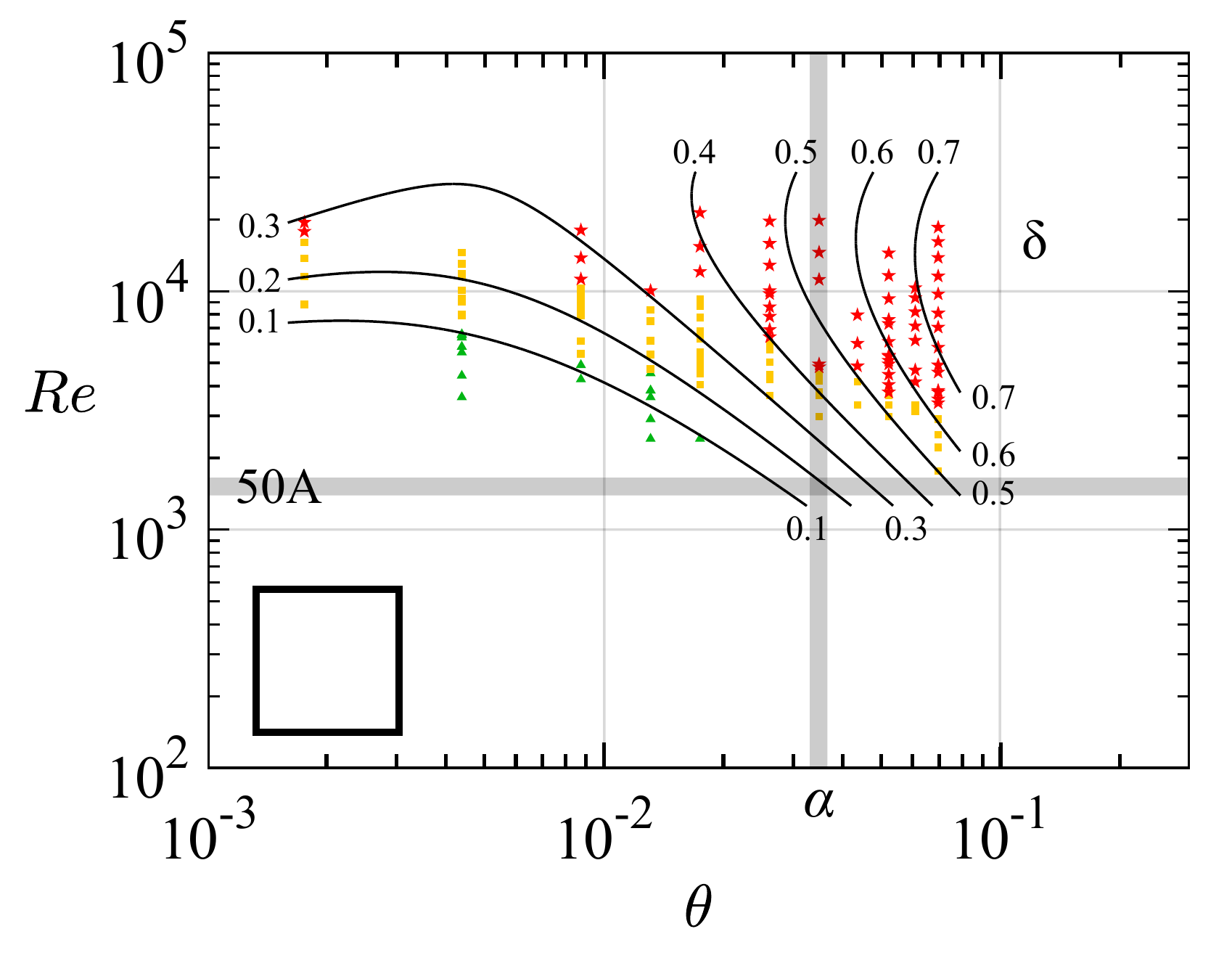}
    \end{subfigure}
    \begin{subfigure}[b]{0.495\textwidth}
          \mbox{\emph{(b)}   \quad HSID } \\ 
        \includegraphics[width=0.95\textwidth]{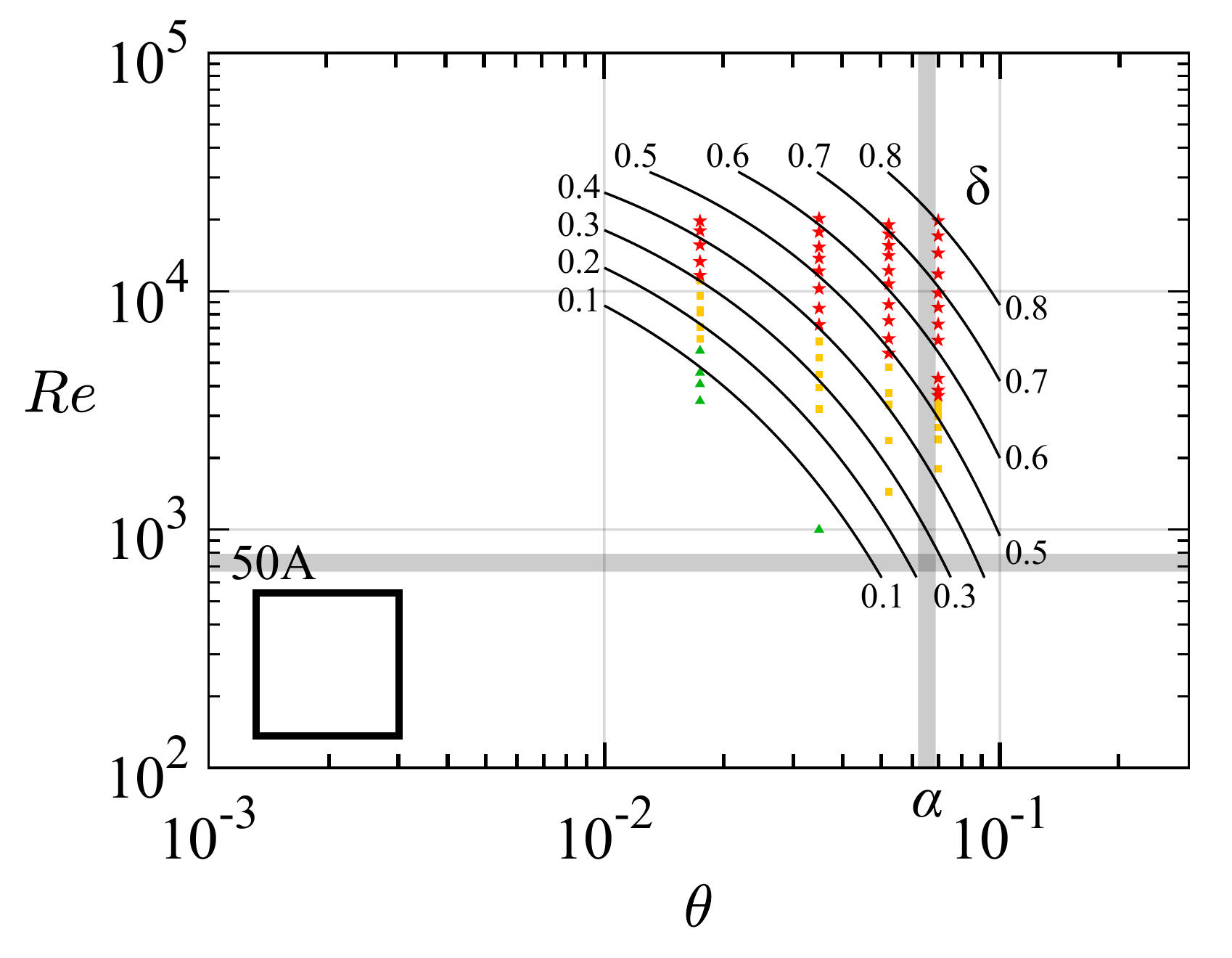}
    \end{subfigure}
\\
    \centering
    \begin{subfigure}[b]{0.495\textwidth}
      \mbox{\emph{(c)}\quad mSID  } \\ 
        \includegraphics[width=0.95\textwidth]{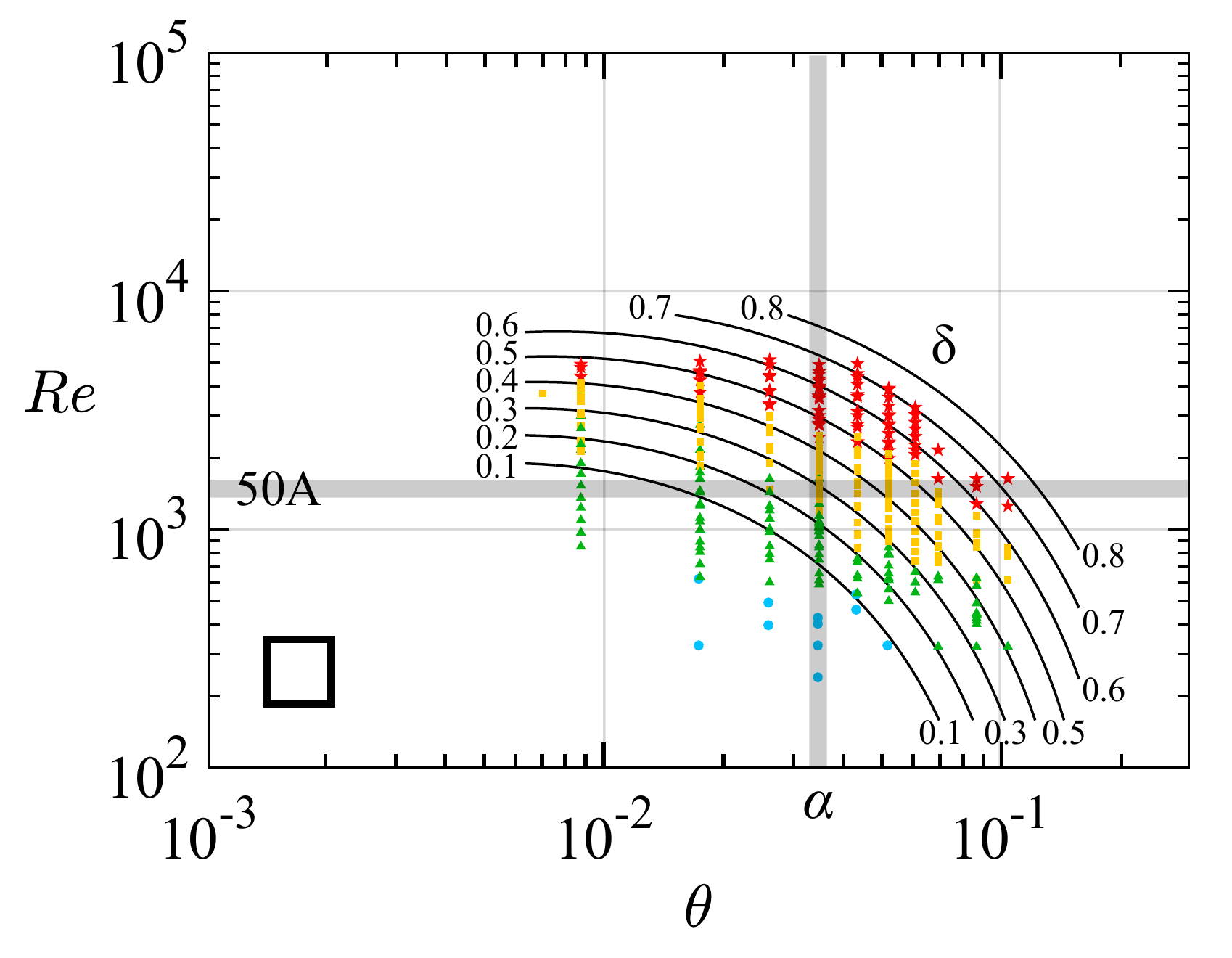}
    \end{subfigure}
    \begin{subfigure}[b]{0.495\textwidth}
            \mbox{\emph{(d)}  \quad  tSID } \\ 
        \includegraphics[width=0.95\textwidth]{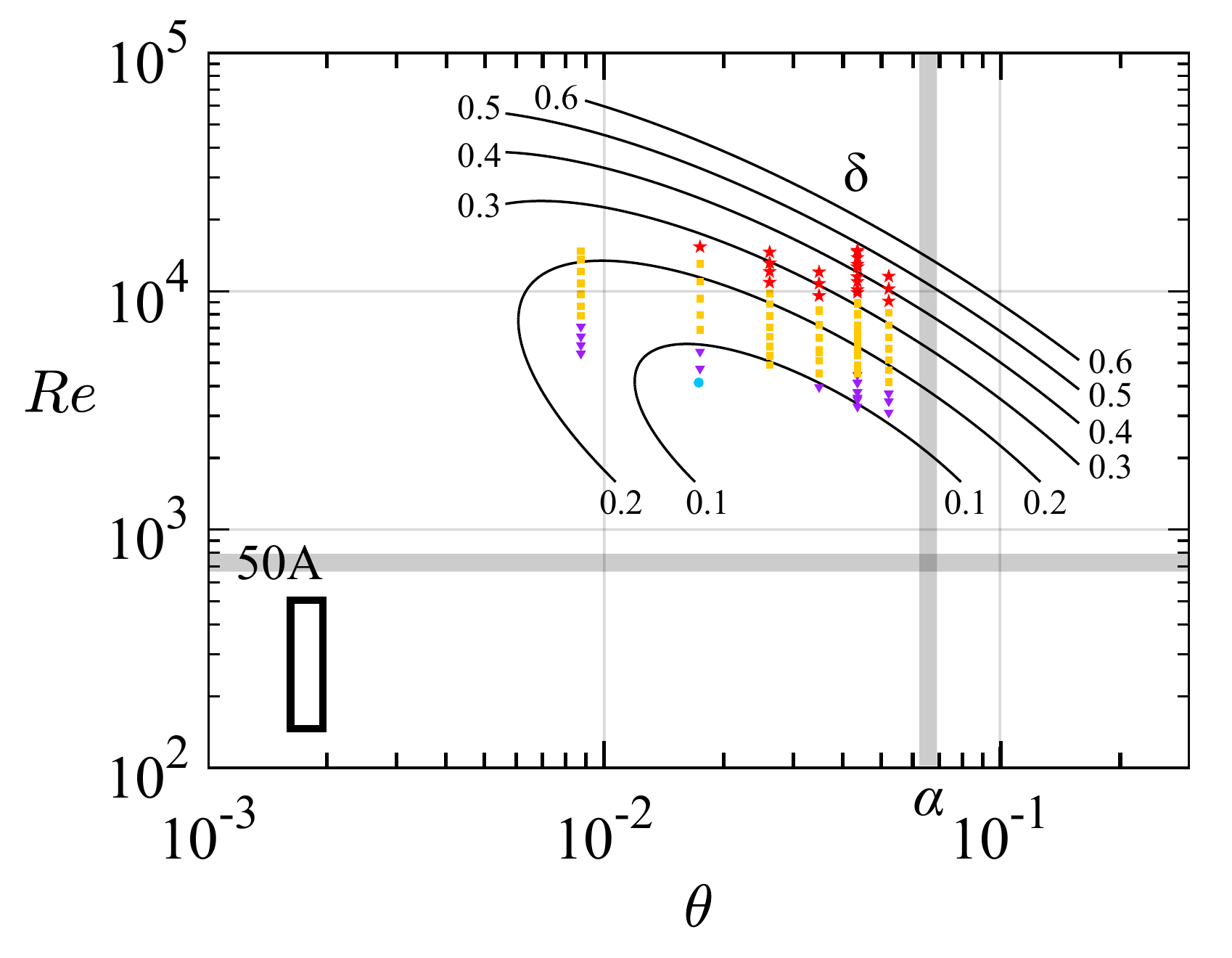}
    \end{subfigure}

    \caption{Interfacial density layer thickness $\delta$ in salt experiments fitted from \emph{(a)} LSID: 115 points ($R^2=0.88$), \emph{(b)} HSID: 58 data points ($R^2=0.97$), \emph{(c)} mSID: 91 data points ($R^2=0.80$), \emph{(d)}  tSID: 87 data points ($R^2=0.75$). Symbol denotes location of the $\delta$ data and colour denotes flow regime. Grey shading denotes $\theta=\alpha$ and $Re=50A$. 
    }
  \label{fig:interfacial-thickness}
\end{figure}

In figure~\ref{fig:delta-diagrams} we plot $\delta$ for our four duct geometries (rows) and three particular angles (representing a subset of our data) $\theta=1^\circ,2^\circ,3^\circ$ (columns). In figure~\ref{fig:interfacial-thickness} we plot a quadratic fit (black contours) to all the available data (represented by the symbols) in the $(\log \theta, \log Re)$ plane following \eqref{MF_contour_fit}. We also added in grey shading the $\theta=\alpha$ and $Re=50A$ values of interest for comparison between panels. In both figures, the colour of the symbol denotes the flow regimes as in figures~\ref{fig:regime-diagrams}-\ref{fig:MF-diagrams}. 

We observe in both figures that $\delta$ monotonically increases with both parameters $\theta$ and $Re$, starting from values as low as $\delta \approx 0.05$ in the $\LL$, $\HH$, and $\WW$ regimes (see figure~\ref{fig:delta-method}\emph{(a)} for an illustration with $\delta = 0.069$), and ending with values as high as $\delta \approx 0.8$ in the $\TT$ regime (see figure~\ref{fig:delta-method}\emph{(a)} for an illustration with $\delta =0.47$). The upper bound corresponds to the turbulent mixing layer filling 80~\% of the duct height, with unmixed fluid only filling the remaining top and bottom 10~\%. We substantiate this value of the lower bound by the thickness of the 99~\% laminar boundary layer resulting from the balance between streamwise advection and vertical diffusion of an initially step-like density field. This calculation  gives, at any point in the duct, $\delta_{99} \approx 10 A^{1/2} (Re \, Pr)^{-1/2} \approx 0.03-0.1$ in the range $Re \in [300,6000]$ where the $\LL,\HH,\WW$ regimes are found.

As already discussed in the regimes and $Q_m$ data, we also observe a clear role of the dimensional parameter $H$ in `shifting' the LSID/HSID/tSID data to higher $Re$ than the mSID data and hindering their overlap. Note that $A$ and $B$ play additional, more subtle roles shown by the differences between the LSID and HSID data and between the HSID and tSID data respectively.

Finally, we observe in figure~\ref{fig:interfacial-thickness}  that there is good agreement between  iso-$\delta$ contours and `iso-regime' curves, or regime transitions (not shown for clarity, but easily visualised by the colours of the symbols). This suggests that $\delta$ is more closely correlated to regimes than $Q_m$ is.

\section{Models and discussion}

In this section, we attempt to explain some of the above observations with three particular classes of models, whose prior success in the literature make them natural candidates to tackle this problem. 

In \S~\ref{sec:model-energ} we attempt to explain the scaling of regime transitions at high $Re \gg 50A$ by generalising the time- and volume-averaged energetics analysis of LPL19. In \S~\ref{sec:model-fric}, we investigate the scaling of regimes and $Q_m$ by developing a frictional two-layer hydraulic model. In \S~\ref{sec:model-mixing}, we  tackle the scaling of $\delta$ in the $\II$ and $\TT$ regimes by a variety of turbulence mixing models.

\subsection{Volume-averaged energetics} \label{sec:model-energ}

We recall from \S~\ref{sec:review} that the simultaneous, volumetric measurements of the density and three-component velocity fields of LPL19 confirmed their theoretical prediction that in forced flows ($\theta\gtrsim \alpha$) the time- and volume-averaged norm of the strain rate tensor (non-dimensional dissipation) followed the scaling $\langle \mathsf{s}^2 \rangle_{x,y,z,t} \sim \theta Re$ \eqref{definition-s2}.  They further decomposed the dissipation into:

\begin{myenumi}
\item[\textnormal{(i)}\hspace{2.6ex}]   a `two-dimensional' component $\mathsf{s}^2_{2d}$ (based on the $x-$averaged velocity $\uu_{2d}\equiv \langle \uu \rangle_x$). LPL19 measured flows in the mSID geometry at $Re<2500$, i.e. $Re \not \gg 50A=1500$, in which case the viscous interfacial and top and bottom wall boundary layers are well or fully developed and $\mathsf{s}^2_{2d} \sim \langle (\partial_{z} u_{2d})^2\rangle_{x,y,z,t}=O(1)$. They indeed observed that $\langle \mathsf{s}^2_{2d}\rangle{x,y,z,t}$ plateaus at $\approx 4$ in the $\II$ and $\TT$ regimes due to the hydraulic limit; 

\item[\textnormal{(ii)}\hspace{2ex}]  a complementary `three-dimensional' part $\mathsf{s}^2_{3d} = \mathsf{s}^2 - \mathsf{s}^2_{2d}$ which, as a consequence of the plateau of $\mathsf{s}^2_{2d}$, takes over in the $\II$ and $\TT$ regime and explains the $\theta Re$ scaling of regime transitions for forced flows in mSID. 

\end{myenumi}

In flows at $Re \gg 50A$ (well above the horizontal grey shading in figures~\ref{fig:regime-Re-sintheta},~\ref{fig:interfacial-thickness}) we expect the 99~\% viscous boundary layers to be of typical thickness $\delta_{99} \sim 10 A^{1/2}Re^{-1/2} \ll 1$, and therefore volume-averaged two-dimensional dissipation to be higher $\mathsf{s}^2_{2d} \sim \langle (\partial_{z} u_{2d})^2\rangle_{x,y,z,t} \sim 10^{-1}A^{-1/2}Re^{1/2} \gg 1$. Therefore, we extend the  prior results of LPL19 that regime transitions correspond to threshold values of
\begin{equation}\label{s2d-s3d-qmResintheta-lowRe}
\langle \mathsf{s}^2_{3d} \rangle_{x,y,z,t}  \sim  \theta Re \quad \text{for} \quad Re < 50A,
\end{equation}
by conjecturing that they correspond to threshold values of
\begin{equation}\label{s2d-s3d-qmResintheta-highRe}
\langle \mathsf{s}^2_{3d} \rangle_{x,y,z,t}  \sim  \theta Re - A^{-1/2} Re^{1/2} \quad \text{for} \quad Re \gg 50A,
\end{equation}
which introduces $A$ and a different exponent to $Re$ into the scaling.

Unfortunately we have little regime data for forced flows at $Re \gg 50A$ (upper right quadrants of each panel in figure~\ref{fig:regime-Re-sintheta}) except in LSID (panel~\emph{(a)}). Nevertheless, it does not appear that this conjectured scaling would be able to explain the observed $\theta Re^2$ scaling. Detailed flow measurements would be required in this geometry to confirm or disprove the above two assumptions that two-dimensional dissipation follows a different scaling, and that regime transitions are tightly linked to three-dimensional dissipation.

Furthermore, we recall that the under-determination of the energy budgets of lazy flows ($\theta<\alpha$, see LPL19 figure~8\emph{(a)}) does not allow us to predict the rate of energy dissipation ($\mathsf{s}^2$) from the rate of energy input ($\sim \theta Re$) and therefore to substantiate the transition scalings in lazy flows (left two quadrants of each panel in  figure~\ref{fig:regime-Re-sintheta}).

\subsection{Frictional two-layer hydraulics}\label{sec:model-fric}

We introduce the fundamentals of this model in \S~\ref{sec:fric-intro}, before examining the physical insight it provides in \S~\ref{sec:fric-insight}, and its implications for the scaling of regime transitions and mass flux in \S~\ref{sec:fric-implication}.

\subsubsection{Fundamentals} \label{sec:fric-intro}

\begin{figure}
\centering
\includegraphics[width=0.99\textwidth]{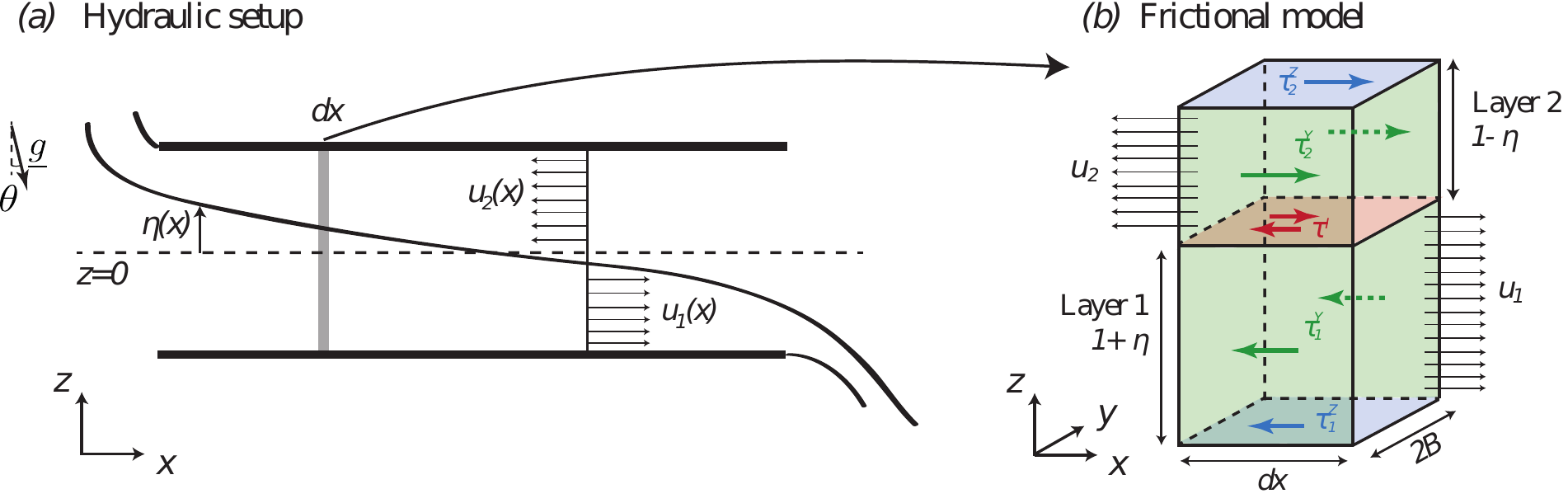}
\caption{Schematics of the \emph{(a)} hydraulic model setup and notation and \emph{(b)} the frictional model with stresses acting on the: top and bottom walls $\tau_{1,2}^Z$ (in blue), side walls $\tau_{1,2}^Y$ (in green) and interface $\tau^I$ (in red) of an infinitesimally small slab of fluid $dx$.}\label{fig:hydraulics-model}
\end{figure}

The two-layer hydraulic model for exchange flows (figure~\ref{fig:hydraulics-model}\emph{(a)}) assumes two layers flowing with non-dimensional velocities $u_1(x)>0$ (lower layer) and $u_2(x)<0$ (upper layer), and separated by an interface of non-dimensional elevation $\eta(x) \in [-1,1]$ above the neutral level $z=0$. 

In the idealised \emph{inviscid} hydraulic model (i.e. in the absence of viscous friction) the conservation of volume and of Bernoulli potential, and the requirement of hydraulic control yield a horizontal and symmetric interface $\eta(x)=0$ for $x \in [-A,A]$ and a volume flux $Q=u_1=-u_2=\nicefrac{1}{2}$ as already mentioned in \S~\ref{sec:scaling-of-vel}  (see appendix~\ref{sec:appendix-frictional-1} for more details).

The \emph{frictional} hydraulic model is of more relevance to SID flows at finite $Re$. This model parameterises the effects of viscous friction while retaining the hydraulic assumptions (hydrostatic, steady, two-layer flow with uniform velocities $u_{1,2}(x)$). Dating back to  \cite{schijf_theoretical_1953,assaf_sea_1974,anati_laboratory_1977}), it was formalised by \cite{zhu_hydraulics_2000,gu_frictional_2001,gu_analytical_2005}, who considered the effects of friction at the interface and bottom wall only, with applications to wide, open, horizontal channels. Here we further develop this model to add the effects of gravitational forcing ($\theta>0$) and friction at the top and side walls. The full development of this model can be found in (L18, \S~5.2) and we offer a summary in appendix~\ref{sec:appendix-frictional}. Some of its conclusions were briefly discussed in LPL19 \S~4.3.1 (e.g. the distinction between lazy/forced flows). Below we provide a concise self-contained presentation of the key results of this model regarding the particular problem of the scaling of regimes and $Q_m$.

As sketched in figure~\ref{fig:hydraulics-model}\emph{(b)}, we consider that each infinitesimal duct sub-volume $dx\times 2B \times 2$ centred around $x$ is subject to horizontal, resistive stresses at the bottom wall $\tau^Z_1(x)$, top wall $\tau^Z_2(x)$ (in blue), side walls $\tau^Y_{1,2}(x)$ (respectively in the bottom and top layers, in green), and interface $\tau^I$ (in red). The inclusion of these stresses in the evolution of Bernoulli potential along the duct (see \S~\ref{sec:appendix-frictional-1}) yields a nonlinear differential equation for the slope of the interface along the duct of the form 
\begin{equation} \label{eta-ODE}
\eta'(x) = \eta'(\eta,Q,\theta,Re, f_Z,f_Y,f_I)
\end{equation}
(see \eqref{deta/dx} for the full expression). Here $f_Z,f_Y,f_I$ are constant friction factors parameterising respectively the top and bottom wall stress, the side wall stress, and the interfacial stress, (they can be determined \emph{a posteriori} from any finite-$Re$ flow profile, see \S~\ref{sec:appendix-frictional-2} and \eqref{fzfyfi-2}). For any set of parameters $\theta,Re, f_Z,f_Y,f_I$, this dynamical equation can be combined with the hydraulic control condition and solved numerically using an iterative method to yield a unique solution for $Q$ and $\eta(x)$ (\S~\ref{sec:appendix-frictional-3}). The volume flux $Q$ generally increases with the forcing $\theta Re$, and decreases with friction  $f_Z,f_Y,f_I,$ and $A$.

\subsubsection{Physical insight}\label{sec:fric-insight}

We now consider the mid-duct slope $\eta'(x=0)$, whose simplified expression shows the balance between the forcing $\theta Re$ and the `composite friction parameter' $F$:
\begin{equation} \label{existence-condition}
\eta'(0) = \frac{\theta Re  - 2QF} {Re(1-4Q^2)} \quad \text{where} \quad F \equiv f_Z(1+2r_Y+8r_I),
\end{equation}
and $r_Y\equiv B^{-1} f_Y/f_Z$ and $r_I \equiv f_I/f_Z$ are respectively the side wall friction ratio and the interfacial friction ratio. 

We further note that our model has three properties: \emph{(i)} the interface must slope down everywhere ($\eta'(x)<0$) since the lower layer accelerates convectively from left to right ($u_1u'_1(x)>0$) and conversely ($u_2 u'_2(x)<0$);  \emph{(ii)} the interface must remain in the duct $|\eta(x=\pm A)|<1$; \emph{(iii)} $\eta'$ always reaches a maximum ($|\eta'|$ reaches a minimum) at the inflection point $x=0$. 

From these properties we deduce that the existence of a solution requires the mid-duct interfacial slope to satisfy
\begin{equation}\label{inequality-slope-1} 
 -A^{-1} <\eta'(0)  < 0,
\end{equation}
and therefore, using \eqref{existence-condition}, we obtain the following bounds:
\begin{equation} \label{inequality-slope-2}
 \quad  \theta Re < 2QF < (1+b) \theta Re \quad \text{where} \quad b(A,\theta,Q) \equiv \frac{1-4Q^2}{A\theta}
\end{equation}
The first inequality in \eqref{inequality-slope-1} comes from property \emph{(ii)} and means that the mid-duct interfacial slope must not be too steep compared to the duct geometrical slope $A^{-1}\approx \alpha$. The second inequality  comes from \emph{(i)} and \emph{(iii)} and means that the mid-duct interfacial slope must be negative for $\eta(x)$ to be negative everywhere.  

When suitably rescaled by $2Q \in [0,1]$, the combined friction parameter $F$ must therefore follow a $\theta Re$ scaling, strictly bounded from below.  The upper bound in \eqref{inequality-slope-2} is loose ($b>0$) in lazy flows, and  tight  ($b\rightarrow 0$) in forced flows ($A\theta \approx \theta/\alpha \gg 1$ and $Q \rightarrow \nicefrac{1}{2}$).

\subsubsection{Implications for regimes and $Q_m$}\label{sec:fric-implication}

\begin{figure}
\centering
\includegraphics[width=0.65\textwidth]{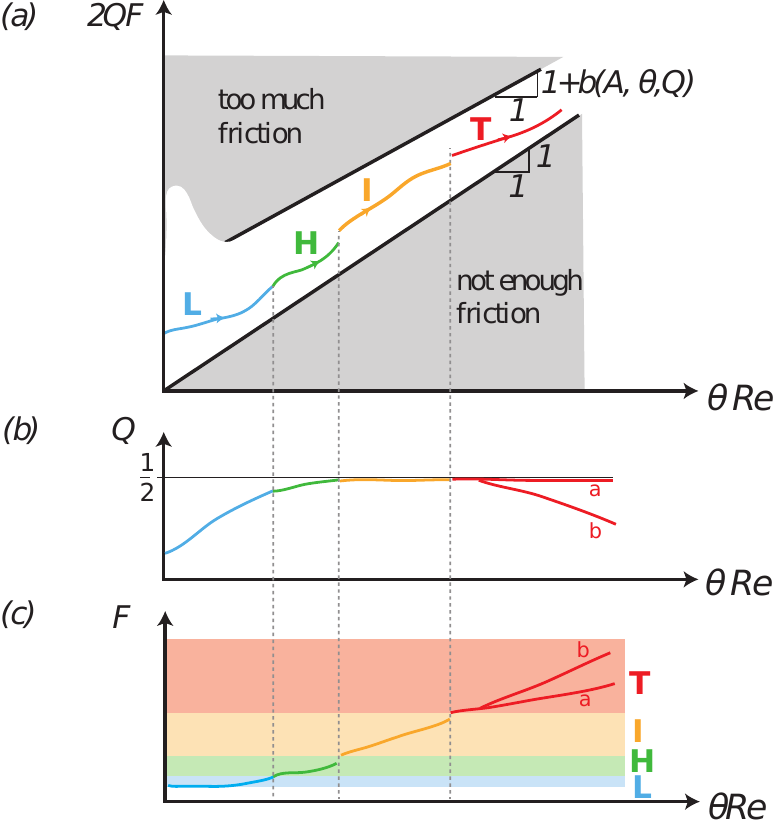}
\caption{Conclusions of the frictional hydraulic model as the `forcing parameter' $\theta Re$ is increased: \emph{(a)}  $2QF$ is bounded above and below by \eqref{inequality-slope-2}; \emph{(b)} volume flux $Q$, and \emph{(c)} composite friction parameter $F$ (a and b in the $\TT$ regime denote two possible scenarios). We conjecture that regime transitions correspond to threshold values of $F$.}\label{fig:hydraulics-hyp}
\end{figure}

Combining the above physical insight with our experimental observations, we conjecture the following behaviour about regimes and $Q_m$, summarised in figure~\ref{fig:hydraulics-hyp}.

\begin{myenumi}
\item[\textnormal{(i)}\hspace{2.6ex}]  At zero or `low'  $\theta Re$ (i.e. at $\theta\approx 0$, since $Re$ must be large for hydraulic theory to hold) due to the inevitable presence of wall and interfacial friction ($F>0$) and the looseness of the upper bound $b$, $2QF$ is typically well above the forcing $\theta Re$. The friction $F$ is independent of $\theta Re$ and the flow is typically laminar ($\LL$ regime). The interface has a noticeable slope all along the duct $\eta'(0)\ll 0$, associated with a small volume flux $Q \ll \nicefrac{1}{2}$ (see \eqref{G*2}).  Such \emph{lazy flows} are underspecified, and the scaling of $Q$ and $F$ with $\theta Re$ is therefore impossible to predict \emph{a priori}.

\item[\textnormal{(ii)}\hspace{2ex}]  At moderate $\theta Re$ ($\theta > 0$): $2QF$ increases above its `default' $\theta = 0$ value. This is achieved, on one hand, through an increase in $Q$ (and therefore $Q_m$), making the flow approach the hydraulic limit (panel~\emph{(b)}), and on the other hand, through an increase in $F$, in particular through laminar interfacial shear ($r_I$), rendering the flow unstable to Holmboe waves above a certain threshold ($\LL \rightarrow \HH$ transition, panel~\emph{(c)}). The phenomenology of this  transition agrees with that proposed by the energetics of LPL19 (see their \S~6.2-6.3). The fact that the $\LL \rightarrow \HH$ (or $\LL \rightarrow \WW$)  transition exhibits different scalings in our different data sets is not presently understood. It may come from the complex, individual roles of $Q$ and $F$ in the precise flow profiles $u(y,z), \rho(z)$ responsible for  triggering the Holmboe instability, and the different scalings of $Q$ and $F$ that could allow $2QF$ to follow a $\theta Re$ scaling.

\item[\textnormal{(iii)}\hspace{1.4ex}]  At high $\theta Re$: the hydraulic limit is reached, the upper bound is tight ($b\approx 0$), the interface is mostly flat ($\eta(x)\approx 0$ everywhere), and the inequality \eqref{inequality-slope-2} becomes $2QF\approx F \approx \theta Re$. In such \emph{forced flows}, the friction parameter $F$ alone must precisely balance the forcing.  Arbitrarily large $\theta Re$ requires  arbitrarily large $F$,  which we conjecture is largely achieved by \emph{turbulent interfacial} friction (increase in $r_I$ responsible for the $\HH\rightarrow\II$ and eventually the $\II\rightarrow\TT$ transition). 
\end{myenumi}
From implication (iii), it is natural to conjecture that these two transitions are also caused by threshold values of the interfacial friction ratio $r_I$, which, as explain in appendix~\ref{sec:appendix-frictional-2}, can be written $r_I = 1+K_I$, where $K_I$ is a turbulent momentum diffusivity (non-dimensionalised by the molecular value $\nu$) parameterising interfacial Reynolds stresses (see \eqref{definition-Km}). Assuming that all wall shear stresses are similar ($r_Y\approx 1$), and that interfacial Reynolds stresses eventually dominate over laminar shear ($K_I \gg 1$), we have $K_I \approx F/(8f_Z)$. For $Re<50A$, fully-developed boundary layers yield $f_Z \sim 1$, implying regime transitions scaling with (ignoring pre-factors)
\begin{equation}\label{Km-lowRe}
K_I \sim  \theta Re \quad \text{for} \quad Re < 50A.
\end{equation}
For $Re \gg 50A$, thin top and bottom wall boundary layer arguments similar to those of \S~\ref{sec:model-energ} yield $f_Z\sim A^{1/2}Re^{-1/2}$, implying regime transitions scaling with
\begin{equation}\label{Km-highRe}
K_I \sim   A^{1/2} \theta Re^{1/2} \quad \text{for} \quad Re \gg 50A.
\end{equation}

Comparing \eqref{Km-lowRe}-\eqref{Km-highRe} to \eqref{s2d-s3d-qmResintheta-lowRe}-\eqref{s2d-s3d-qmResintheta-highRe} we see that the $Re< 50A$ scaling obtained with frictional hydraulics is identical to that obtained by the energetics. However, the $Re \gg 50A$ scaling is different, and unfortunately it does not allow us to explain the regime transitions data (a $\theta Re^{1/2}$ or $\theta^2 Re$ scaling is never observed). In addition, direct estimations of friction coefficients using  three-dimensional, three-component velocity measurements in all flow regimes (L18, \S~5.5) suggest \emph{a posteriori} that the assumption that $K_I \gg 1$ might only hold \emph{beyond} the $\II \rightarrow \TT$ transition, undermining its usefulness to predict the $\HH \rightarrow \II$ and $\II \rightarrow \TT$  transitions. 

From implication (ii), we understand why the volume flux $Q$, and hence the mass flux $Q_m$, both increase with $\theta$ and $Re$ in the $\LL$ and $\HH$ regimes, as observed in \S~\ref{sec:results-Qm}. However lazy flows are under-specified; only one equation governs both the volume flux and friction  ($2QF \sim \theta Re$), which does not allow us to obtain the value of the exponent $\gamma$ in the scaling $Q \sim \theta Re^\gamma$. From implication (iii), we conjecture two potential reasons for the decrease of the mass flux $Q_m$ in the $\TT$ regime (labelled `a' and `b' in panels~\emph{b,c}). In scenario `a', $Q_m$ decreases due to increasing mixing despite the volume flux $Q$ staying relatively constant ($2QF\sim F \sim \theta Re$). In scenario `b', $Q_m$ decreases partly due to mixing, and partly due to a decrease in $Q$ (compensated by $F$ increasing faster than $\theta Re$).  Accurate $Q$ and $Q_m$ data obtained by volumetric measurements of velocity and density in L18 (figure 5.12\emph{(b)}) support scenario `b' up to $\theta Re=132$, but additional data are required to draw general conclusions.

Finally,  because this frictional hydraulic model assumes a two-layer flow without any form of mixing, it does not allow us to discuss the behaviour of the interfacial thickness $\delta$. We thus discuss several models that explicitly address mixing in the next section.

\subsection{Mixing models}\label{sec:model-mixing}

The importance and difficulty of modelling interfacial mixing in exchange flows has long been recognised \citep{helfrich_time_1995,winters_role_2000}. However, despite the existence of hydraulic models for multi-layered or continuously-stratified flows \citep{engqvist_self-similar_1996,hogg_continuously_2004}, to date there exists no `three-layer' hydraulic model allowing for the exchange of momentum or mass between the two counter-flowing layers suitable to our problem (which would violate most hydraulic assumptions). Below we review some experimental, numerical, and theoretical work most relevant to the scaling of $Q_m$, $\delta$, and their relation to fundamental stratified turbulence properties such as diapycnal diffusivity and mixing efficiency.

\subsubsection{Turbulent diffusion models}

\cite{cormack_natural_1974} tackled natural convection in a shallow ($A\gg1$) cavity with differentially heated walls. This problem is analogous to SID flows in the limit of maximum `interfacial' thickness  ($\delta=1$) in which turbulent  mixing dominates to such an extent that the exchange flow is only weakly stratified in the vertical (i.e. $\langle |\partial_z \rho| \rangle_z < 1$ because $|\rho(z=\pm1)| < 1$) and becomes stratified in the horizontal (i.e. $|\partial_x \rho(z=\pm1)| > 0$ and mean isopycnals are no longer horizontal).  In their model, the horizontal hydrostatic pressure gradient is balanced only by a uniform vertical turbulent diffusion with constant $K_T$. Using the terminology of \S~\ref{sec:scaling-of-vel}, this balance could be called the hydrostatic-mixing (or `HM') balance where `mixing' plays a similar role to `viscosity' in the `HV' balance of \S~\ref{sec:scaling-of-vel}. \cite{cormack_natural_1974} solved this problem analytically and found:
\begin{subeqnarray} \label{cormack}
    Q &=& \frac{5}{384}(A K_T)^{-1}, \\
    Q_m &=& 4 A K_T + \frac{31}{1451520}(A K_T)^{-3},
\end{subeqnarray} 
where we assumed a turbulent Prandtl number of unity for simplicity (i.e. the density equation has the same turbulent diffusivity). The above equations are adapted from equations (19) and (20) of \cite{hogg_hydraulics_2001} (in their review of the results of \cite{cormack_natural_1974}) to match our slightly different definitions of $Q,Q_m,A$ and especially our definition of $K_T$ as being non-dimensionalised by the inertial scaling $\sqrt{g'H}H/2$ (giving $K_T = (4Gr_T)^{-1/2}$ where $Gr_T$ is their `turbulent Grashof number'). We also contrast the the uniform diffusivity $K_T$ in this model and the interfacial diffusivity $K_I$ in the frictional hydraulics model of \S~\ref{sec:fric-implication}, which have different roles and different non-dimensionalisation ($\sqrt{g'H}H/2$ \emph{vs} $\nu$, hence  `$K_T=K_I/Re$').
The predictions \eqref{cormack} were verified numerically and experimentally in two papers of the same series \citep{cormack_natural_1974b,imberger_natural_1974}, but only hold in the `high-mixing' limit of $A K_T > \nicefrac{1}{15}$ below which inertia becomes noticeable and the assumptions start to break down (at $A K_T < \nicefrac{1}{25}$, $Q$ and $Q_m$ even exceed the hydraulic limit...). 

\cite{hogg_hydraulics_2001} built on the above results and developed a model with linear velocity and density profiles within an interfacial layer of thickness $\delta<1$ and a uniform turbulent momentum and density diffusivity $K_T$. This models the `IHM' balance, i.e. the transition between the \cite{cormack_natural_1974}  $A K_T> \nicefrac{1}{15}$ high-mixing limit (the `HM' balance where turbulent diffusion dominates over inertia, $\delta=1$, and \eqref{cormack} holds) and the $A K_T \rightarrow 0$ hydraulic limit (the `IH' balance where inertia dominates over mixing, $\delta=0$, and $Q=Q_m=\nicefrac{1}{2}$ holds).  \cite{hogg_hydraulics_2001} argued  that $\delta$ would increase diffusively during the `duct transit' advective timescale $A$, and integrated the linear velocity and density profiles across the interfacial layer to find
\vspace{-0.5cm}

\begin{subeqnarray} \label{hogg}
    \delta &\approx& 5(A K_T)^{1/2}, \\
    Q &\approx& \frac{1}{2}-\frac{5}{4}(A K_T)^{1/2}, \slabel{hogg-2} \\ 
    Q_m &\approx& \frac{1}{2}-\frac{5}{3}(A K_T)^{1/2}, \slabel{hogg-3}
\end{subeqnarray} 
where the prefactors $5,\nicefrac{5}{4},\nicefrac{5}{3}$ come from the imposed matching with the high-mixing solution \eqref{cormack}. \cite{hogg_hydraulics_2001} validated these  predictions with large eddy simulations and found good quantitative agreement for  $Q,Q_m,\delta$ across the range $A K_T \in [\nicefrac{1}{2000},\nicefrac{1}{15}]$, below which convergence to the inviscid hydraulic limit was confirmed.

In order to use these models to explain the scaling of $Q_m$ and $\delta$ with $A,B,\theta,Re,Pr$, we need to \emph{(i)} extend them to the more complex `IHGM' balance of  SID flows in the $\II$ and $\TT$ regimes in which gravitational forcing is present ($\theta>0$); \emph{(ii)} have a model for the scaling of $K_T$ on input parameters (the above models prescribed $K_T$ as an input parameter, but it is \emph{a priori} unknown in the SID). To do so, we propose to use insight gained by the energetics and frictional hydraulics models. 

First, following the results of LPL19 and \S~\ref{sec:model-energ} on the average rate of turbulent dissipation, it is tempting to model $K_T$ using a turbulence closure scheme like the mixing length or $K-\epsilon$ model, yet these require either a lenghtscale or the turbulent kinetic energy, which are both unknown (only the \emph{rate} of dissipation is known, see \eqref{s2d-s3d-qmResintheta-lowRe}-\eqref{s2d-s3d-qmResintheta-highRe}). 

Second, following the frictional hydraulics results of \S~\ref{sec:model-fric}, we may conjecture that the `Reynolds stresses' interfacial diffusivity $K_I$ in the $\II$ and $\TT$ regimes may play a similar role to the uniform turbulent diffusivity in the present model. Recalling that by definition $K_T = K_I/Re$, combining the scalings \eqref{Km-lowRe}-\eqref{Km-highRe}  with \eqref{hogg} would suggest:
\begin{subeqnarray} \label{wild-conjecture}
    \delta \sim  \frac{1}{2} - Q_m &\sim& (A \theta)^{1/2} \quad \quad\quad \quad  \, \text{for}\quad Re < 50A, \slabel{wild-1} \\
      \delta  \sim \frac{1}{2} - Q_m  &\sim&  (A^3 \theta^2 Re^{-1})^{1/4} \quad \text{for} \quad Re \
      \gg 50A.  \slabel{wild-2}
\end{subeqnarray} 
Unfortunately these scalings are not consistent with the observations of figures~\ref{fig:regime-Re-sintheta}-\ref{fig:interfacial-thickness}: $\delta$ is clearly a function of $Re$ for $Re<50A$ (less so at high $Re$ where the $A\theta$ scaling has indeed been observed by K91), and  $\delta$ is clearly not a decreasing function of $Re$ for $Re \gg 50A$.

\subsubsection{Previous mixing efficiency measurements and models}

In this section we discuss two studies of the interfacial layer thickness $\delta$ and its relation to the Richardson number and mixing efficiency as a basis for the development of a more suitable model for SID flows in the next section.

\cite{prastowo_mixing_2008} studied exchange flows through short ($A\approx 2-3$), wide ($B\gg 1$), horizontal ($\theta=0$) contractions.  Their measurements suggest an approximately constant interfacial thickness $\delta \approx 0.23-0.25$ across the range $Re \in [10^4,10^5]$, in rough agreement with previously quoted estimates for shear-driven mixing flows (e.g. \cite{sherman_turbulence_1978}, p.~275 and references therein). They support this observation with `equilibrium' or `marginally stable' Richardson number arguments that the gradient Richardson number should be maintained near the Miles-Howard linear stability threshold, a phenomenon commonly observed subsequently in the observational literature on shear-driven mixing \citep{thorpe_marginal_2009,smyth_marginal_2013}.  Assuming  a linear profile for $u(z)$ and $\rho(z)$ across the mixing layer yields $Ri_g \approx \delta \approx 0.25$.

\cite{prastowo_mixing_2008} also measured the time-averaged mixing efficiency in their exchange flow using density profile measurements in the reservoirs at the end of the experiments, defined as $\mathcal{M}\equiv (P_{f}-P_{r})/(P_i-P_r) \in [0,1]$, where $P_i$ the initial potential energy in the system (before the exchange flow starts), $P_{f}$ is the final measured potential energy in the system, and $P_{r}$ is the `reference' or `minimum' potential energy obtained by adiabatic (`no-mixing') rearrangement of fluid parcels from the initial conditions (i.e. $P_i-P_r$ is the initially available potential energy).  They found collapse of the $\mathcal{M}$ data with $A Re$ and $\mathcal{M} \rightarrow 0.11$ for $A Re \rightarrow 10^5$ (using our notation). Finally, they supported this observation and linked $\mathcal{M}$ to $\delta$ by estimating mixing efficiency as the ratio of potential energy gain to kinetic energy deficit caused by the presence of a linear mixing layer, which yielded $\mathcal{M} \approx  Ri_g/2 \approx  \delta/2 \approx 0.125$.

\cite{hughes_mixing_2016} studied horizontal lock exchange gravity currents, which for some of their life cycle, behave similarly to exchange flows. They measured $\delta\approx 0.33$ in the range $Re\in [10^4,10^5]$. Using similar measurements to \cite{prastowo_mixing_2008}, they found $\mathcal{M} \rightarrow 0.08$ asymptoting from below as $Re \rightarrow 10^5$. They supported this asymptotic value using a simple mixing model based on idealised linear profiles in the mixing layer, which yielded $\mathcal{M} = (2\delta^2/3)(1-2\delta/3)(1-\delta/2)^{-2} \approx \delta^2 \approx 0.08$.

However, we have seen that exchange flows in inclined ducts have $\delta$ monotonically increasing not only with $A$ and $Re$, but also with $\theta$. In addition, much higher values of $\delta \gg 0.3$ (up to 0.8, and even 1 in K91) can be achieved even at moderate values of $\theta$ of a few $\alpha$ and $Re<10^4$. Therefore, the above models supporting values of $\delta = 0.2-0.3$ and $\mathcal{M} = 0.08-0.12$ in the $\TT$ regime disagree with our data, despite \emph{(i)} the similarity of SID flows to the flows assumed above (shear-driven mixing flows with the same `IH' velocity scaling $-1 \lesssim u \lesssim 1$) and \emph{(ii)} the fact that these models would apparently not be modified by the presence of gravitational forcing ($\theta>0$).

\subsubsection{New mixing efficiency model}

\begin{figure}
    \centering
        \includegraphics[width=0.95\textwidth]{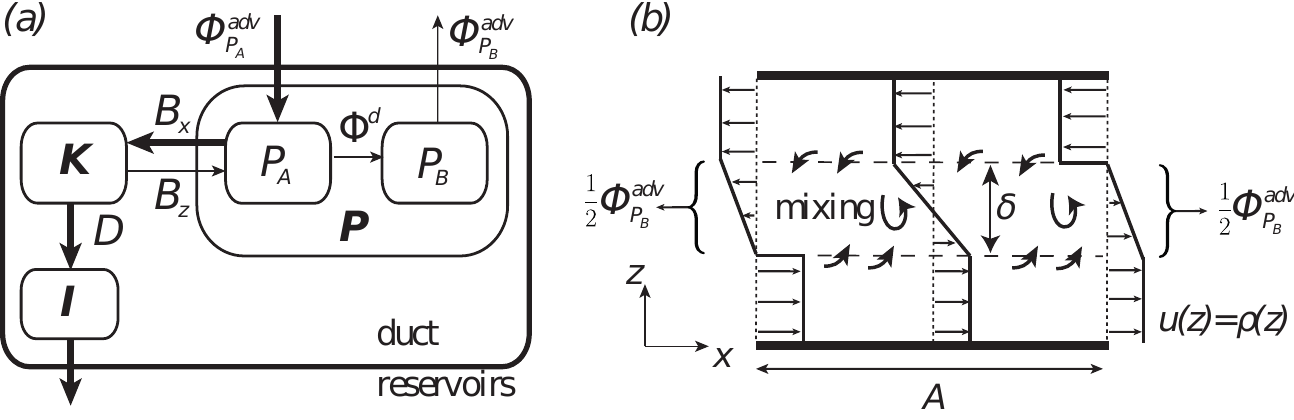}
    \caption{Mixing model for SID flows: \emph{(a)} Time- and volume-averaged energetics model developing on that in (LPL19, figure~8\emph{(b)}) by subdividing the potential energy reservoir as $P=P_A+P_B$. We also show  the kinetic energy $K$, internal energy $I$, and all relevant fluxes: horizontal buoyancy flux $B_x$, vertical buoyancy flux $B_z$, viscous dissipation $D$, diapycnal flux $\Phi^d$ and advective fluxes with the external reservoirs $\Phi^{\textrm{adv}}_{P_{A}},\Phi^{\textrm{adv}}_{P_{B}}$. The direction of the arrows denotes the net (time-averaged) transfer, and the thickness of the arrows denotes the expected magnitude of the fluxes (with the expectation that $\Phi^{\textrm{adv}}_{P_{A}}\approx B_x \approx D$ and $B_z \approx \Phi^d \approx \Phi^{\textrm{adv}}_{P_{B}}$).  \emph{(b)} Simplified flow model in the duct to estimate the mixing rate from $\Phi^{\textrm{adv}}_{P_{B}}$ and link it to $\delta$. The in-flow of unmixed fluids from the external reservoirs and the out-flow of mixed fluid back into them are modelled by the broken line profiles $u(z)=\rho(z)$ drawn at the left and right ends of the duct (consistent with the typical mid-duct profile  drawn, equal to $u=\rho=\pm 1$ above and below the mixing layer and $u=\rho=-2z/\delta$ in the mixing layer, assumed elsewhere in the literature). }
  \label{fig:mixing_model}
\end{figure}

To address this, we propose a different toy model of mixing based on the energetics framework of LPL19. As sketched in figure~\ref{fig:mixing_model}\emph{(a)}, we consider that the duct is composed of three volume-averaged energy reservoirs (in bold): potential energy $P$, kinetic energy $K$, and internal energy $I$ (heat). We further decompose the potential energy reservoir into an available potential energy $P_A$, and a background potential energy $P_B$ (such that $P=P_A+P_B$), as is customary in the study of mixing (see e.g. \cite{winters_available_1995}).  

As explained in LPL19 (see their \S~4.1-4.3 and figure~8\emph{(b)}), forced flows have, to a good approximation, the following steady-state energetics: the external fluid reservoirs provide an advective flux of potential energy into the duct, which we identify here as being an advective flux of \emph{available} potential energy $\Phi^{\textrm{adv}}_{P_{A}}\approx Q_m \theta/8$, which is then converted to kinetic energy by the horizontal buoyancy flux $B_x$, and to heat by the viscous dissipation $D \approx (2/Re)\langle \mathsf{s}^2 \rangle_{x,y,z,t}$. When turbulent mixing is neglected, all these fluxes have equal magnitude, and $D \approx (1/8) Q_m \theta$. When turbulent mixing is included, a net vertical buoyancy flux $B_z$ converts part of $K$ back to $P_A$, and a net irreversible diapycnal flux $\Phi^d$ converts part of $P_A$ to $P_B$, at a steady-state rate equal to the advective flux of $P_B$ out of the duct, back into the external reservoirs $|\Phi^{\textrm{adv}}_{P_{B}}|=|\Phi^d|$. The mixing efficiency quantifies the percentage of total time- and volume-averaged power throughput $\Phi^{\textrm{adv}}_{P_{A}}$ that is spent to irreversibly mix the density field inside the duct
\begin{equation}\label{M-in-SID}
    \mathcal{M}= \frac{\Phi^d}{D+\Phi^d} = \frac{\Phi^{\textrm{adv}}_{P_{B}}}{\Phi^{\textrm{adv}}_{P_{A}}} .
\end{equation}
It is expected that $\mathcal{M}\ll 1$ in such  flows, as represented by the respective thickness of the arrows in figure~\ref{fig:mixing_model}\emph{(a)}, representing  the order of magnitude of the fluxes.

As sketched in figure~\ref{fig:mixing_model}\emph{(b)}, we propose piecewise-linear flow profiles $u(z)=\rho(z)$ at either end of the duct as a minimal model to estimate the magnitude of $\Phi^{\textrm{adv}}_{P_{B}}$ as a function of the interfacial layer thickness $\delta$, and eventually link it to input parameters $A,\theta,Re$. We consider that fluid comes from the external reservoirs into the duct \emph{unmixed} below  (resp. above) the interfacial mixing layer at the left (resp. right) end of the duct, and leaves the duct \emph{mixed} with a linear profile, going from 0 at the bottom (resp. top) edge of the mixed layer to $-1$ (resp. 1) at the top (resp. bottom) edge of the mixed layer. (In more central sections of the duct, mixing smoothes out the discontinuities at the edges of the mixing layer present at the ends, and we expect the continuous linear profile drawn in the centre, but it is irrelevant to the following calculations). The outflow of mixed fluid creates the following net flux of background potential energy out of the duct:
\begin{equation}\label{phi_adv_B}
    \Phi^{\textrm{adv}}_{P_B} = \frac{1}{4A} \langle z \rho u \rangle_z |_{L-R} = \frac{2}{4A\delta} \int^{\delta/2}_{-\delta/2} z\big( z+\frac{\delta}{2}\big)^2 \, \d z = \frac{\delta^3}{24A},
\end{equation}
where $|_{L-R}$ denotes the difference between the values at the left and right boundary, and the prefactor $1/(4A)$ comes from the non-dimensionalisation of the energy budget equations (see LPL19, equation (4.14a)). From \eqref{M-in-SID}-\eqref{phi_adv_B} and $\Phi^{\textrm{adv}}_{P_{A}}\approx Q_m \theta/8$, we  now deduce:
\begin{equation}\label{delta_from_mixing}
   \delta \approx (3A\theta Q_m \mathcal{M})^{1/3}.
\end{equation}

Encouragingly, this estimation has the potential to be consistent with our data in the SID. Assuming that $Q_m\approx 0.5$ throughout most of the $\II$ and $\TT$ regimes, we conjecture that most of the dependence on $Re$ observed in the $\delta$ data of figure~\ref{fig:delta-diagrams} is due to the underlying monotonic increase of $\mathcal{M}(Re)$, which is \emph{a priori} unknown, but consistent with the observations of \cite{prastowo_mixing_2008} and \cite{hughes_mixing_2016}. The observation of K91 and figure~\ref{fig:delta-diagrams}\emph{(a-b)} that $\delta$ primarily scales with the group $A \theta$ at $Re \gg 500 A$ (as sketched in figure~\ref{fig:lit_sum}\emph{(c)}) would suggest that $\mathcal{M}$ asymptotes to a constant value at high $Re$, which is also consistent with the observations of \cite{prastowo_mixing_2008} and \cite{hughes_mixing_2016} at $A Re>10^5$ and $Re > 10^5$ respectively. Assuming their high-$Re$ asymptotic value of $\mathcal{M}\approx 0.1$, we obtain 
\begin{equation}\label{delta_from_mixing_const_M}
   \delta\rightarrow 0.5\Big(\frac{\theta}{\alpha}\Big)^{1/3}.
\end{equation}
This gives, for example, $\delta \approx 0.4$ when $\theta/\alpha \approx 1/2$. This value agrees with the K91 data (see \S~\ref{sec:review-delta}) and our LSID data (figure~\ref{fig:delta-diagrams}\emph{(a)}, at $Re>10^4$ and $A Re>10^5$). However, this value does not agree well  with our HSID, tSID, and mSID $\delta$ data (figure~\ref{fig:delta-diagrams}\emph{(b-d)}, in which $\delta$ remains dependent on $Re$. This is presumably due to the lower values of $A$ and/or $Re$ in these data sets, which remain below the asymptotic values of $Re>10^5$ and $A Re > 10^5$. In other words, we believe that our $\delta$ data and \eqref{delta_from_mixing_const_M} are consistent and provide further (albeit indirect) evidence for the monotonic increase of $\mathcal{M}$ with $Re$.


\clearpage

\section{Conclusions}

\subsection{Problem and approach}

In this paper, we investigated buoyancy-driven exchange flows taking place in inclined rectangular ducts (figure~\ref{fig:setup}). We focused on the behaviour of three key dependent variables: the qualitative flow regime (laminar, wavy, intermittently turbulent, or fully turbulent), the non-dimensional mass (or buoyancy) flux $Q_m$, and the non-dimensional thickness of the  interfacial layer $\delta$ as the five non-dimensional input parameters were varied: the duct longitudinal aspect ratio $A$, spanwise aspect ratio $B$, tilt angle $\theta$, Reynolds number $Re$, and Prandtl number $Pr$. 

Dimensional analysis (figure~\ref{fig:balance}) and the experimental literature (figure~\ref{fig:lit_sum} and table~\ref{tab:review}) showed that the rich dynamics of these sustained stratified shear flows were accessible for a wide range of $Re$ and for $\theta$ of at most a few duct aspect ratios $\alpha = \tan^{-1} (A^{-1})$.  Our focus on `long' ducts ($A \gg 1$)  allowed us to explore these dynamics while keeping $\theta$ small enough to remain relevant to (largely-horizontal) stably-stratified geophysical flows and turbulence, which are our ultimate motivation. 

To  overcome the limitations of past studies of the problem, we presented extensive experimental results for all three variables of interests (regimes, $Q_m$, and $\delta$) in the $(\theta,Re)$ plane for five different data sets, between which $(A,B,Pr)$ were varied systematically (table~\ref{tab:sid_geo_param}).

\subsection{Experimental results}

First, our data (figures~\ref{fig:regime-Re-sintheta}-\ref{fig:interfacial-thickness}) confirmed the conclusions of past studies: that increasingly disordered and turbulent regimes are found as $A,\theta,Re$ are increased, that $Q_m$ is non-monotonic in $\theta$ and $Re$, and that $\delta$ is monotonic in $A,\theta,Re$. Second, our data confirmed beyond doubt the previously-supposed existence and importance of at least one additional elusive non-dimensional input parameter related to the dimensional height of the duct $H$, because our regime, $Q_m$, and $\delta$ data at the same $A,B,\theta,Re,Pr$ but different $H$ do not collapse.  Third, our data highlighted the non-trivial dependence of all variables on all five parameters $A,B,\theta,Re,Pr$. Regime transition, iso-$Q_m$, and iso-$\delta$ curves are not only shifted in the $(\theta,Re)$ plane at different $A,B$, or $Pr$, but they also generally exhibit different power law scalings in $\theta$ and $Re$ at different $A,B,Pr$.  

Given the breadth of our observations summarised above, and the relative richness of our data in the $(\theta,Re)$  plane compared to the few values of $A,B,Pr$ studied, we focused specifically on the very last observation above, i.e. on the various scalings of the form $\theta Re^\gamma = $ const. governing the regime transitions curves and the major axis of hyperbolas best fitting $Q_m$ in the $(\log \theta, \log Re)$ plane.  
Even within this specific focus, we discovered that $\gamma$ not only   varies between data sets (at different $A,B,Pr$) but that it also varies within a given data set (at fixed $A,B,Pr$): \emph{(i)} $\gamma$ is generally different for the regime data ($\gamma =1$ or 2) and the $Q_m$ data ($0.3<\gamma<2.6$) implying that regime and $Q_m$ are not well correlated (whereas regimes and $\delta$ are); and \emph{(ii)} in one regime data set, $\gamma$ even takes two different values (of 1 and 2) in different regions of the $(\theta,Re)$ plane.

\subsection{Modelling results and outlook}

To provide a modelling framework, particularly to understand the above observation \emph{(ii)}, we split the $(\theta,Re)$ plane into four quadrants delimited by $\theta=\alpha$ (the `lazy/forced flow boundary', based on the respective dominance of hydrostatic/gravitational forcing) and $Re=50A$ (the `low/high $Re$ boundary', based on whether or not wall boundary layers are fully developed across the duct or remain thin). To explain the above observations \emph{(i)}-\emph{(ii)}, we discussed three families of candidate models.

In \S~\ref{sec:model-energ} we considered the volume-averaged energetics framework of \cite{lefauve_regime_2019} (LPL19). LPL19 physically explained the $\theta Re=$ const. scaling of regime transitions of forced ($\theta\gtrsim \alpha$), low-Re ($Re \lesssim 50A$), salt-stratified ($Pr=700$) flows as being caused by threshold values of the three-dimensional kinetic energy dissipation (equation \eqref{s2d-s3d-qmResintheta-lowRe}). We extended their argument to high-$Re$ ($Re \gg 50A$) flows, by accounting for the scaling of two-dimensional, laminar boundary layer dissipation. However, the resulting predicted scaling in $\theta Re - A^{-1/2}Re^{1/2}$ (equation \eqref{s2d-s3d-qmResintheta-highRe}) did not agree with any of our regime data. Detailed measurements of dissipation in these high-$Re$ flows (not found in LPL19) would be valuable to understand why this is the case, but they are very challenging to perform due to the required spatio-temporal resolution.

In \S~\ref{sec:model-fric} we developed a two-layer frictional hydraulics model of SID flows (figure~\ref{fig:hydraulics-model}) from \cite{gu_analytical_2005} and showed that the existence of a solution imposed a lower and upper bound on the product of the volume flux by a parameter quantifying wall and interfacial friction (equation \eqref{inequality-slope-2}). This model explained the qualitative behaviour of $Q_m$ with $\theta Re$, and the fact that regimes and $Q_m$ could have different scalings (figure~\ref{fig:hydraulics-hyp}). This model also provided a quantitative scaling for the interfacial friction parameter and, in turn, for regime transitions, based on our conjecture that regime transitions were directly linked to interfacial turbulent stresses. Although the resulting low-$Re$ scaling in $\theta Re$ (equation \eqref{Km-lowRe}) was identical to that predicted by the energetics model and correct (at least for $Pr=700$), the high-$Re$ scaling in $A^{1/2}\theta Re^{1/2}$  (equation \eqref{Km-highRe}) did not agree with our regime data. 

Neither the energetics nor the frictional hydraulics model could predict the observed scalings in $\theta Re$ or $\theta Re^2$ observed in lazy flows ($\theta \lesssim \alpha$) because these flows are under-specified in either model (they have more unknowns than equations). In addition, scalings laws deduced from  plots in the $(\log \theta,\log Re)$ plane break down for lazy flows at slightly negative angles ($-\alpha \lesssim \theta \lesssim 0$), which we largely ignored in this paper.

In \S~\ref{sec:model-mixing}, we focused on the scaling of $\delta$ by modelling turbulent mixing. We first considered a model with constant turbulent diffusivity imposed throughout the domain \citep{cormack_natural_1974,hogg_hydraulics_2001}. We attempted to link this diffusivity to input parameters following insights gained from frictional hydraulic theory, but the resulting scalings (equation \eqref{wild-conjecture}) did not agree with our data. We then explained why previous measurements and models of $\delta$ in related stratified shear flows \citep{prastowo_mixing_2008,hughes_mixing_2016} were inconsistent with our results on exchange flows in inclined ducts. We thus developed a new model that explicitly represents the rate of mixing in the energy budget analysis of LPL19, and quantifies this mixing as a function of known input parameters and an unknown mixing efficiency $\mathcal{M}$ using a simplified flow profile (figure~\ref{fig:mixing_model}, equation~
\eqref{phi_adv_B}). The resulting expression for $\delta$ (equations \eqref{delta_from_mixing}-\eqref{delta_from_mixing_const_M}) is qualitatively consistent with our observations, but it involves $\mathcal{M}$ (not measured in these experiments) whose scaling on $Re$ is critical. Our model and  data indirectly support previous observations of \cite{prastowo_mixing_2008} and \cite{hughes_mixing_2016} that $\mathcal{M}$ monotonically increases with $Re$ to reach asymptotic values of $\mathcal{M}\approx 0.1$ at very high $Re$, but direct measurements of $\mathcal{M}$ would be desirable to confirm this.


While these models have allowed us to make significant progress by providing useful physical insights and partial quantitative results regarding scaling laws in $A,\theta,Re$,  our experimental observations have raised an even larger number of questions which remain open. Among these  are the elusive existence of an important sixth non-dimensional input parameter linked to the duct height $H$, the influence of the spanwise aspect ratio $B$ and Prandtl number $Pr$, and the scaling of the mixing efficiency $\mathcal{M}$.

\vspace{0.5cm}

\noindent \textbf{Acknowledgements}

\vspace{0.2cm}
AL is supported by an Engineering and Physical Sciences Research Council (EPSRC) Doctoral Prize Fellowship. We also acknowledge funding from EPSRC under the Programme Grant EP/K034529/1 `Mathematical Underpinnings of Stratified Turbulence' (MUST) and from the European Research Council (ERC) under the European Union's Horizon 2020 research and innovation Grant No 742480  `Stratified Turbulence And Mixing Processes' (STAMP). We thank Colin Meyer and Simon Vincent for collecting the regime and mass flux data in the LSID and HSID geometries in 2012-2014, and we thank Dr Ling Qu visiting from the South China Sea Institute of Oceanology (SCSIO, Guangzhou, China) for collecting part of the regime and mass flux data in the mSID and tSID in 2015-2016. \AL{We also thank Dr Darwin Kiel from Coanda Research and Development (Burnaby, Canada) for his reading of the manuscript, his suggestions,  and his permission to reproduce some of his results}. Finally, we are grateful for the invaluable experimental support of Prof Stuart Dalziel, Dr Jamie Partridge and the technicians of the G. K. Batchelor Laboratory.

\clearpage

\appendix{
\numberwithin{figure}{section} 
\numberwithin{table}{section}

\section{Summary table of the literature review}
\label{sec:appendix_table}

\begin{small}
\begingroup
\vfill
\centering 
\begin{sideways}
  \begin{threeparttable}
\centering
\renewcommand*{\arraystretch}{1.0}
\caption{Summary of the literature review. For each paper, we specify the scale of the apparatus (duct height or pipe diameter $H$), the parameters that were either fixed or whose variation was not studied, those that were varied and studied, the key conclusions about the scaling of transitions between flow regimes (based on empirical or physical arguments), mass flux $Q_m$, and interfacial layer thickness $\delta$. }
\label{tab:review}
\vspace{0.8cm}
\begin{tabular}{l l l l l l l l l l l}
Name  & \ $H$ (mm)  & \ Fixed params.  & \ Varied params. & \  Regimes & \ Mass flux $Q_m$ & \ Interfacial thickness $\delta$  \\
\midrule     
MR61  & \ 150 & \ \begin{tabular}{@{}l@{}l@{}l@{}}$A= 6$\\ $B=\nicefrac{1}{3}$ \\ $Re \approx 2\times 10^3$ \\ $Pr=700$ \end{tabular} & \ $\theta \in$ ? & \ \begin{tabular}{@{}l@{}}$\sim Re \, Q^2$ \\ (empirical) \end{tabular} & \ $-$ & \ $-$   \\ \midrule
LT75 & \ \begin{tabular}{@{}l@{}l@{}}$150-1140$ \\ (pipe) \end{tabular} & \ \begin{tabular}{@{}l@{}}$B=1$ \\ $\theta=0^\circ$  \end{tabular} & \ \begin{tabular}{@{}l@{}l@{}} $A\in[0.5,20]$ \\ $Re \in [15,75]\times 10^3$  \\ $Pr=1, \, 700$\end{tabular}  & \ $-$ & \ \begin{tabular}{@{}l@{}l@{}} Constant in $A,Re,Pr$ \\ at $\theta=0^\circ$ \\ $Q_m\approx 0.23$\end{tabular}   & \ $-$ \\ \midrule
MT75 & \ \begin{tabular}{@{}l@{}l@{}}$50$, $150$ \\ (pipe) \end{tabular} & \ \begin{tabular}{@{}l@{}}$B=1$  \\ $Pr=700$ \end{tabular} & \ \begin{tabular}{@{}l@{}} $A\in[0.5,18]$ \\ $Re =2, \, 20 \times 10^3$ \\ $\theta \in [0^\circ,90^\circ]$ \end{tabular}  & \ $-$ & \ \begin{tabular}{@{}l@{}l@{}l@{}l@{}} Non-monotonic in $A, \,\theta$\\  $Q_m \approx 0.3$ at $\theta=0^\circ$ \\  $Q_m \approx 0.4$ at $\theta \approx \nicefrac{\alpha}{2}$  \\ $Q_m \rightarrow 0$ as $\theta \gg \alpha$ \\ \AL{Dependence on $Re$} \end{tabular}   & \ $-$ \\ \midrule
W86 & \ \begin{tabular}{@{}l@{}l@{}}$14-25$ \\ (pipe) \end{tabular} & \ \begin{tabular}{@{}l@{}l@{}} $B=1$ \\ $\theta =0^\circ$ \\ $Pr=700$ \end{tabular} & \ \begin{tabular}{@{}l@{}} $A=1.6,\,3.5,\,9.6 $ \\ $Re \in [0.2,3]\times 10^3$\end{tabular} & \ \begin{tabular}{@{}l@{}}$\sim Re$ \\ (empirical) \end{tabular}  & \  \begin{tabular}{@{}l@{}l@{}} Monotonic in $A^{-1}Re$ \\ $Q_m \approx 0.15$ at $Re\approx 20A$ \\ $Q_m \approx 0.35$ at $Re\approx 500A$ \end{tabular} & \ $-$ \\ \midrule 
K91 & \ 50, 100 & \ \begin{tabular}{@{}l@{}l@{}}$B=1,\, 2, \,4$ \\ \AL{$Re \in [2,15]\times 10^3$} \\ $Pr=700$ \end{tabular} & \ \begin{tabular}{@{}l@{}l@{}}$A=1,\, 2,\, 4, \, 8$ \\ $\theta \in [-45^\circ,90^\circ]$ \end{tabular} & \ \begin{tabular}{@{}l@{}} $\sim A\,\tan\theta$ \\ (empirical) \end{tabular}  & \ \begin{tabular}{@{}l@{}l@{}l@{}} \AL{Non-monotonic in $A, \,\theta$} \\ Collapse with $A\tan\theta$ \\ \AL{Independence on $Re$} \end{tabular}  & \ \begin{tabular}{@{}l@{}l@{}} Monotonic in $A,\, \theta$ \\ Max. $\delta=1$ at $\theta \approx 2\alpha$ \\ Collapse with $A\tan\theta$ \end{tabular} \\ \midrule
ML14 & \ 100 & \ \begin{tabular}{@{}l@{}}$B=1$ \\  $Pr=700$ \end{tabular} & \ \begin{tabular}{@{}l@{}l@{}}$A=15, 30$ \\  $\theta \in [-1^\circ,4^\circ]$ \\ $Re \in [1,20]\times 10^3$ \end{tabular}  & \ \begin{tabular}{@{}l@{}} $\sim A \, \theta \, Re^2$  \\ (empirical) \end{tabular}  & \ \begin{tabular}{@{}l@{}} Monotonic in $\theta$ \\ $Q_m \rightarrow 0.5$ as $\theta>\nicefrac{\alpha}{2}$ \end{tabular}  & \ $-$  \\ \midrule
LPL19 & \ 45 & \  \begin{tabular}{@{}l@{}l@{}}$A=30$ \\ $B=1$ \\ $Pr=700$\end{tabular} & \  \begin{tabular}{@{}l@{}}$\theta \in [-1^\circ,6^\circ]$ \\ $Re \in [0.3,6]\times10^3 $\end{tabular} & \ \begin{tabular}{@{}l@{}l@{}l@{}} $ \sim \theta \, Re^2$ as $\theta\lesssim \alpha$  \\  (empirical) \\$\sim \theta \, Re$ as $\theta\gtrsim \alpha$  \\ (physical)  \end{tabular}   & \   Non-monotonic in $\theta, \,Re$   & \ $-$ \\ \hline

\end{tabular}

  \end{threeparttable}
\end{sideways}
\vfill

\endgroup

\end{small}

\section{Unpublished data from \cite{kiel_buoyancy_1991}} \label{sec:kiel-data}
   
In figure~\ref{fig:kiel} we reproduce some of the unpublished mass flux and interfacial thickness data in \cite{kiel_buoyancy_1991} (K91) discussed in \S~\ref{sec:review-Qm}-\S~\ref{sec:review-delta}. In panel~\emph{(a)} K91 observed relative independence of  $Q_m$ on $Re$ at large enough values $Re>400A=1600$ \AL{(his figure~4.10)}. In panel~\emph{(b)} we show his reproduction (his \AL{figure~2.6}) of the $Q_m(A,\theta)$ in \cite{mercer_experimental_1975} (MT75,  \AL{figure~5}), and in panels \emph{(d,f)}, his $Q_m(A,\theta)$ and $\delta(A,\theta)$ data \AL{(his figures~5.2 and 5.13 respectively)}. In panels \emph{(c,e,f)} we show the collapse with his proposed $Ri_G$ criterion \eqref{definition-RiG}.

\begin{figure}
    \centering
        \includegraphics[width=\textwidth]{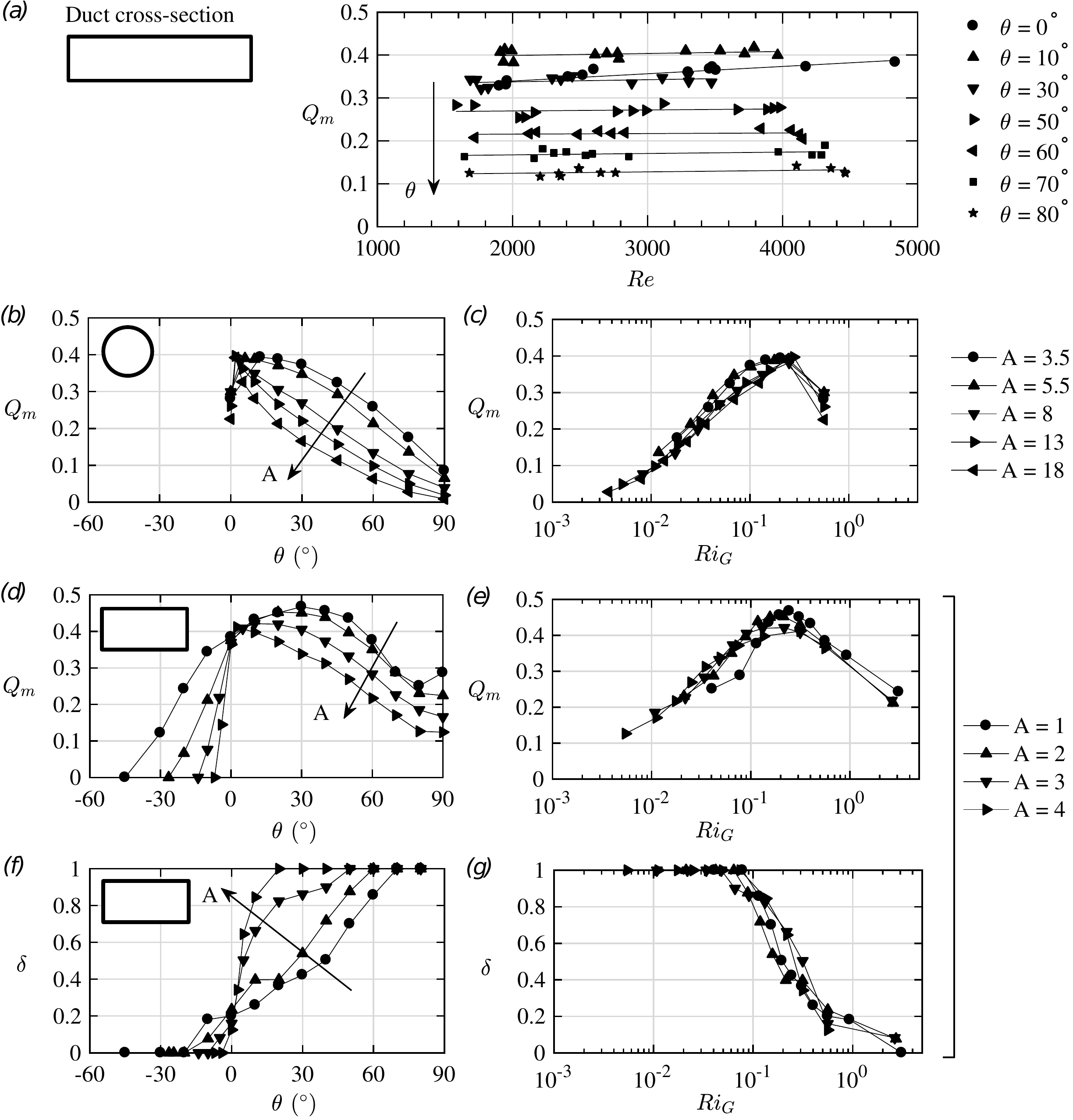}
    \caption{Unpublished experimental data in \cite{kiel_buoyancy_1991}, reproduced with his kind permission. \emph{(a)} Apparent independence of $Q_m(\theta,Re)$ on $Re$ at $A=B=4$. Left panels: \emph{(b)} MT75's $Q_m(A,\theta)$ data in a circular pipe, \emph{(d)} K91's $Q_m(A,\theta)$ data and \emph{(f)} $\delta(A,\theta)$ data, both at $B=2$. Right panels \emph{(c,e,g)}: collapse of the data in the respective left panel with $Ri_G$ (see \eqref{definition-RiG}). These data have been converted to follow our notation and non-dimensionalisation.}
  \label{fig:kiel}
\end{figure}

\section{Experimental methodology} \label{sec:method}

\subsection{Flow regimes} \label{sec:method_regimes}

Regimes were largely determined by shadowgraph observations over a subsection of the length of the duct, following the qualitative description of each regimes provided by ML14 (see their \S~3.1 and figure~3). A schematic of the shadowgraph setup can be found in L18, \S~2.1. 

In the mSID data set, 48 out of 360 regime identifications were not made by shadowgraph, but rather by direct visualisation of the density field by planar laser induced fluorescense (PLIF), since more detailed measurements of the velocity and density fields (incompatible with simultanenous shadowgraph) have been performed in this geometry \citep{lefauve_structure_2018,partridge_versatile_2019,lefauve_regime_2019}.

All raw video data, including those obtained by other experimenters (acknowledged at the end of the paper), were reprocessed in an effort to ensure that regimes were identified as consistently as possible across all five data sets of table~\ref{tab:sid_geo_param} (especially in the cases where the distinction between regimes can be subtle).

\subsection{Mass flux} \label{sec:method_Qm}

Mass fluxes were determined, as in ML14, by measuring the average initial (`i') and final (`f') density in each reservoir: reservoir `1', initially at density $\rho_1^{i}=\rho_0+\Delta \rho/2$ and finally at a well mixed density $\rho_1^{f}$ and `2', initially at $\rho_0-\Delta \rho/2$ and finally at $\rho^f_2$, giving the following two estimations
\begin{equation}  \label{Qm_measurements_formula}
\tilde{Q}_{m, 1}  =   \frac{-(\rho_1^{f}-\rho_1^{i})V_1}{\Delta \rho(H^2/2)\sqrt{g'H} T} \quad \textrm{and} \quad \tilde{Q}_{m, 2}  =    \frac{(\rho_2^{f}-\rho_2^{i})V_2}{\Delta \rho(H^2/2)\sqrt{g'H} T},
\end{equation} 
where $V_1,V_2$ are the (typically approximately equal) volumes of fluid in the respective reservoirs, and the tilde on $\tilde{Q}_m$ stresses the fact that they are non-dimensional (despite all quantities on the right side of the $=$ sign being dimensional).   Experiments in which both estimates differed by more than $(Q_{m,1}-Q_{m,2})/(Q_{m,1}+Q_{m,2}) > 10~\%$ were rejected (typically due to an initial misadjustement of the free surfaces resulting in a net volume flux $\langle u \rangle_{x,y,z,t}\neq 0$). All data shown in this paper thus have near-zero net volume flux, and we only use the average value $Q_m \equiv (Q_{m,1}+Q_{m,2})/2$, . 

We recall that $T$ in \eqref{Qm_measurements_formula} is the (dimensional) duration of an experiment. The determination of the relevant $T$ was made carefully but remains subject to intrinsic uncertainties which affect $Q_m$ as we explain next. The duct is opened at time $t^a$ initiating a gravity current lasting until the  exchange flow is considered fully established by shadowgraph visualisations at time $t^b$. The exchange flow of interest continues until the levels of the discharged fluids approach the ends of the duct, at which point one end of the duct is closed at time $t^c$, shortly before the other end of the duct is closed at $t^d$. To avoid under- and over-estimations of $Q_m$ by the intervals $t^d-t^a$ and $t^c-t^b$ (respectively), we choose to use the average of the two $T=(t^d-t^a+t^c-t^b)/2$, and to use error bars to indicate the magnitude of the resulting uncertainty (the difference between the over- and under-estimation). Note that error bars tend to be larger at high $Re$ (figure~\ref{fig:delta-diagrams}) because the overall duration $T$ of an experiment is inversely proportional to the  magnitude of the dimensional exchange velocities (scaling with $\sqrt{g' H}$, and hence with $Re$)  due to the finite size of the reservoirs. A smaller duration $T$ increasing the relative duration of initial transients (typically fixed) and therefore the uncertainty about $T$. 

Note that measurements of $Q_m$ in temperature-stratified experiments (mSID \heat data set) could not be performed due to the practical impossibility to control the heat loss occurring through the boundaries of the reservoir and the free surface.

For more details on these measurements, see L18, \S~2.2.

\subsection{Interfacial layer thickness} \label{sec:method_delta}

The thickness of the interfacial density layer $\delta$ was estimated from shadowgraph images. To a reasonable approximation, the refraction of near-parallel light beams by inhomogeneities in the density field results in a recorded greyscale light intensity $I(x,z)$ proportional to the second vertical derivative of the density field  integrated in the spanwise direction $I(x,z) \propto \int_{-B}^B \partial_{zz}\rho \, \d y$ (for a full derivation and discussion of the approximations, see L18, \S~2.1). This makes shadowgraphy particularly well-suited to detect the average location of large-scale curvatures in the density field, which are precisely the edges of the interfacial density layer. 
\begin{figure}
\centering
        \includegraphics[width=\textwidth]{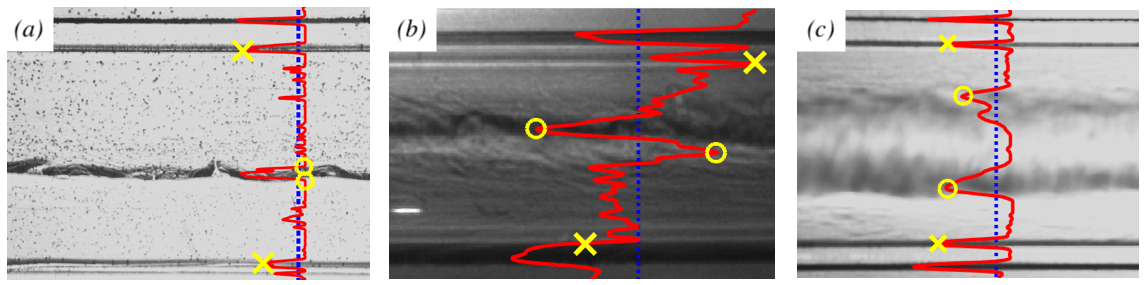}
    \caption{Example of the determination of $\delta$ from shadowgraph snapshots in the \emph{(a)} $\HH$ regime (LSID) where $\delta=0.069$; \emph{(b)} $\II$ regime (mSID), where $\delta=0.14$; \emph{(c)} $\TT$ regime (LSID), where $\delta=0.47$. At a randomly chosen streamwise position (dotted blue line), the greyscale intensity $I(z)$ (solid red curve) is displayed using a convenient horizontal scale. The positions of the interfacial density layer and of top and bottom walls  are carefully clicked by hand (identified by the yellow circles and crosses respectively) and $\delta$ is determined as the ratio of the spacing between the pair of circles and crosses. }
  \label{fig:delta-method}
\end{figure}

Due to the nature of shadowgraph images, and to its sensitivity to air bubbles or scratches on the walls of the reservoirs, the identification of minima and maxima of $I(z)$ could only be semi-automated according to the following strict methodology, illustrated in figure~\ref{fig:delta-method}:
\begin{myenumi}
\item[\textnormal{(i)}\hspace{2.6ex}] A random sample of typically three to five snapshots per movie were selected and averaged (although in rare cases only one still image was available); 
\item[\textnormal{(ii)}\hspace{2.2ex}] A randomly-generated location in the streamwise direction was selected (dotted blue lines) and they greyscale intensity profile $I(z)$ at this particular $x$ location was superimposed onto the image (solid red curves);
\item[\textnormal{(iii)}\hspace{1.7ex}] The profile $I(z)$ was carefully interpreted, and the local extrema representing the top and bottom duct boundaries (yellow crosses) and edges of the interface to measure (circles) were carefully selected by a click. 
\item[\textnormal{(iv)}\hspace{1.8ex}] The ratio of pixel distances between the selected edges of the density interface and the top and bottom walls was computed to yield $\delta$.
\end{myenumi}
All images were processed by the first author to ensure consistency, and yielded a total of 351 values of $\delta$ for all four duct geometries (table~\ref{tab:sid_geo_param}). 

Note that measurements of $\delta$ in temperature-stratified experiments (mSID \heat data set) could not be performed since the refractive index of water is a weaker function of temperature than salinity at comparable density differences, resulting in insufficient contrast and thus noisier $I(z)$ (though sufficient to determine the flow regime).

\section{Frictional two-layer hydraulic model} \label{sec:appendix-frictional}

In this section we give  details of the two-layer frictional hydraulic model introduced in \S~\ref{sec:model-fric} and sketched in figure~\ref{fig:hydraulics-model}. This model is based on  \cite{gu_frictional_2001,gu_analytical_2005} but includes non-zero tilt angles and a wider range of frictional stresses suited to the SID.
We cover the model formualtion in \S~\ref{sec:appendix-frictional-1}, the parameterisation of frictional effects in \S~\ref{sec:appendix-frictional-2}, and the solution to the full problem in
\S~\ref{sec:appendix-frictional-3}.

\subsection{Model formulation}\label{sec:appendix-frictional-1}

The frictional hydraulic model appears at first inconsistent because it is based on  velocities that are uniform in the cross-sectional plane ($\partial_{y,z} u_{1,2}=0$), while implicitly acknowledging and parameterising the effects of viscous stresses resulting from $\partial_{y,z} u_{1,2} \neq 0$. This model is however consistent provided that the departure from hydrostaticity is small (vertical and spanwise accelerations are negligible) and that viscous stresses are localised in relatively narrow boundary layers at the walls and interface ($Re\gg 50A$), rather than rather than through the whole volume ($Re< 50A$).  

Following standard hydraulic practice,  the effective `hydraulic' velocities $u_{1,2}(x)$ that will be used to compute the total Bernoulli head (kinetic energy) of each layer need to be defined in a way that accounts for the non-uniformity of the underlying `real' velocity profile in the SID  $u(x,y,z)$
\begin{equation}\label{definition-alpha}
u_{1,2}(x) \equiv \sqrt{\lambda_{1,2}(x)} \ \langle u(x,y,z) \rangle_{y,z_{1,2}},
\end{equation}
where $\langle \cdot \rangle_{z_{1,2}}$ denotes averaging over the lower/upper layer ($z \in [-1,\eta]$ and $z \in [\eta,1]$ respectively, see figure~\ref{fig:hydraulics-model}\emph{(a)}), and the velocity distribution coefficient $\lambda_{1,2}$  (also called kinetic energy correction coefficient or Coriolis coefficient) is defined as
\begin{equation} \label{alpha1}
\lambda_{1,2}(x) \equiv \frac{\langle u^3(x,y,z)  \rangle_{y,z_{1,2}}}{\langle u(x,y,z) \rangle_{y,z_{1,2}}^3} >1,
\end{equation}
respectively in the lower and upper layer (see e.g. \citealp[\S~2.7-2.8]{chow-open-1959} and \citealp[\S~3.2.2]{chanson-hydraulics-2004}). The greater the non-uniformity of the velocity profile $u$, the larger  $\lambda$ is. For the SID flows considered in this paper, volumetric velocity measurements showed that $\lambda$ varies over a relatively relatively small range  $1<\lambda \lesssim 2$ (see L18, \S~5.5.2). To simplify the following discussion, and since the effects of $\lambda$ are not central here (they quantitative rather than qualitative), we make the approximation that $\lambda_{1,2}(x) \approx 1$, effectively assuming that  $u_{1,2}(x) =  \langle u(x,y,z) \rangle_{y,z_{1,2}}$ in the following.

First, the conservation of Bernoulli potential in two-layer hydraulic flows is commonly expressed using the so-called `internal energy' of the system
\begin{equation}
    E(x) \equiv \eta(x) + u_2^2(x) - u_1^2(x).
\end{equation}

Second, the conservation of volume and zero-net flux conditions are expressed all along the duct as 
\begin{equation} \label{cons_volume}
   u_1(x)(1+\eta(x)) =  -u_2(x)(1-\eta(x)) = Q.
\end{equation}

The third important ingredient of two-layer hydraulics is the condition of hydraulic control, which requires that the composite Froude number $G$ is unity at sharp changes in geometry, i.e. at the duct ends \citep{armi_hydraulics_1986,lawrence_hydraulics_1990}:
\begin{equation}\label{non-dim-composite}
G^2(x) \equiv 2 \Big( \frac{u_1^2}{1+\eta}+\frac{u_2^2}{1-\eta}\Big) = 4 Q^2   \frac{1+3\eta^2}{(1-\eta^2)^3} = 1 \ \ \text{at} \ \ x=\pm A,
\end{equation}
where the second equality uses \eqref{cons_volume} and the third equality is the control condition.

In horizontal, frictionless ducts, $E(x) = 0$, hence $\eta=0$ and $u_1=-u_2=Q=\nicefrac{1}{2}$ all along the duct.

When the combined effects of a small positive tilt angle $\theta>0$ and  frictional stresses are added, the slope of the internal energy becomes
\begin{equation} \label{dEdx-friction}
E'(x)  = \eta'(x) (1-G^2(x)) = \theta - S(x)
\end{equation}
(this is the two-layer equivalent of single-layer ideas found in \cite[\S~4.4-4.5]{henderson_open_1966}). By analogy with the topographic slope $\theta$, the `frictional' slope $S(x)$  is computed by a balance of all the stresses acting on an infinitesimal slice of thickness $dx$ (figure \ref{fig:hydraulics-model}\emph{(b)}):
\begin{equation} \label{sum-stresses}
 S(x)   = \frac{ \sum_{\substack{\textrm{stresses} j \\ \textrm{layer 1}}} \tau^j_1 A_1^j}{V_1 } +  \frac{ \sum_{\substack{\textrm{stresses} j\\ \textrm{layer 2}}} \tau^j_2 A^j_2}{V_2 }.
\end{equation}
The subscript $i=1,2$ represents respectively the bottom and top layers, the superscript $j=Z,Y,I$ represents the origins of the stresses in the model: top and bottom wall stresses ($Z$, shown in blue in the figure),  side wall stresses ($Y$, in green) and  interfacial stresses ($I$, in red), $A_i^j$ represents the surface area over which the respective stresses act, and $V_{i}$ the volume of each layer. Note that the interfacial stresses have equal magnitudes on either sides of the interface $|\tau_1^I|=|\tau_2^I|\equiv\tau^I$. Following figure \ref{fig:hydraulics-model}\emph{(b)} and after elementary algebra, the balance in \eqref{sum-stresses} can be rewritten as:
\begin{eqnarray} \label{stangamma}
 S(x) = \frac{1}{1+\eta} \tau_1^Z + 2B^{-1} \tau_1^Y +  \frac{1}{1+\eta} \tau^I + \frac{1}{1-\eta} \tau_2^Z + 2B^{-1} \tau_2^Y +  \frac{1}{1-\eta} \tau^I.
\end{eqnarray}
where all the stresses in this equation and henceforth are norms and have positive values. For further details about the development of this model from first principles, see L18, \S~5.2.

\subsection{Parameterisation of shear stresses} \label{sec:appendix-frictional-2}

We now tackle the relation between the stresses $\tau_i^j$ and the underlying `real' flow profiles $u(x,y,z)$. We start by considering the bottom wall stress of the lower layer $\tau_{1}^Z$ in order to introduce the key concepts and definitions, before extending them to the other stresses. Using non-dimensional variables for $\tau_{1}^Z$ and $u(x,y,z)$, we first write the \emph{dimensional} equation for this stress as a simple function of the local shear
\begin{equation} 
\Big(\frac{\Delta U}{2}\Big)^2 \tau_{1}^Z(x)   =  \nu \frac{\Delta U/2}{H/2}  \Big\langle \Big| \frac{\p u(x,y,z)}{\p z}\Big|_{z=-1}\Big\rangle_y
\end{equation}
where the $\Delta U/2$ and $H/2$ factors come from non-dimensionalising $\tau_1^Z, u,  z$, and simplify to
\begin{equation}\label{tau1-z-1}
\tau_{1}^Z(x) = \frac{1}{Re} \Big\langle \Big| \frac{\p u(x,y,z)}{\p z}\Big|_{z=-1}\Big\rangle_y.
\end{equation}
In order to correctly parameterise $\tau_{1}^Z(x)$ and all other relevant stresses using well-defined, constant friction coefficients, we follow the following five steps.

\begin{enumerate}
    \item  \ First, we define the cross-sectional `shape' ($y-z$ dependence) of the local velocity profile in the lower layer as 
\begin{equation}\label{u1hat}
\hat{u}_1(x,y,z) =  \frac{u(x,y,z)}{u_1(x)},
\end{equation}
such that $\langle \hat{u}_1(x,y,z) \rangle_{y,z_1} = 1$. This decomposition allows to rewrite \eqref{tau1-z-1} as
\begin{equation}
\tau_{1}^Z(x) = \frac{1}{Re} u_1(x)\Big\langle \Big| \frac{\p \hat{u}_1(x,y,z)}{\p z}\Big|_{z=-1} \Big\rangle_{y},
\end{equation}
which is an exact expression for the local shear stress that does not require any assumptions about the value of the velocity gradient or flow profile.

\item \  Second, we define a `layer-rescaled' coordinate $\hat{z}_1$ as 
\begin{equation}
\hat{z}_1 \coloneqq \frac{z}{1+\eta} = z \frac{Q}{u_1(x)},
\end{equation}
in which layer 1 always has thickness one ($\hat{z}_1 \in [-1, 0]$), giving us
\begin{equation} \label{tau1-Z}
\tau_{1}^Z(x) = \frac{1}{Re} \frac{u^2_1(x)}{Q} \Big\langle  \Big| \frac{\p \hat{u}_1(x,y,\hat{z}_1)}{\p \hat{z}}\Big|_{\hat{z}_1=-1} \Big\rangle_{y}.
\end{equation}

\item \ Third, we define a constant, bottom friction parameter $f_{Z_1}$ to parameterise the stress:
\begin{equation} \label{tau1-Z-2}
\tau_{1}^Z(x)   =  \frac{f_{Z_1}}{Re}  \ \frac{u_{1}^2(x)}{Q} \quad \text{with} \quad f_{Z_1} \equiv   \Big\langle \Big| \frac{\p \hat{u}_1(x,y,\hat{z}_1)}{\p \hat{z}}\Big|_{\hat{z}_1=-1} \Big\rangle_{x,y}. 
\end{equation}
We note that despite the rescaling of $u(x,y,z)$ by $u_1(x)$ and the stretching of $z$ to $\hat{z}_1$ such that the interface is located at $\hat{z}_1(x)=0$,  $\hat{u}_1$ still has a weak residual $x$ dependence. Since for simplicity, we choose to model $f_{Z_1}$ as independent of $x$, the velocity gradient $\p \hat{u}_1(x,y,z)/\p \hat{z}|_{\hat{z}_1=-1}$ must now technically be averaged not only over $y$ but over $x$ and $y$, as shown in \eqref{tau1-Z-2}.  We also note that the $u_1^2(x)/Q$ factor in \eqref{tau1-Z-2} results from the product of  $u_1(x)$ (by definition of $\hat{u}_1$) by $u_1(x)/Q$ (by definition of $\hat{z}_1$). Physically, this quadratic dependence corresponds to the vertical shear being enhanced not only by the magnitude of $u_1$, but also by the enhanced vertical gradient due to the thinner layers where $u_1$ is larger. This $u_1^2(x)/Q$ scaling will be found in the interfacial stress $\tau^I$ too. However, the equivalent formulation to \eqref{tau1-Z} for the side wall stress in layer 1, $\tau_{1}^Y$, is
\begin{equation} \label{tau1-Y}
\tau_{1}^Y(x) = \frac{1}{Re} u_1(x) \Big\langle \Big|\frac{\p \hat{u}_1(x,y,z)}{\p y}\Big|_{y=\pm 1} \Big\rangle_{z_1}, 
\end{equation}
where we assume identical shear at $y=\pm 1$. We emphasise that since the $y$ derivative does not experience any rescaling due to the layer thickness, it follows a $u_1(x)$ scaling (as opposed to $u^2_1(x)/Q$ for $z$ derivatives).

\item \ Fourth, we generalise the above definitions of $\hat{u}_1$ and $\hat{z}_1$ to both layers by defining a global $\hat{u}$ as
\begin{equation}
\hat{u}(x,y,z) \coloneqq \left\{
\begin{array}{l}
 \dfrac{u(x,y,z)}{u_1(x)}  \quad \textrm{for} \quad z \in [-1,\eta], \\ 
\vphantom{.}\\ 
 \dfrac{u(x,y,z)}{u_2(x)}  \quad \textrm{for} \quad z \in [\eta,1], 
\end{array}
\right.
\end{equation}
and a global $\hat{z}$ as
\begin{equation}
\hat{z} \coloneqq \left\{
\begin{array}{l}
 \dfrac{z}{1+\eta}= z \dfrac{Q}{u_1(x)} \quad \textrm{for} \quad z \in [-1,\eta], \\ 
\vphantom{.}\\ 
 \dfrac{z}{1-\eta} =  z \dfrac{Q}{u_2(x)} \quad \textrm{for} \quad z \in [\eta,1] .
\end{array}
\right.
\end{equation}

\item \ Fifth, we consider the role of turbulence at the interface, caused by Reynolds stresses which we parameterise, by analogy with \eqref{tau1-z-1}, as follows
\begin{equation} \label{definition-Km}
\langle -\hat{u}'w' \rangle_{x,y,z_I,t} = \frac{1}{Re} K_I \Big\langle \frac{\p \langle \hat{u} \rangle_{xyt}}{\p \hat{z}} \Big\rangle_{z_I},
\end{equation}
where $\uu' \equiv \uu - \langle \uu \rangle_t$ is the perturbation around the temporal mean and $K_I$ the turbulent momentum diffusivity non-dimensionalised by the molecular value $\nu$.
Under these conditions, the total (molecular $+$ turbulent) interfacial  stress $\tau^I$ can be expressed precisely as:
\begin{equation}
\tau^I(x) = \frac{1+K_I}{Re}  \frac{(u_1(x)-u_2(x))^2}{Q} \Big\langle \Big| \frac{\p \hat{u}(x,y,\hat{z})}{\p \hat{z}}\Big| \Big\rangle_{y,\hat{z}_I},
\end{equation}
where $\hat{z}_I$ denotes averaging over the interfacial mixed layer. 

\end{enumerate}

Based on the five above steps, we propose the following  parameterisation of frictional effects in the hydraulic model
\begin{subeqnarray}\label{definition_fzfyfi}
\tau_{1,2}^Z(x)   & = &  \frac{f_Z}{Re}  \ \frac{u_{1,2}^2(x)}{Q}, \slabel{def-fz}\\
\tau_{1,2}^Y(x)   & = &   \frac{f_Y}{Re} \ |u_{1,2}(x)|, \slabel{def-fy} \\
\tau_{1}^I (x) = \tau_{2}^I(x)  & = &   \frac{f_I}{Re} \ \frac{(u_{1}(x)-u_2(x))^2}{Q}.\slabel{def-fi}
\end{subeqnarray}
where the vertical, spanwise and interfacial friction parameters are, respectively,
\begin{subeqnarray}\label{fzfyfi-2}
f_Z &\equiv &  \Big\langle \Big| \frac{\p \hat{u}(x,y,\hat{z})}{\p \hat{z}}\Big|_{\hat{z}=\pm 1} \Big\rangle_{xy}, \\
f_Y &\equiv &  \Big\langle \Big| \frac{\p \hat{u}(x,y,\hat{z})}{\p y}\Big|_{y\pm 1} \Big\rangle_{xz}, \\
f_I &\equiv & (1+K_I) \Big\langle \Big| \frac{\p \hat{u}(x,y,\hat{z})}{\p \hat{z}}\Big| \Big\rangle_{x,y,z_I}.
\end{subeqnarray}
The $y$ and $\hat{z}$ derivatives at $y,\hat{z} =\pm 1$  should be very similar, and the average of the two is implied. The three parameters can be computed from three-dimensional, three-component velocity measurements, as was done in L18, \S~5.5.

\subsection{Key equations and solution method} \label{sec:appendix-frictional-3}

We can now rewrite the frictional slope $S(x)$ in \eqref{stangamma} using  \eqref{fzfyfi-2} and \eqref{cons_volume} as 
\begin{equation} \label{S-tangamma}
Re \, S(x) = \frac{2Qf_Z}{(1-\eta^2)^3}  \Big\{ (1+3\eta^2) + 2\frac{f_Y}{f_Z}B^{-1} (1-\eta^2)^2 + 8\frac{f_I}{f_Z} \Big\} .
\end{equation}
By combining this expression for $S(x)$ with the expression for the composite Froude number $G^2(x)$ in \eqref{non-dim-composite} we finally obtain an expression for the differential equation governing the evolution of the interfacial slope $\eta'(x)$ in \eqref{dEdx-friction}
\begin{equation} \label{deta/dx}
\eta' (x) = \frac{ \theta Re (1-\eta(x)^2)^3  - 2 Q f_Z \{ 1+3\eta^2(x) + 2r_Y(1-\eta^2(x))^2 + 8 r_I \}}{Re \{ (1-\eta(x))^2)^3 - 4Q^2(1+3\eta^2(x)) \} }, 
\end{equation}
where  the spanwise friction ratio $r_Y$ and interfacial friction ratio $r_I$  are defined under \eqref{existence-condition}. This equation was simplified to \eqref{eta-ODE} for the discussion in \S~\ref{sec:model-fric}.

The idea behind the solution to this kind of problem can essentially be found in \cite{gu_analytical_2005}. However, contrary to their model (which had no tilt angle and no top and side wall friction $\theta=f_{Z_1}=f_Y=0$), our model does not allow us to find an analytical solution to \eqref{deta/dx}. We must therefore resort to an iterative numerical approach which we briefly outline below.

By symmetry of the problem (guaranteed under the Boussinesq approximation), $\eta$ is an odd function of $x$. We impose the boundary condition $\eta(0)=0$ and need only solve \eqref{deta/dx} in half of the domain (say $x \in [0,A]$).  

However, since  the volume flux $Q$ in \eqref{deta/dx} is a priori unknown, we must solve a coupled problem imposing the additional condition of hydraulic control at each duct end (denoted by the superscript $^*$)
\begin{equation} \label{G*2}
G^{*2} \equiv G^2(-\eta^*) = 4Q^2 \frac{1+3\eta^{*2}}{(1-\eta^{*2})^3} =1 \quad \Longrightarrow  \quad Q = \frac{1}{2} \sqrt{\frac{(1-\eta^{*2})^3}{1+3\eta^{*2}}},
\end{equation}
where $\eta^*$ is the result of the forward integration of \eqref{deta/dx}
\begin{equation} \label{definition-eta^*}
\eta^* \equiv \eta(-A) = -\eta(A) = - \int^A_0 \eta'(Q,\theta,f_Z,r_Y,r_I) \, \d x > 0.
\end{equation}

The coupled problem for $\eta(x)$ and $Q$ for any given set of forcing and friction parameters $(\theta,Re,f_Z,r_Y,r_I)$ can then be solved by the following iterative algorithm (illustrated in L18, figure~5.4).
\begin{myenumi}
\item[\textnormal{(i)}\hspace{2.6ex}] Guess $Q$;
\item[\textnormal{(ii)}\hspace{2.2ex}] Integrate numerically  \eqref{deta/dx} from $x=0$ to $A$ to get $\eta^*$ as in \eqref{definition-eta^*}; 
\item[\textnormal{(iii)}\hspace{1.7ex}] Get the $Q$ corresponding to this $\eta^*$ by the criticality condition \eqref{G*2};
\item[\textnormal{(iv)}\hspace{1.8ex}] Compare this $Q$ with the initial guess and update the guess;
\item[\textnormal{(v)}\hspace{2.4ex}] Repeat until convergence of $Q$.
\end{myenumi}
This model and its solution were validated using parameters $(\theta,Re,f_Z,r_Y,r_I)$ from an experiment in the $\LL$ regime, and quantitative agreement with $\eta(x)$ and $Q$ measurements was found L18, \S~5.5.3.

}

\bibliographystyle{jfm}
\bibliography{AL_references_2019_02.bib}

\clearpage

\tableofcontents

\end{document}